\documentclass[a4paper, 11pt, twoside]{report}
\usepackage{psfig}
\newcommand{\be}{\begin{equation}}
\newcommand{\ee}{\end{equation}}

\newcommand{\bea}{\begin{eqnarray}}
\newcommand{\eea}{\end{eqnarray}}

\newcommand{\befig}{\begin{figure}}
\newcommand{\efig}{\end{figure}}

\newcommand{\inv}{\frac{1}}

\newcommand{\fm}{\mbox{fm}}
\newcommand{\MeV}{\mbox{MeV}}
\newcommand{\GeV}{\mbox{GeV}}

\begin{document}
%
%
% ___ The Title Page ________________________________________________
%
\pagestyle{empty}
\vspace*{-1.8cm}
\begin{Huge}
\begin{center}
{\bf Bremsstrahlung\\
out of the\\
Quark-Gluon Plasma}\\
\end{center}
\end{Huge}
\vspace*{2.cm}
\begin{Large}
\begin{center}
Diplomarbeit\\
\vspace*{3.cm}
vorgelegt von\\
Frank Daniel Steffen\\
aus Wattenscheid
\end{center}
\vspace*{1.5cm}
\begin{center}
Institut f\"ur Theoretische Physik \\
der Justus-Liebig-Universit\"at Giessen
\end{center}
\vspace*{1.5cm}
\centerline{Giessen, 1999}
\end{Large}

\cleardoublepage
%
%
% ___ The Abstract __________________________________________________
%
\enlargethispage*{100cm}
\begin{abstract}
A systematic investigation of hard thermal photon spectra from central ultra-relativistic heavy ion collisions is presented with emphasis on the effects of bremsstrahlung processes in the quark-gluon plasma (QGP). Bremsstrahlung photon production in the quark-gluon plasma has recently been considered within the Braaten-Pisarski method in thermal QCD, where rates have been found that exhibit the same order in the coupling constants as those describing the lowest order processes, Compton scattering and $q\bar{q}$-annihilation. The impact of these bremsstrahlung photon production rates on the thermal photon spectra is studied systematically within a simple, well understood one-fluid hydrodynamical model that describes an only longitudinally expanding fireball (1+1 Bjorken scaling hydrodynamics). A first-order phase transition is implemented in which QGP (simulated by an ideal massless parton gas of two-flavors) ``hadronizes'' according to the Gibbs criteria and Maxwell construction into a hot hadronic gas (HHG) (simulated by an ideal massless pion gas). It is found that the bremsstrahlung processes enhance the thermal photon yield from the QGP by about one order of magnitude over the complete considered $p_{\perp}$-range independent of the choice of the model parameters. This results in an enhancement of the total thermal photon yield which is most significant for parameter sets that support a highly contributing QGP phase. The influence of each model parameter on the thermal photon spectra is examined carefully and a thorough understanding of the model is obtained. Experimental upper limits on direct photon production in fixed target $200\;A\cdot\GeV$ $S + Au$ collisions at the CERN SPS are also considered and used to extract upper limits for the initial temperature of the QGP, where the QGP bremsstrahlung processes are found to make a difference of about 15 to 20 MeV depending on the temperature at which the phase transition is assumed.
%again the impact of the QGP bremsstrahlung processes is checked.
%which turns out to be the crucial model parameter. The difference between the maximum values found with and without the QGP bremsstrahlung processes again illustrates the impact of the QGP bremsstrahlung processes. 
In comparison with other theoretical studies, the importance of reaction features not described in the simple model are estimated and interesting elements for a future extension of this systematic investigation are identified, which will be of great interest in prospect of the upcoming experiments at the BNL RHIC and the CERN LHC.
\end{abstract}

\cleardoublepage
%
%
% ___ The Dedication ________________________________________________
%
%
% dedication.tex
%
\vspace*{3.in}
\begin{Large}
\begin{center}
To Natascha
\end{center}
\end{Large}

\cleardoublepage
%
%
% ___ Table of Contents _____________________________________________
%
\pagenumbering{roman}
\pagestyle{plain}
\tableofcontents
\cleardoublepage
%
%\addcontentsline{toc}{chapter}{List of Figures}
%\listoffigures
%\cleardoublepage
%
%\addcontentsline{toc}{chapter}{List of Tables}
%\listoftables
%\cleardoublepage
%
%
% ___ The Chapters __________________________________________________
%
\pagenumbering{arabic}
\pagestyle{plain}
%
%
%
%
%----------------------------------------------------------
\chapter{Introduction}
\label{Introduction}
%----------------------------------------------------------
%
%
This thesis presents a {\em systematic investigation of thermal photons} produced in ultra-relativistic heavy ion experiments. The primary goal of these experiments is the discovery of the quark-gluon plasma which is the deconfined state of strongly interacting matter predicted by quantum chromodynamics. If nature allows the existence of this state, physicists face two challenges. First, they have to produce the quark-gluon plasma, and second, they must clearly identify that it indeed has been produced. For the first task high energy nucleus-nucleus collisions are the ideal means, for the second task many signals have been proposed. In this thesis the quality of thermal photons as a potential signature for the quark-gluon plasma will be examined theoretically. A systematic study of thermal photon emission will deliver insights into the space-time development of the fireball formed in ultra-relativistic heavy ion collisions. Within a simple hydrodynamical model for the fireball evolution the effects of the {\em most recent thermal photon rates} on the photon spectra will be analyzed. Of course, experimental photon data will also be inspected and compared with theoretical results. Due to decisive experiments underway the thermal photon investigation illustrated in this thesis will be of interest at least for the next decade.
%In the research connected with this thesis, the quality of thermal photons as a potential signature for the quark-gluon plasma was examined theoretically. A systematic study of thermal photon emission delivered insights on the space-time development of the fireball formed in ultra-relativistic heavy ion collisions. Within a simple hydrodynamical model for the fireball evolution, the effects of the most recent thermal photon rates on the photon spectra were analyzed. Of course, also experimental photon data was inspected and compared with theoretical results. Because decisive experiments are underway, the thermal photon investigation illustrated in this thesis will be of interest at least for the next decade.

This chapter will provide the motivation for the systematic investigation of thermal photon production in ultra-relativistic nucleus-nucleus reactions, and it will give background information on heavy ion physics at ultra-relativistic energies. In Sec.~\ref{Quantum_Chromodynamics}, the accepted theory of strong interactions, quantum chromodynamics, will be reviewed. This theory predicts quark-gluon plasma as the state of nuclear matter at high temperatures and high densities. Section~\ref{The_Quark-Gluon_Plasma} will describe this state of matter and its presumed appearance in nature. In the subsequent section, Sec.~1.3, the pursued production of quark-gluon plasma in the laboratory will be discussed. Then Sec.~\ref{Signatures_of_the_Quark-Gluon_Plasma} will center on proposed signatures for the quark-gluon plasma. Because the focus of this thesis is on photons, the emphasis will be on distinguishing properties of electromagnetical probes. The experimental situation, present and future perspectives, will be the topic of Sec.~\ref{The_Experimental_Situation}. Based on the background information gathered in these preceding sections, the motivation for the systematic study of thermal photons will be addressed in Sec.~\ref{Motivation_for_a_Systematic_Investigation_of_Thermal_Photons}. An overview of this thesis will conclude Chap.~\ref{Introduction}.
%
%
%----------------------------------------------------------
\section{Quantum Chromodynamics}
\label{Quantum_Chromodynamics}
%----------------------------------------------------------
%
%
In our present understanding of nature, {\em quantum chromodynamics} ({\em QCD}) is the theory of strong interactions~\cite{QCD}. It is a non-Abelian gauge theory which is based on the SU(3) color gauge group. Thus, the fundamental principle is local gauge invariance under SU(3)-transformations. For keeping up this symmetry, gauge fields are crucial. The gauge quanta of QCD are the {\em gluons}, massless bosons of spin one. Gluons are exchanged between particles that carry {\em color}, the quantum number of QCD. Such gluon sources are the {\em quarks}, massive particles of spin one-half. Today, six different quark types or ``flavors'' have been observed, the up (u), charm (c), and top (t) quarks having electrical charge $+2e/3$, and the down (d), strange (s), and beauty (b) quarks having electrical charge $-e/3$. Every quark, independent of its flavor, comes in one of three colors, e.g., red, blue, or green. However, not only quarks can emit and absorb gluons, the gluons themselves carry one color and one anti-color. Because a gluon cannot be color-neutral, there are as many as eight gluons with different color ``charge.'' Consequently, the gauge fields of QCD interact among themselves. It is this distinguishing property that made QCD the accepted theory of strong interactions. 

There are two phenomena characteristic for the strong force: {\em asymptotic freedom} and {\em confinement}. High energy deep inelastic scattering experiments revealed asymptotic freedom: at close distances, quarks behave like free particles. QCD can derive this observation from first principles. As long as there are no more than 16 flavors, QCD is asymptotically free: the coupling constant becomes weak at high energy. This decrease of the strong coupling constant at high energy implies an increase at low energy. It is a hypothesis inferred from these considerations, that quarks or gluons cannot be observed as isolated particles. This hypothesis, called confinement, matches the experimental fact that neither an isolated quark nor an isolated gluon have ever been detected. Only color-neutral particles can be found as isolated objects in our physical world. 

Baryons and mesons are color-neutral particles made up of quarks and gluons. In a baryon, the combination of three constituent quarks, each with a different color, leads to a color-neutral state. In a meson, the color-neutral state is realized in the combination of a colored quark and an anti-quark bearing the corresponding anti-color. Of course, one can think of various other combinations that lead to color-neutral objects, such as glueballs or mesonic molecules. These exotics have not been detected clearly, however, their observation would confirm QCD as the theory of strong interactions, a fact that motivates an area of current research in nuclear physics~\cite{GSI_1998}.
%
%
%
%----------------------------------------------------------
\section{The Quark-Gluon Plasma}
\label{The_Quark-Gluon_Plasma}
%----------------------------------------------------------
%
%
There is another exotic state of matter predicted by QCD, the {\em quark-gluon plasma} ({\em QGP}) \cite{MUELLER_1985,HWA_1990,HWA_1995,QUARK_MATTER}. According to QCD, the strong coupling constant decreases at high energy. Thus, in an environment of extremely high energy density, quarks and gluons are expected to form a relativistic weakly-interacting parton gas: a QGP. The search for QGP is one of the central topics in strong interaction physics, and also this thesis aims at reviewing the QGP formation in heavy ion collisions.

The exploration of the QGP gains much attention because it is considered a key to various fundamental questions. For example, the standard model of cosmology assumes the existence of a QGP phase in the early universe. It is believed that this color deconfined state of the universe underwent a phase transition about 10$^{-6}$ seconds after the big-bang, where the quarks and gluons became confined in baryons and mesons. This process, called {\em hadronization}, is directly connected to confinement and far from being understood. It is manifested in high-energy physics experiments as jet formation, but can only be reproduced in simplified theoretical models~\cite{TRAXLER_1999}. In the universe we see today, QGP could still be present, since supernovae and neutron stars provide extreme astrophysical environments which favor the creation and existence of QGP~\cite{SCHERTLER_1998}.

\begin{figure}
        \centerline{\psfig{figure=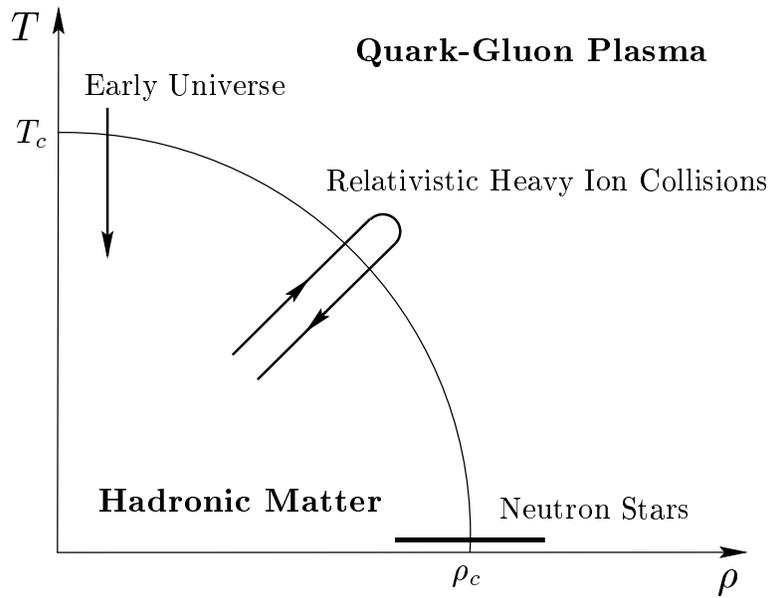,clip=,width=4.in}}
        \caption[Phase Diagram of Nuclear Matter]{Phase Diagram of Nuclear Matter. At high temperatures~$T$ and high densities~$\rho$, nuclear matter is expected to be in the QGP state. By lowering the temperature and the density, one should cross the critical boundary where quarks and gluons become confined into hadrons. Below the phase transition values of temperature, $T_c$, and density, $\rho_c$, we experience nuclear matter in its hadronic state.}
\label{Fig_Phase_Diagram_of_Nuclear_Matter}
\end{figure}

Presently, nuclear matter under extreme conditions can be produced and studied in heavy ion collisions~\cite{WONG_1994,CSERNAI_1994}. These experiments are ideal means to compress and heat up nuclear matter in the laboratory, where they give a unique tool to determine the nuclear equation of state. If the QCD predictions are valid, this nuclear equation of state should contain a deconfinement phase transition to QGP at high temperatures and densities as is illustrated in Fig.~\ref{Fig_Phase_Diagram_of_Nuclear_Matter}. In fact, physicists are confident to produce QGP in ultra-relativistic heavy ion collisions if nature allows its existence. 
%
%
%
%----------------------------------------------------------
\section{Ultra-Relativistic Heavy Ion Collisions}
\label{Ultra-Relativistic_Heavy_Ion_Collisions}
\begin{figure}
        \centerline{\psfig{figure=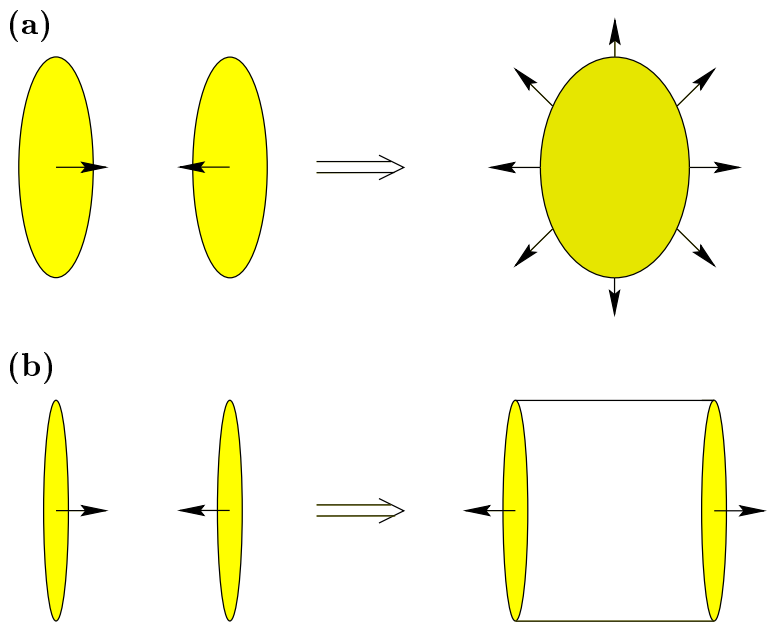,clip=,width=4.in}}
        \caption[Ultra-Relativistic Heavy Ion Collisions]{Ultra-Relativistic Heavy Ion Collisions. According to the center-of-mass energy $\sqrt{s}$, one distinguishes the stopping (a) and the transparent region (b). The two scenarios shown are, of course, only the limiting cases. For example, in high-energy SPS heavy ion experiments a scenario in between these pictured seems realistic.}
\label{Fig_Ultra-Relativistic_Heavy_Ion_Collisions}
\end{figure}

%----------------------------------------------------------
%
%
Ultra-relativistic heavy ion collisions are performed at the Brookhaven National Laboratory (BNL) and at the European Center for Nuclear Research (CERN), where heavy ions are accelerated in high-energy proton accelerators before they are aimed at fixed targets. In this way the BNL Alternating Gradient Synchrotron (AGS) has provided $^{28}Si - ^{197}\!\! Au$ and $^{197}\!Au - ^{197}\!\! Au$ collisions with a center-of-mass energy of up to $\sqrt{s} = 5\;A\cdot\GeV$. At CERN, the Super Proton Synchrotron (SPS) is still producing ultra-relativistic heavy ion collisions in fixed-target experiments. center-of-mass energies of up to $\sqrt{s} = 20\;A\cdot\GeV$ are reached for $^{32}S - ^{208}\!Au$ and $^{208}Pb - ^{208}\!\! Pb$ reactions. In the experiments at AGS and the medium-energy experiments at SPS, a complete stopping of the baryonic projectile constituents in the middle of the reaction zone was observed. The production of a short-lived QGP is expected because the significant overlap of the baryons presumably causes a screening of the color confining potential. Experiments in this energy region, the {\em stopping region}, are considered ideal for delivering insights into the nature of neutron and hybrid stars for which astrophysicists predict similar baryon dense conditions. A schematical drawing of a central heavy ion collision in the stopping region is shown in Fig.~\ref{Fig_Ultra-Relativistic_Heavy_Ion_Collisions}~(a).

A different energy region, the {\em transparent region}, is possibly already considered in the high-energy nucleus-nucleus collisions at SPS, and it will definitely be studied in the next generation of ultra-relativistic heavy ion experiments. By applying the collider principle, the Relativistic Heavy Ion Collider (RHIC) at BNL will achieve $\sqrt{s} = 200\;A\cdot\GeV$ with $^{197}\!Au$~projectiles. At the moment, this dedicated heavy ion accelerator is in its testing phase. In the year 2005, even higher energies will be available with the completion of the Large Hadron Collider (LHC) at CERN. This device will collide $^{208}Pb$~projectiles up to center-of-mass energies of $\sqrt{s} = 5500\;A\cdot\GeV$. Because of the high energies pursued in these upcoming experiments, the accelerated heavy ions will suffer extreme Lorentz contraction, therefore, the overlap of the projectiles will not last long enough to stop the nuclei significantly. The heavy ions will instead be approximately transparent keeping much of their initial energy~\cite{BJORKEN_1983,GYULASSY_1984}. However, there will be a strong color field between the emerging baryons of the initial projectiles, which will polarize the vacuum and cause parton pair production. In this transparent energy region, physicists expect the production of a quasi baryon free QGP. As the QGP phase of our early universe was also baryon free, heavy ion reactions at these energies should help in understanding the QGP phase of our early universe. Figure~\ref{Fig_Ultra-Relativistic_Heavy_Ion_Collisions}~(b) illustrates a central ultra-relativistic nucleus-nucleus collision in the transparent region.

This thesis concentrates only on transparent heavy ion reactions because the central region of the longitudinally expanding fireball is ideally suited for the study of thermal photon yields. A significant stopping of the baryons of the initial projectiles and consequently a finite baryon density in the central region is already expected in SPS high-energy nucleus-nucleus collisions. However, for simplicity, we also treat high-energy SPS heavy ion experiments as being in the transparent region and keep in mind that this might only be a fair approximation.

\begin{figure}
        \centerline{\psfig{figure=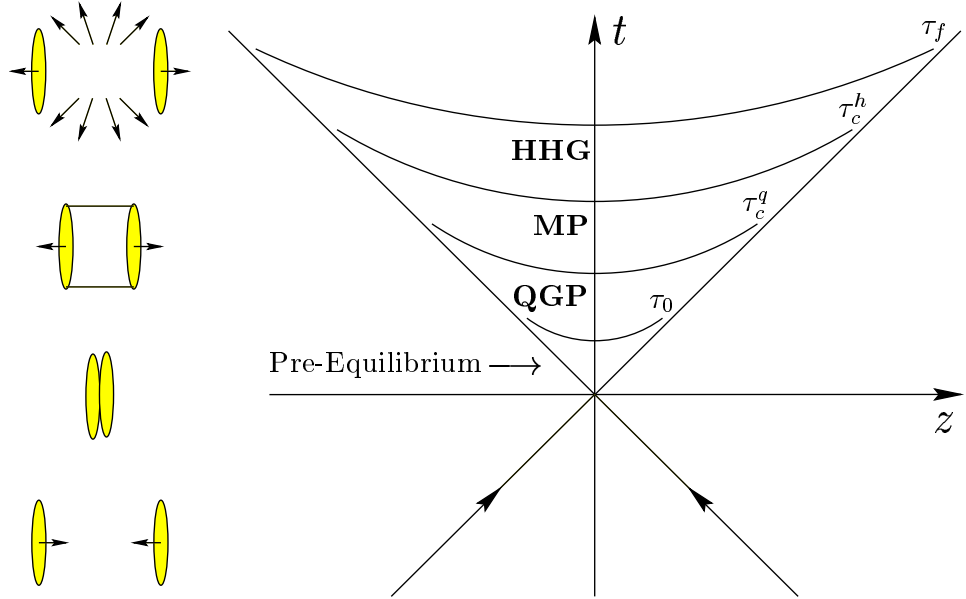,clip=,width=5.in}}
        \caption[Space-Time Evolution of an Ultra-Relativistic Heavy Ion Collision]{Space-Time Evolution of an Ultra-Relativistic Heavy Ion Collision in the Transparent Region. The Minkowski-diagram in the right half of the figure drafts the fireball behavior along the beam-axis labeled by~$z$  as seen in center-of-mass time~$t$. The left half of the figure sketches the spatial evolution of the heavy ion reaction on the same time-scale. The projectiles are almost light-like moving practically on the light-cone. At the point of maximum overlap, $(t,z) = (0,0)$, the initial nuclei are hardly stopped. Then, a strongly interacting continuum is produced by the strong color field of the receding projectile fragments. The hyperbolas in the upper time-like part of the diagram characterize domains of constant proper time. These domains mark the starting and ending points of the different collision phases as stated in the text.}
\label{Fig_Space-Time_Evolution}
\end{figure}

A common picture for the space-time evolution of a transparent ultra-relativistic heavy ion collision~\cite{SATZ_1994} is illustrated in Fig.~\ref{Fig_Space-Time_Evolution}. The maximum overlap of the colliding nuclei, which are highly Lorentz contracted, defines proper time $\tau = 0$~fm in the considered system. By multiple scatterings among the initial partons produced through vacuum polarization, the system goes into thermal equilibrium at {\em initial time}~$\tau_0$. In the transparent energy region, the {\em initial temperature}~T$_0$ is expected to be sufficiently high for supporting the existence of a QGP. As the nuclei emerge, the system expands and cools. At time~$\tau_c^q$ the {\em transition temperature}~T$_c$ is reached, where the phase transition sets in. Assuming a first-order phase transition, a mixed phase (MP) of constant temperature T$_c$ follows, in which quarks and gluons become confined into hadrons. The system cools further when all the quark matter has transformed into hadronic matter. This completion of the phase transition takes place at $\tau_c^h$. The produced hadrons do still interact among themselves forming a hot hadronic gas (HHG). When the system arrives at the {\em freeze-out temperature}~T$_f$, the hadrons will stream as free particles out of the collision zone.
%
%
%
%----------------------------------------------------------
\section{Signatures of the Quark-Gluon Plasma}
\label{Signatures_of_the_Quark-Gluon_Plasma}
%----------------------------------------------------------
%
%
Provided nature allows the existence of QGP, the challenge of ultra-relativistic heavy ion experiments will not only be the production but also the clean identification of this deconfined state. An ideal signature would be an observable physical phenomenon that can only be explained by assuming the production of a QGP phase. This signature should not be understandable in any realistic model not embedding the deconfinement phase transition. Thus, potential signals need to be examined in both phase transition and no phase transition scenarios.

Many phenomena have been proposed as signals for QGP formation~\cite{MUELLER_1992}. They can be grouped into hadronic and electromagnetic observables. Hadronic signatures are, e.g., {\em strangeness enhancement}, {\em J/$\psi$-suppression}, and {\em detection of strangelets}, which are exotic objects that contain several strange quarks. Also, thermodynamic signals can be considered hadronic because thermodynamic variables, such as temperature~$T$, entropy density~$s$, and energy density~$\varepsilon$, are inferred respectively from average transverse momentum, multiplicity and transverse energy distributions of hadrons. In general, hadronic observables can be measured easily due to the copious production of hadrons in ultra-relativistic heavy ion reactions. However, with the exception of jets, hadrons cannot carry direct information about the early collision stage because they interact until the system undergoes freeze-out, as mentioned in the previous section. Therefore, observed hadrons carry primarily information on the freeze-out stage of the collision. For electromagnetic observables, {\em photons} and {\em dileptons}, the situation looks different~\cite{RUUSKANEN_1992}. Since electromagnetic mean free paths are much larger than the transverse size of the fireball, most photons and dileptons produced in the reaction reach the detector with no final state interaction. In this way even the earliest and hottest phase of the collision can be probed. However, in experiments electromagnetic signals are hard to detect. The overall rate of electromagnetic probes is small and needs to be extracted out of large backgrounds from hadronic decay processes. Nevertheless, we concentrate on electromagnetic signals: employing a well-understood model for the space-time evolution of the fireball, we investigate systematically the production of {\em thermal photons} that have momenta above the fireball temperature, $p_{\gamma} > T$.
%
%
%
%----------------------------------------------------------
\section{The Experimental Situation}
\label{The_Experimental_Situation}
%----------------------------------------------------------
%
%
Most important within a systematic theoretical study is a comparison with experimental results. For thermal photon investigations experiments provide direct photon yields. {\em Direct photons} are those produced inside the strongly interacting continuum present before freeze-out. In their extraction, a clean separation of decay photons is crucial. These photons mainly originating from Dalitz decays $\pi^0 \rightarrow \gamma\gamma$ and $\eta \rightarrow \gamma\gamma$ dominate the observed photon spectra. The measurement of direct photon spectra was a main goal in the experiment WA80 at the CERN SPS. In fact, the only reliable direct photon data has been published by the WA80 collaboration~\cite{WA80_1996,CERES_1996_KAMPERT_1997}, where upper limits on the direct photon production in $200\;A\cdot\GeV$ $S + Au$ collisions were determined on a statistical basis. At present, the WA80 experiment has been upgraded to WA98, which is examining $158\;A\cdot\GeV$ $Pb + Pb$ collisions at the CERN SPS. Because of the larger projectiles higher yields not only of direct photons but unfortunately also of resonance decay photons are expected. However, with the sophisticated WA98 photon spectrometer LEDA (LEadglass Detector Array) the WA98 results for direct photon production should be more precise than those from WA80. Today, the results of the WA98 direct photon analysis are eagerly awaited and will be published in the very near future. For results from RHIC and LHC, we have to be patient. PHENIX, one of the two large-scale detectors at RHIC, will start data taking in fall 1999. With high granularity and excellent particle identification capabilities, it will be well suited to extract the direct photon signal in the high-multiplicity environment predicted for the RHIC experiments~\cite{PHENIX_1998}, but it might take about five years until the PHENIX collaboration can provide more than preliminary direct photon production data. Then, in 2005, the CERN LHC will start operating. At this accelerator, there will be ALICE (A Large Ion Collider Experiment), a dedicated high energy heavy ion physics experiment. ALICE will be equipped with PHOS, a state-of-the-art electromagnetic calorimeter designed for photon physics in high multiplicity reactions~\cite{ALICE_1995_96}. With the experience collected in WA80, WA98, and PHENIX, ALICE will give presumably the cleanest extraction of direct photons, yet we will need to wait another decade for these results.
%
%
%
%----------------------------------------------------------
\section{Motivation for a Systematic Investigation of Thermal Photons}
\label{Motivation_for_a_Systematic_Investigation_of_Thermal_Photons}
%----------------------------------------------------------
%
%
Our research ultimately aims at the discovery of the QGP and the connected confirmation of QCD by considering hard thermal photon production in ultra-relativistic heavy ion collisions. For a theoretical prediction of the photon spectra measured experimentally, cross sections of the elementary photon emitting reactions, or photon production rates, need to be calculated and integrated over the space-time history of the fireball. Recently, thermal photon rates have been revised for the QGP phase in the framework of finite temperature QCD~\cite{AURENCHE_1998}. An astonishing result has been found: because of medium effects, bremsstrahlung processes contribute in the same order of the coupling constant as the Compton scattering and $q\bar{q}$-annihilation processes calculated in~\cite{KAPUSTA_1991,BAIER_1991}. This leads to a significantly enhanced photon production rate. In fact, bremsstrahlung processes become the dominant source for hard thermal photons in the QGP phase. Earlier systematic investigations implemented for the QGP state only the Compton scattering and $q\bar{q}$-annihilation rates~\cite{ARBEX_1995,ALAM_1996,SOLLFRANK_1997,CLEYMANS_1997}, which means thermal photon emission of the QGP phase was underestimated. A further investigation with emphasis on the effects of the new rates seemed very necessary, and it was this consideration that triggered the research for this thesis. Because experiments cannot distinguish in which stage of the fireball a detected photon was emitted, thermal photons from the HHG phase were also included in our study. We extract these photon spectra employing the most recent parameterization of the production rate in hadronic matter~\cite{XIONG_1992,XIONG_1992_HAGLIN}. 
For the dynamics of the fireball, many models can be found and more are under construction. The variety reflects the ongoing discussions and uncertainties on the space-time evolution of an ultra-relativistic heavy ion collision. Since our prime interest is on the effect of the rates due to bremsstrahlung processes in comparison to the rates used before (Compton scattering, $q\bar{q}$-annihilation), we employ only a simple hydrodynamic model. The physics basis of this model is well understood and the effects of the rates will not be covered by fancy features of the reaction dynamics. However, the price for good understanding are the severe limitations one has to bear in mind, e.g., the model will not describe any transverse expansion or any non-equilibrium behavior. Within this model that has a decent number of parameters, we perform the systematic study on thermal photon spectra. Direct constraints on the parameter set will be inferred from WA80 photon data. More severe constraints are expected from comparisons with future experimental data of WA98, PHENIX, and ALICE. With these upcoming constraints we are confident that our systematic investigation will contribute in the verification of the QGP state.
%Further, we will outline qualitatively the effects of a more complicated fireball dynamics by comparison with other investigations.
%
%
%
%----------------------------------------------------------
\section{Overview of Thesis}
\label{Overview_of_Thesis}
%----------------------------------------------------------
%
The next chapter will describe the simple model we use for the space-time evolution of the fireball. 
%in our extraction of thermal photon spectra
A short review on relativistic hydrodynamics and Bjorken initial conditions will be given followed by a discussion of the nuclear matter equation of state.
%This chapter will also illustrate the two scenarios considered, a phase transition scenario and a no phase transition scenario. 
Chapter~\ref{Photons} will be dedicated to photon production in hot thermalized, strongly interacting environments. We will enumerate and investigate photon production processes embedded in our calculations. Having studied the current thermal photon rates, the results of our systematic investigation will be presented in Chap.~\ref{Systematic_Investigation:_Thermal_Photons}. This part will address separately the influence of each model parameter. Comparisons with other works and results from the analysis of experimental data on direct photon production will be given in Chap.~\ref{Analysis_of_Experimental_Data_on_Direct_Photon_Production}.

\cleardoublepage
%
%
%
%----------------------------------------------------------
\chapter{A Simple Model for Ultra-Relativistic Heavy Ion Collisions}
\label{A_Simple_Model_for_Ultra-Relativistic_Heavy_Ion_Collisions}
%----------------------------------------------------------
%
%
%
For a systematic investigation of thermal photons as a potential signature of QGP, one must convolute the rates with the space-time history of the nucleus-nucleus collision. The elementary photon rate, $E_{\gamma}\, dN/(d^4x\, d^3p_{\gamma})$, is the number of photons~$dN$ with energy~$E_{\gamma}$ emitted per unit volume per unit time within the three-momentum interval [$\vec{p}_{\gamma}$, $\vec{p}_{\gamma} + d^3p_{\gamma}$]. For thermal photons, this rate depends, of course, on the temperature~$T$ of the emitting space-time point,
\be
E_{\gamma}\, \frac{dN}{d^4x\, d^3p_{\gamma}} = E_{\gamma}\, \frac{dN}{d^4x\, d^3p_{\gamma}} \left( T(x) \right).
\ee
Since photons cannot be traced back experimentally to their origin in space-time, only energy and three-momentum are measured. Consequently, for a comparison with the detected photon spectrum, the rates need to be integrated over the space-time evolution of the heavy ion reaction,
\be
E_{\gamma}\, \frac{dN}{d^3p_{\gamma}} = \int d^4x\: E_{\gamma}\, \frac{dN}{d^4x\, d^3p_{\gamma}} \left( T(x) \right).
\label{integration_over_space-time_evolution}
\ee
The integration limits and the temperature field $T(x)$ are necessary for the evaluation of this four-dimensional integral. While some integration limits are determined by the accelerator and the projectile nuclei, others can be considered parameters. However, the temperature field~$T(x)$ should be delivered as a feature of some reaction model. We obtain this field within a simple hydrodynamical picture of an ultra-relativistic heavy ion collision that is described in this chapter.

We use relativistic one-fluid hydrodynamics because it provides a simple but, in our view, realistic framework to study thermal photon production in high-energy nucleus-nucleus reactions. It is our strategy to employ a model that can be well understood and maybe even solved analytically. Within this approach, we find very clearly the influences of the photon rates and the model parameters on the thermal photon spectrum, which will be presented in Chap.~4. The influence of the nuclear equation of state (EOS) can also be checked because it is a crucial input of a hydrodynamical calculation.

Our idea of a relativistic heavy ion reaction embedding a deconfinement phase transition was already discussed in Sec.~\ref{Ultra-Relativistic_Heavy_Ion_Collisions}. It is this picture we refer to whenever we talk of a phase transition scenario. On the presented space-time development, the use of a hydrodynamical simulation seems reasonable. A stringent condition for a system to be described by hydrodynamics is {\em local thermal equilibrium}. This allows a hydrodynamic simulation of the fireball evolution only between initial time~$\tau_0$ and freeze-out time~$\tau_f$. Non-equilibrium behavior present before $\tau_0$, also called thermalization time, cannot be modeled. In addition, the phase after freeze-out in which the particles are heading freely towards the detector is not suited for a hydrodynamical description. However, because the emphasis is on thermal photon spectra, no thermal photon yields are lost from these restrictions. Thermal photons can, of course, only be emitted from a thermalized system.

In the subsequent section, we introduce relativistic hydrodynamics. To perform a hydrodynamical calculation, important ingredients are necessary as the EOS and initial conditions. In the simulation, we use Bjorken initial conditions, which allow a scaling ansatz of the four-velocity field~$u^{\mu}(x)$ that describes the hydrodynamic flow. Together with a simple EOS, this scaling ansatz enables an analytical derivation of the temperature field $T(x)$. While Bjorken initial conditions are covered in Sec.~\ref{Bjorken_Initial_Conditions}, the EOS will be addressed in Sec.~\ref{The_Equation_of_State}. We complete the description of the employed hydrodynamical model by discussing two scenarios, one with and the other without a deconfinement phase transition.
%
%
%
%----------------------------------------------------------
\section{Relativistic Hydrodynamics}
\label{Relativistic_Hydrodynamics}
%----------------------------------------------------------
%
%
This section discusses the essentials of relativistic hydrodynamics~\cite{BLAIZOT_1990}. 
%For a more detailed presentation we refer the reader to the article of J. P. Blaizot and J. Y. Ollitrault, {\it Hydrodynamics of Quark-Gluon Plasmas}, published in~\cite{HWA_1990}. 
Within relativistic hydrodynamics, strongly interacting matter is considered a relativistic fluid. Because we neglect any dissipative effects as heat transfer or viscosity, it is even treated as a {\em perfect} or {\em ideal relativistic fluid}. The use of relativity is required because the particles in the collision system have velocities close to the velocity of light. There are two sorts of velocities considered in hydrodynamical calculations; collective (macroscopic) velocities and thermal (microscopic) velocities. Thermodynamic functions, such as {\em energy density}~$\varepsilon$ and {\em hydrostatic pressure}~$P$, characterize the thermal motion of the microscopic constituents of the continuum. Thus, they are a measure for the thermal velocities. The collective velocities describe the hydrodynamic flow in the form of the four-velocity field
\be
        u^{\mu} = \gamma \left( 1, \vec{v} \right)
\ee
with
\be
        \gamma = \inv{\sqrt{1 - \vec{v}\,^2}},
\ee
where $\vec{v}$ is the spatial flow velocity vector. By checking
\be
       u_{\mu}u^{\mu} = 1,
\ee
one can easily see that~$u^{\mu}$ is a time-like unit vector. We follow Landau's approach in attaching the hydrodynamic flow to the flow of energy~\cite{LANDAU_1959}. In other words,~$u^{\mu}$ is always tangential on the world lines of the energy flow. This defines the local rest frame~($LR$) of the fluid as the frame in which the energy flux vanishes. The collective velocity in this frame has the form
\be
        u^{\mu}_{(LR)} = \left( 1, 0, 0, 0\right).
\ee

\pagebreak
The hydrodynamic equations of motion are basic conservation laws. One is local conservation of energy-momentum
\be
        \partial_{\mu} T^{\mu \nu}(x) = 0. 
\label{energy_momentum_conservation}
\ee
In the above expression, $T^{\mu \nu}$ denotes the energy-momentum tensor, which can be written for perfect fluids as
\be
        T^{\mu \nu}(x) = \left[ \varepsilon(x) + P(x) \right] u^{\mu}(x) u^{\nu}(x) - P(x) g^{\mu \nu}
\ee
with the metric tensor $g^{\mu \nu} = \mbox{diag}(1, -1, -1, -1)$. Further, for any conserved scalar quantity, i.e. baryon number, a continuity equation holds locally for the corresponding volume density $\rho$,
\be
         \partial_{\mu} \left[ \rho (x) u^{\mu}(x) \right] = 0.
\label{scalar_quantity_conservation}
\ee

Another important conservation law can be derived by contracting Eq.~(\ref{energy_momentum_conservation}) with~$u_{\nu}(x)$. Applying the first law of thermodynamics on the contracted equation exhibits local conservation of entropy
\be
        \partial_{\mu} \left[ s(x) u^{\mu}(x) \right] = 0,
\label{entropy_conservation}
\ee
where $s$ denotes {\em entropy density}.

In order to solve the equations of motion, (\ref{energy_momentum_conservation}) and (\ref{scalar_quantity_conservation}), an additional equation of the form $P = P(\varepsilon, \rho)$, the EOS, is needed. Together with the EOS, the equations of motion form a closed system. However, before solving, initial conditions must be specified.
%
%
%
%----------------------------------------------------------
\section{Bjorken Initial Conditions - The Bjorken Model}
\label{Bjorken_Initial_Conditions}
%----------------------------------------------------------
%
%
In this thesis, energy regions are investigated in which nuclear transparency is present. One-fluid hydrodynamics cannot describe this ``leading baryon'' effect, and two-fluid~\cite{AMSDEN_1978} and three-fluid~\cite{CSERNAI_1982} models are too technical to fit our strategy. Therefore, sticking to the simple one-fluid model, the transparent behavior of the colliding nuclei has to be implemented in the initial conditions. Concentrating on the {\em central region} in which the baryon density is small and can be neglected and ignoring the fragmentation regions that contain most of the baryons close to initial velocity greatly facilitates the solution of the hydrodynamic equations.

Following Bjorken's approach~\cite{BJORKEN_1983}, we describe only the {\em longitudinal expansion} of systems produced in {\em central} ultra-relativistic heavy ion collisions. Any transverse expansion is neglected! Thus, we employ a 1~+~1~dimensional model considering an only longitudinally expanding tube of strongly interacting matter. Within this model the tube radius appears as an integration limit in Eq.~(\ref{integration_over_space-time_evolution}). Because only central collisions with zero impact parameter are studied, we take for this radius the projectile radius as given by the simple phenomenological formula
\be
        R_A = 1.3 \mbox{ fm } A^{1/3}.
\label{R_A}
\ee

Because ``fluid'' cells within the longitudinally expanding tube move with relativistic velocities, it is sensible to describe the fluid cells in the variables, {\em proper time} $\tau$, and {\em rapidity} $y$. For an expansion only parallel to the beam-direction, these variables have the following form,
\begin{eqnarray}
        \tau & = & \sqrt{t^2 - z^2}, \\
           y & = & \inv{2} \log \frac{t+z}{t-z},
\end{eqnarray}
where~$t$ and~$z$ are the center-of-mass frame coordinates for time and longitudinal position of the corresponding fluid cell, respectively. In this reference frame, the origin, $(t, x, y, z) = (0, 0, 0, 0)$, was chosen as the point where the two incident nuclei have maximum overlap. We showed already in Fig.~\ref{Fig_Space-Time_Evolution} curves of constant proper time as hyperbolas in a Minkowski-diagram illustrating the space-time evolution of a heavy ion collision. The rapidity~$y$ is the variable that indicates the position of the fluid cells on these hyperbolas. The proper time~$\tau$ is just the variable coinciding with the local time~$t_{(LR)}$ in the rest frame of the considered fluid element.

Of course, the initial conditions should also be given in these variables. Motivated by the observation of a ``central-plateau'' structure in nucleon-nucleon and nucleon-nucleus collisions, Bjorken asserted the initial conditions being invariant under Lorentz boosts along the beam-axis. We follow this assertion which means that the initial conditions imposed at initial proper time~$\tau_0$ do {\em not} depend on rapidity~$y$. Since the hydrodynamic equations are Lorentz covariant, this symmetry is preserved for the complete hydrodynamical evolution. More specific, thermodynamical quantities, as energy density~$\varepsilon$, pressure~$P$, temperature~$T$ or entropy density~$s$, remain independent of rapidity~$y$,
\be
        \varepsilon = \varepsilon(\tau), \;
        P = P(\tau), \;
        T = T(\tau), \;
        s = s(\tau).
\ee

In this picture of an only longitudinal expansion exhibiting the symmetry discussed above, a {\em scaling ansatz} for the four-velocity can be made,
\be
        u^{\mu} (t, z) = \inv{\tau} \left( t, 0, 0, z\right).
\label{scaling_ansatz}
\ee
With this ansatz and conservation of energy-momentum~(\ref{energy_momentum_conservation}), the basic differential equation of Bjorken's hydrodynamical model,
\be
        \frac{d\varepsilon}{d\tau} + \frac{\varepsilon + P}{\tau} = 0,
\label{Bjorken_model_equation}
\ee
can be derived. Further, by inserting~(\ref{scaling_ansatz}) into~(\ref{entropy_conservation}) the equation for entropy conservation gets the form
\be
        \frac{ds}{d\tau} + \frac{s}{\tau} = 0,        
\ee
or equivalently
\be
        \frac{d}{d\tau} \left( s\tau \right) = 0.
\label{expansion_time_scale_provider}
\ee
The two latter equations imply that the entropy per rapidity slice~$dS/dy$ is constant of the motion~\cite{BJORKEN_1983},
\be
        \frac{d}{d\tau} \left( \frac{dS}{dy} \right) = 0,
\label{entropy_per_rapidity_conservation}
\ee
which can be used to infer the initial entropy density, $s_0 = s(\tau_0)$, from the measured multiplicity distribution~$dN/dy$.
%
%
%----------------------------------------------------------
\section{The Equation of State}
\label{The_Equation_of_State}
%----------------------------------------------------------
%
As emphasized in the preceding sections, the EOS is an important ingredient in any hydrodynamic calculation. Together with the EOS the hydrodynamic equations become deterministic. For strongly interacting matter, the EOS should be derived directly from the QCD Lagrangian. Because of the nonperturbative character of QCD at large spatial distances, this task is far from being trivial and an analytical derivation of the EOS from first principles seems impossible. Instead, one performs computer simulations of QCD on a discrete lattice of space and time. These {\em lattice gauge calculations}, which are built on first principles, work in the nonperturbative regime of QCD. The EOS of strongly interacting matter can thus be inferred quantitatively from lattice QCD~\cite{LAERMANN_1996,UKAWA_1998}. In fact, this numerical approach provided a first quantitative prediction of the deconfinement phase transition~\cite{CREUTZ_1988}. However, we employ only a model EOS that displays features of lattice QCD results. We construct EOS's separately for both deconfined and confined matter. Assuming a {\em first-order phase transition}, we match the two EOS's by {\em Maxwell construction}. Within this approach of an idealized EOS, we again follow the philosophy of keeping the physics basis of our simulation well understood.

Because only the central region of ultra-relativistic heavy ion collisions is investigated, we neglect baryon density $\rho_B (x)$ completely and assume chemical equilibrium. By considering an only longitudinally expanding fireball, we are left with one hydrodynamical equation which is the basic differential equation of Bjorken's hydrodynamical model~(\ref{Bjorken_model_equation}). Consequently, for solving the hydrodynamical calculation, an EOS of the form $P = P(\varepsilon)$, or equivalently $\varepsilon = \varepsilon(P)$, needs to be specified. A bag model EOS for the QGP phase and an equally simple EOS for the hot hadronic gas (HHG) phase is presented. Finally, the issue of the phase transition, which we implement as a first-order transition, is addressed.

%Then, two EOS's for the hot hadronic gas (HHG) phase are discussed, a simple one and a more realistic one. Finally, we address the issue of the phase transition, which we implement as a first order transition by Maxwell construction.
%
%
%----------------------------------------------------------
\subsection{Quark-Gluon Plasma - The Ideal Massless Parton Gas}
%----------------------------------------------------------
%
%
A prerequisite for the formation of a QGP is an extremely high energy density, $\varepsilon > 1\;\GeV/\fm^3$. In this environment, the QCD coupling constant tends to zero, a phenomenon already mentioned as asymptotic freedom, and quarks and gluons form to a good approximation a noninteracting relativistic quantum gas. This can be seen in lattice QCD calculations, where the energy density follows the Stefan-Boltzmann law for temperatures higher than about $2 T_c$. Because in a QGP, light quarks, as u, d, and maybe s, would dominate, quark masses can be neglected and the description of the QGP as an {\em ideal massless parton gas} seems reasonable. In calculating the thermodynamic quantities of this ideal massless parton gas, one needs to regard the different statistics that quarks and gluons obey. While Fermi-Dirac statistics governs the thermodynamics of quarks, gluons follow Bose-Einstein statistics. The nature of the vacuum in which an ideal parton gas can exist must also be taken into account. This is simplest done by giving this perturbative QCD vacuum a constant energy density~$B$, known as {\em bag constant}. Typical values lie in the region around $B^{1/4}=200$~MeV. The bag constant appears also with a different sign in the pressure at the boundary of the QGP, where it describes phenomenologically confinement of the partons within the QGP ``bubble''~\cite{BAG_MODEL_REVIEWS}. Under the above considerations, the following 
%thermodynamic quantities 
Stefan-Boltzmann expressions can be derived for zero quark chemical potential, $\mu_q = 0$,
\begin{eqnarray}
                  P_q & = & g_q\, \frac{\pi^2}{90}\, T^4 - B,
\label{P_q} \\
        \varepsilon_q & = & g_q\, \frac{\pi^2}{30}\, T^4 + B,
\label{varepsilon_q} \\
                  s_q & = & g_q\, \frac{2\pi^2}{45}\, T^3,
\label{s_q}
\end{eqnarray} 
where QGP as the referred state of matter is indicated in the subscript~$q$. $g_q$~is the effective number of degrees of freedom. For QGP with $N_c$~colors and $N_f$~flavors,
\be
        g_q = 2\, (N_c^2 - 1) + \left( \frac{7}{8} \right) 4\, N_c\, N_f
\ee
and with the standard value $N_c = 3$, one gets $g_q = 37$ for a two-flavored and $g_q = 47.5$ for a three-flavored QGP. Finally, from the expressions for pressure~(\ref{P_q}) and energy density~(\ref{varepsilon_q}), one can directly read off the {\em bag model EOS}
\be
        \varepsilon_q = 3 P_q + 4 B.
\label{bag_model_EOS}
\ee
%
%
%----------------------------------------------------------
\subsection{Hot Hadronic Matter - The Ideal Massless Pion Gas}
%----------------------------------------------------------
%
%
While in the limit of high temperatures strongly interacting matter can be considered an ideal massless parton gas, an equally simple description can be realized for low temperatures. The picture of hadronic matter as an {\em ideal massless pion gas} should be a decent approximation at temperatures $T_c > T > m_{\pi}$. Taking into account finite hadron masses and also more massive hadrons than pions would definitively give a HHG EOS that is closer to reality. However, since finite mass states are Boltzmann-suppressed, the higher number of hadrons considered would be compensated to some extend. Due to this fact and the numerical demand necessary to get the EOS in this more realistic picture, we apply the ideal massless pion gas where the EOS can be derived analytically. This system governed by Bose-Einstein statistics, of course, also exhibits Stefan-Boltzmann expressions for the thermodynamic quantities,
\begin{eqnarray}
                  P_h & = & g_h\, \frac{\pi^2}{90}\, T^4,
\label{P_h} \\
        \varepsilon_h & = & g_h\, \frac{\pi^2}{30}\, T^4,
\label{varepsilon_h} \\
                  s_h & = & g_h\, \frac{2\pi^2}{45}\, T^3,
\label{s_h}
\end{eqnarray}
where the index $h$ specifies the considered state of matter as HHG and $g_h$ denotes the effective number of degrees of freedom. For a hadronic gas consisting of only pions,
\be
        g_h =3.
\ee
The equation of state embedded in Eq.~(\ref{P_h}) and~(\ref{varepsilon_h}) is the well known {\em ideal gas EOS} for {\em massless} particles,
\be
        \varepsilon_h = 3 P_h.
\label{ideal_gas_EOS}
\ee
%
%
%----------------------------------------------------------
\subsection{The Phase Transition}
%----------------------------------------------------------
%
%
The order of the deconfinement phase transition is still a matter of ongoing research~\cite{LAERMANN_1996}. However, we implement a {\em first-order phase transition} in our ``phase transition scenario'' that implies a mixed phase in which QGP and HHG coexist. During this mixed phase the ideal massless parton gas ``hadronizes'' continuously into the ideal massless pion gas at a hadronization rate that can be characterized in the volume fraction of the QGP,
\be
        \lambda(\tau) = \frac{V_q(\tau)}{V_{tot}(\tau)} = \frac{V_q(\tau)}{V_q(\tau) + V_h(\tau)},
\label{volume_fraction_of_QGP}
\ee
where~$V_q$ and~$V_h$ are the spatial volumes occupied by QGP and HHG, respectively, and $V_{tot}$~denotes the total spatial volume of the fireball. The quantity~$\lambda$ equals one during the pure QGP phase and at the onset of the hadronization, $\tau_0<\tau\leq\tau_q^c$, and equals zero during the pure HHG phase and at the offset of the hadronization, $\tau_h^c\leq\tau<\tau_f$
%, 
%\be
%        \lambda(\tau) = \left\{ \begin{array}{cc}
%                                1  & \mbox{for } \tau_0   <    \tau \leq \tau_q^c \\
%                                0  & \mbox{for } \tau_h^c \leq \tau <    \tau_f,
%                               \end{array}
%                       \right.
%\label{properties_of_lambda}
%\ee
%as is illustrated in Fig.~\ref{Fig_QGP_Volume_Fraction}
.

\begin{figure}
        \centerline{\psfig{figure=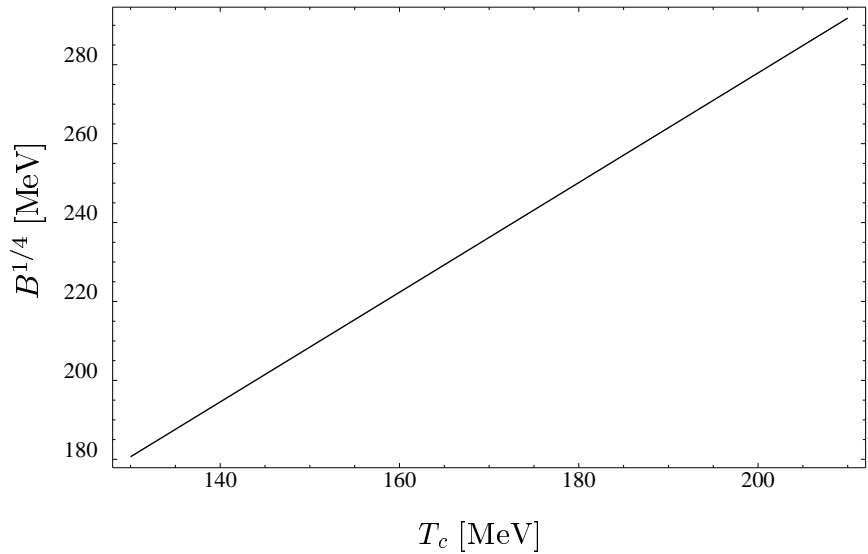,clip=,width=4.5in}}
\caption[The Dependence of the Critical Temperature on the Bag Constant]{The Dependence of the Critical Temperature on the Bag Constant. Expression~(\ref{T_c}) is shown graphically for $g_q = 37$ (two-flavored QGP) and $g_h = 3$ (ideal massless pion gas). In Chap.~\ref{Systematic_Investigation:_Thermal_Photons} we will treat~$T_c$ as a parameter. Because~$T_c$ is in fact determined by~$B$, we only imply the more fundamental setting of~$B$.}
\label{Fig_Bag_Constant}
\end{figure}

\begin{figure}
        \centerline{\psfig{figure=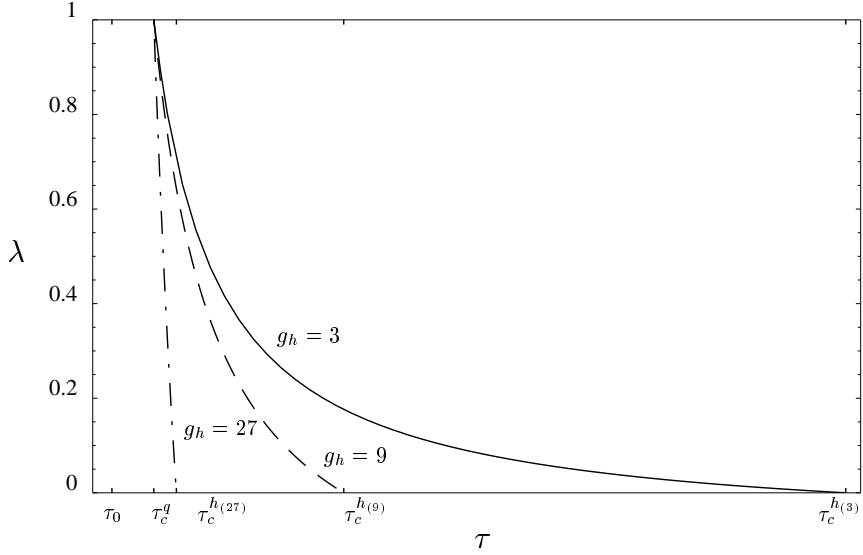,clip=,width=4.5in}}
\caption[Evolution of the QGP Volume Fraction]{Evolution of the QGP Volume Fraction. The proper time dependence of the QGP volume fraction $\lambda$ is illustrated for $g_h = 3$,~$9$, and~$27$. The only other parameter important for the shape of~$\lambda$ is the effective number of degrees of freedom present in the QGP, which was set to the value of two-flavored quark matter, $g_q = 37$. Because we consider an adiabatic expansion, the entropy contained in the QGP must be fully transferred to the constituents of the HHG. For a high number of effective hadronic degrees of freedom, e.g., $g_h =27$, which is, of course, far from reality while neglecting hadron masses, the entropy of the QGP phase can be carried away rapidly by the HHG. This looks different for the ideal massless pion gas, where the effective number of hadronic degrees of freedom is small,~$g_h =3$, and consequently, the mixed phase lasts relatively long.}
\label{Fig_QGP_Volume_Fraction}
\end{figure}

The thermodynamic properties at the critical boundary are determined by the {\em Gibbs criteria},
\begin{eqnarray}
        & T_c^q = T_c^h = T_c,&
\label{thermal_Gibbs_criteria}\\
        & P_c^q = P_c^h = P_c,&
\label{mechanical_Gibbs_criteria}
\end{eqnarray}
which express thermal and mechanical equilibrium, respectively. The superscripts mark the state of matter whose thermodynamic quantity is considered; $q$ indicates the QGP state and $h$ the HHG state. With the above equations and the expressions~(\ref{P_q}) and~(\ref{P_h}), the {\em critical} or {\em transition temperature} can be obtained in terms of the bag constant~$B$ and the effective degrees of freedom, $g_q$ and $g_h$,
\be
        T_c = \sqrt[4]{\frac{90 B}{(g_q - g_h) \pi^2}}.
\label{T_c}
\ee
Figure~\ref{Fig_Bag_Constant} illustrates this relation for $g_q = 37$ (two-flavored QGP) and $g_h = 3$ (ideal massless pion gas). For example, using the reasonable value $B^{1/4}=200$~MeV, one obtains $T_c = 144$~MeV. In a systematic study one should vary the bag constant~$B$ as a fundamental parameter, however, we will vary the parameter~$T_c$, which is equivalent because of Eq.~(\ref{T_c}).
While temperature and pressure remain constant at the critical boundary, energy and entropy density decrease. The evolution of the entropy density is given simply by Eq.~(\ref{expansion_time_scale_provider}), which can be restated as
\be
        s(\tau) = \frac{s(\tau_0)\,\tau_0}{\tau}.
\label{entropy_density_evolution}
\ee
In fact, this equation determines the time scale of the {\em complete} hydrodynamic expansion. A different expression for the entropy density in the mixed phase~$s_c$ can be obtained by Maxwell construction,
\be
        s_c(\tau) = \lambda(\tau)\, s_q(\tau_c^q) + \left[1-\lambda(\tau) \right]\, s_h(\tau_c^h),
\label{s_c}
\ee
which can be used together with Eqs.~(\ref{s_q}), (\ref{s_h}), and~(\ref{entropy_density_evolution}) in deriving an explicit form of the QGP volume fraction
\be
        \lambda(\tau) = \left( \frac{g_q}{g_q - g_h} \right) \left( \frac{T_0}{T_c} \right)^3 \left( \frac{\tau_0}{\tau} \right) - \left( \frac{g_h}{g_q - g_h} \right).
\label{lambda}
\ee
It can easily be checked that this equation reproduces the properties of~$\lambda$ discussed above. The proper time dependence of~$\lambda$ is illustrated for different numbers of effective degrees of freedom in Fig.~\ref{Fig_QGP_Volume_Fraction}. As a consequence of entropy conservation, the hadronization rate rises with a rising number of effective hadronic degrees of freedom~$g_h$. Conservation of entropy also governs through~$\lambda$ the evolution of the critical energy density $\varepsilon_c$, for which the expression
\be
       \varepsilon_c(\tau) = \lambda(\tau)\, \varepsilon_q(\tau_c^q) + \left[1-\lambda(\tau) \right]\, \varepsilon_h(\tau_c^h)
\label{varepsilon_c}
\ee
is obtained by performing again a Maxwell construction.
%
%
%
%----------------------------------------------------------
\section{The Phase Transition Scenario}
\label{The_Phase_Transition_Scenario}
%----------------------------------------------------------
%
%
In the preceding sections, we illustrated qualitatively our idea of an ultra-relativistic heavy ion collision. Since we extract thermal photon yields from a model which is based on this picture, we now present a quantitative discussion.

Our simulation starts as soon as the fireball produced in a central collision of two ultra-relativistic heavy nuclei is in local thermal equilibrium. At this initial time $\tau_0$, we have an initial temperature $T_0$. In the scenario with a phase transition, this temperature is assumed sufficiently high for the existence of QGP, which means the hydrodynamical description starts in the QGP phase. Inserting the bag model EOS~(\ref{bag_model_EOS}) in the basic differential equation obtained with Bjorken initial conditions~(\ref{Bjorken_model_equation}), one gets the following solution for this phase,
\be
        \varepsilon_q(\tau) = \left[ \varepsilon_q(\tau_0) - B \right] \left( \frac{\tau}{\tau_0}\right)^{-4/3} + B. 
\label{solution_q}
\ee
With Eq.~(\ref{varepsilon_q}), the evolution of temperature follows directly from this solution,
\be
        T_q(\tau) = T_0 \left( \frac{\tau_0}{\tau}\right)^{1/3}.
\label{T_q}
\ee
When the decreasing temperature reaches the critical temperature as given by Eq.~(\ref{T_c}), the formation of HHG begins. In this mixed phase that lasts until all QGP has hadronized, the temperature remains at $T_c$ and the evolution of the energy density is given by Eq.~(\ref{varepsilon_c}). At the completion of the phase transition, the fireball consists purely of hadronic matter. For the pure HHG phase, the basic hydrodynamical equation~(\ref{Bjorken_model_equation}) must be solved together with the EOS for the ideal massless pion gas~(\ref{ideal_gas_EOS}) to obtain
\be
        \varepsilon_h(\tau) = \varepsilon_h(\tau_c^h) \left( \frac{\tau}{\tau_c^h}\right)^{-4/3} 
\label{solution_h}
\ee
and
\be
        T_h(\tau) = T_c \left( \frac{\tau_c^h}{\tau}\right)^{1/3}.
\label{T_h}
\ee
The hydrodynamical treatment is only justified for fluid-like matter. Because of the ongoing expansion, the hadronic matter gets more and more dilute. We choose a freeze-out temperature $T_f$ as the stopping point of our hydrodynamical simulation. At this freeze-out temperature, it is assumed that the thermal interactions in the pion gas vanish and that the pions are moving as free particles out of the collision zone.

\begin{figure}
        \centerline{\psfig{figure=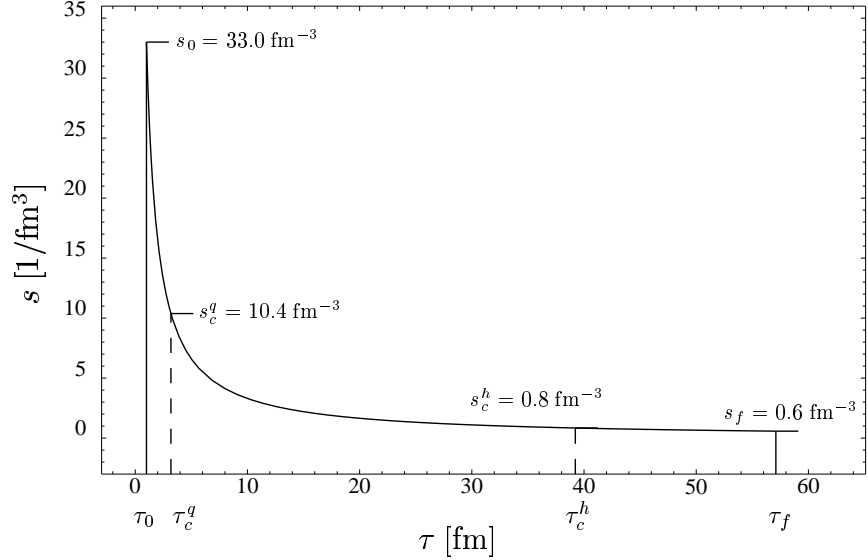,clip=,width=4.5in}}
\caption[Entropy Density Evolution]{Entropy Density Evolution. The proper time dependence of the entropy density is presented for the model parameters, $g_q = 37$, $g_h = 3$, $\tau_0 = 1\;$fm, $T_0 = 250\;$MeV, and $T_f = 150\;$MeV. The values at the onset and offset of the phase transition were obtained for $T_c =170\;$MeV. Ignoring the marks related to the phase transition, this diagram is also valid for the no phase transition scenario that has the same value of~$s_0 \tau_0$. Assuming an identical thermalization time of $\tau_0 = 1\;$fm, the corresponding no phase transition scenario must start with the initial entropy density of the above phase transition scenario. This demands an initial temperature of $T_0 = 578\;$MeV if the ideal massless pion gas, $g_h = 3$, is used to describe the purely hadronic scenario.}
%Assuming an identical thermalization time but a different number of effective degrees of freedom in the scenario without the phase transition, one needs to adjust the initial temperature to keep the initial entropy density at the same value, e.g., the ideal massless pion gas, $g_h = 3$, demands $T_0 = 578\;$MeV.}
%In fact, for a direct comparison of one scenario with and another without a phase transition, one should not choose identical initial temperatures but instead identical initial entropy densities.}
\label{Fig_Entropy_Density_Evolution}
\end{figure}

\begin{figure}
        \centerline{\psfig{figure=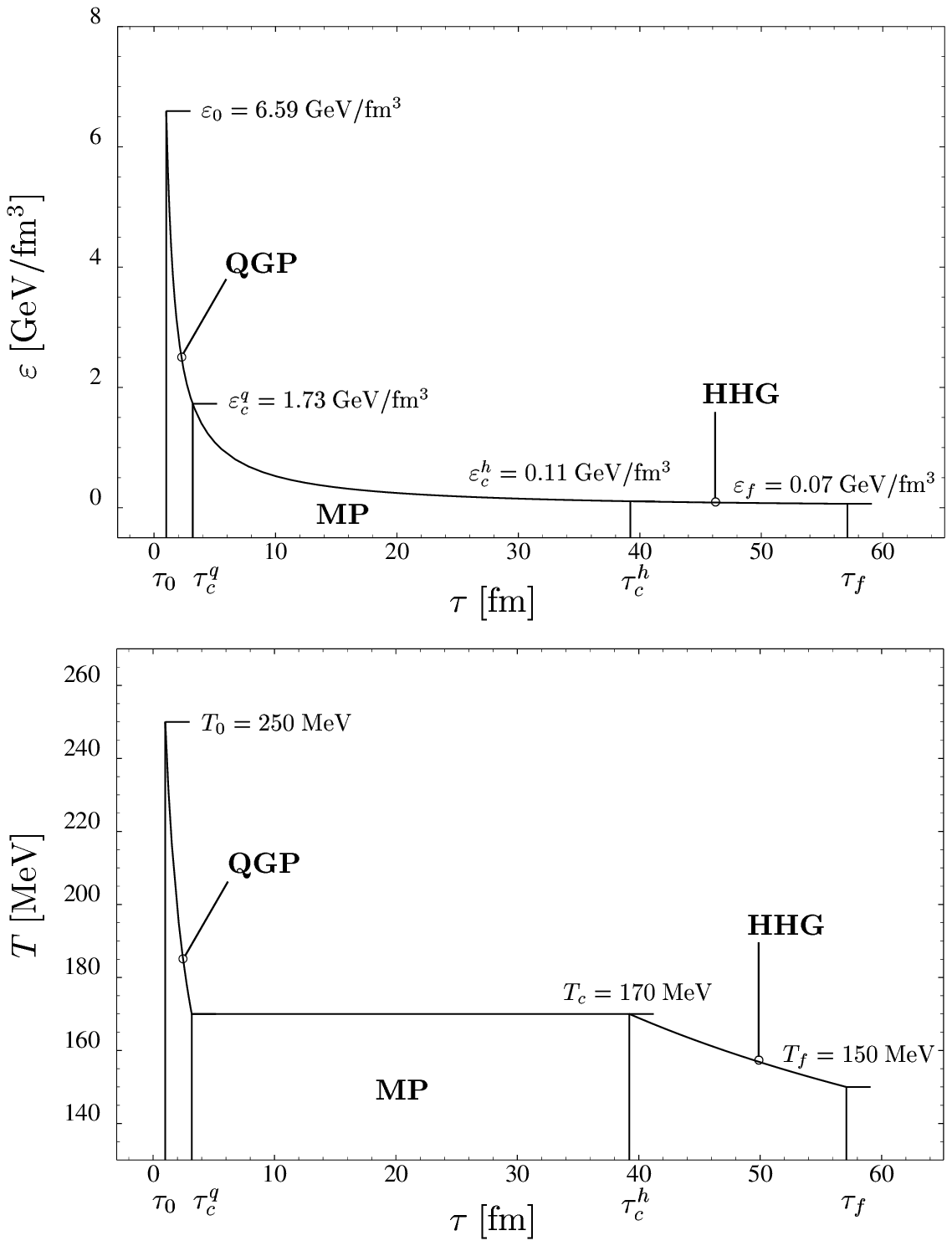,clip=,width=4.5in}}
\caption[Energy Density and Temperature Evolution in the Phase Transition Scenario]{Energy Density and Temperature Evolution in the Phase Transition Scenario. The model parameters were set to typical values, $g_q=37$, $g_h=3$, $\tau_0=1\;\fm$, $T_0=250\;\MeV$, $T_c=170\;\MeV$, and $T_f=150\;\MeV$.}
\label{Fig_PTS_Energy_Density_Temperature_Evolution}
\end{figure}

\newpage
For a closer illustration of the phase transition scenario, we set the model parameters to typical values,
\begin{eqnarray}
           g_q & = & 37\;\;\;\mbox{(two-flavored QGP)},
\nonumber\\
           g_h & = & 3\;\;\;\mbox{(ideal massless pion gas)},
\nonumber\\
        \tau_0 & = & 1\; \mbox{fm},
\nonumber\\
           T_0 & = & 250\; \mbox{MeV},
\nonumber\\
           T_c & = & 170\; \mbox{MeV},
\nonumber\\
           T_f & = & 150\; \mbox{MeV}.
\label{pts_model_parameters}
\end{eqnarray}
With these settings, the time scale of the phase transition scenario is determined by the entropy density evolution that is shown in Fig.~\ref{Fig_Entropy_Density_Evolution}. Because we assume an adiabatic expansion, this graph does not look different for a scenario without a phase transition that has the same value of~$s_0 \tau_0$. In Fig.~\ref{Fig_PTS_Energy_Density_Temperature_Evolution}, the proper time dependence of the energy density~$\varepsilon$ and the temperature~$T$ is illustrated. On the temperature plot, one can see clearly the different stages of the collision. As a consequence of the difference in the effective number of degrees of freedom, the temperature decrease in the pure QGP phase is obviously steeper than in the pure HHG phase.
%
%
%----------------------------------------------------------
\subsection*{Lifetimes of QGP, Mixed and HHG Phase}
%----------------------------------------------------------
%
%
In our systematic study of photon yields which will be presented in Chap.~\ref{Systematic_Investigation:_Thermal_Photons}, we will be interested in the contributions from the different collision stages. Therefore, it is important to consider also the lifetimes of these stages. The pure QGP phase starts at~$\tau_0$, which is a parameter of the model, and ends at
\be
        \tau_c^q = \left( \frac{T_0}{T_c}\right)^3 \tau_0
\label{tau_c^q}
\ee
when the transition temperature~$T_c$ is reached according to Eq.~(\ref{T_q}). The above proper time point marks also the beginning of the mixed phase that lasts until~$\lambda = 0$. With Eq.~(\ref{lambda}), one thus gets 
\be
        \tau_c^h = \frac{g_q}{g_h} \left( \frac{T_0}{T_c}\right)^3 \tau_0
\label{tau_c^h}
\ee 
for the ending of the mixed phase and the beginning of the pure HHG phase. The pure HHG phase ends when the freeze-out temperature~$T_f$ is reached which happens at
\be
        \tau_f = \frac{g_q}{g_h} \left( \frac{T_0}{T_f}\right)^3 \tau_0.
\label{tau_f}
\ee
as can be seen from Eq.~(\ref{T_h}).
Finally, using Eqs.~(\ref{tau_c^q}), (\ref{tau_c^h}), and~(\ref{tau_f}), one finds for the lifetimes the following expressions
\begin{eqnarray}
  \Delta \tau_{q} & = & \tau_{0} \left[ \left( \frac{T_{0}}{T_{c}} \right) ^{3} - 1 \right],
\label{lifetime_q} \\
  \Delta \tau_{c} & = & \tau_{0} \left( \frac{T_{0}}{T_{c}} \right) ^{3} \left[ \frac{g_{q}}{g_{h}} - 1 \right],
\label{lifetime_c} \\
  \Delta \tau_{h} & = & \tau_{0} \left( \frac{T_{0}}{T_{c}} \right) ^{3} \frac{g_{q}}{g_{h}}                                                               \left[ \left( \frac{T_{c}}{T_{f}} \right) ^{3} - 1 \right].
\label{lifetime_h}
\end{eqnarray}
In Chap.~\ref{Systematic_Investigation:_Thermal_Photons}, we will refer several times to these expressions and also discuss their~$\tau_0$, $T_0$, $T_c$, and $T_f$ dependence.
%
%
%
%----------------------------------------------------------
\section{The No Phase Transition Scenario}
\label{The_No_Phase_Transition_Scenario}
%----------------------------------------------------------
%
%
To decide if a phase transition occurred, one needs to compare experimental data with theoretical predictions. In theoretical calculations, two scenarios always need to be addressed, a ``phase transition scenario'' and a ``no phase transition scenario.'' A clear indication for the existence of QGP will be an experimental result that only coincides with theoretical predictions obtained in a calculation {\em with} a phase transition. The signature cannot be considered clean if it can also be deduced in a reasonable calculation {\em without} a phase transition. Thus, also in this thesis we do not only consider a scenario that exhibits a phase transition but also one that is purely hadronic (no phase transition scenario).

The hydrodynamic description starts at~$\tau_0$ with a massless pion gas of temperature~$T_0$ because no QGP is produced in the no phase transition scenario. Thus, the subsequent energy density and temperature evolution is as outlined for the HHG phase of the phase transition scenario with the recognition that~$\tau_c^h$ and~$T_c$ need to be replaced by~$\tau_0$ and~$T_0$, respectively. To be explicit, we get
\be
        \varepsilon_h(\tau) = \varepsilon(\tau_0) \left( \frac{\tau}{\tau_0}\right)^{-4/3} 
\label{solution_hh}
\ee
and
\be
        T_h(\tau) = T_0 \left( \frac{\tau_0}{\tau}\right)^{1/3}.
\label{T_hh}
\ee
As the phase transition scenario, this simulation ends with the freeze-out of the pions at temperature~$T_f$.

For a meaningful direct comparison of a scenario with and another without a phase transition, 
%the same heavy ion collision must be considered. In our picture, entropy is conserved throughout the complete reaction. Since the initial volume,~$V(\tau_0)$, depends only on the properties of the initial projectiles and the centrality of the collision, both scenarios need to start with the same initial entropy density~$s_0$. 
we follow the approach of~\cite{SRIVASTAVA_1994} by assuming identical values of the entropy~$S$ and the thermalization time~$\tau_0$ in both scenarios. Because the entropy is a constant of the motion, as is expressed in Eq.~(\ref{entropy_per_rapidity_conservation}), this implies identical initial entropy densities~$s(\tau_0)$ and accordingly also an identical evolution of the entropy density, which can be seen on the relation
\be
        s(\tau) \; = \; \frac{S}{2\pi\,R_A^2\,\tau},
\ee
where the denominator describes the total fireball volume $V_{tot}(\tau_0)$ as obtained in the Bjorken model. Thus, the no phase transition scenario with $g_h$~effective degrees of freedom must have an initial temperature of
\be
        T_0^{h} = \left( \frac{g_q}{g_h} \right)^{1/3} T_0^{q}
\label{T_0_relation}
\ee
to be appropriate for the direct comparison\footnote{To be appropriate for the direct comparison with the phase transition scenario means here to start with the same initial entropy density as the phase transition scenario.} with the phase transition scenario that starts with a QGP of $g_q$~effective degrees of freedom at $T_0^{q}$. For example, the no phase transition scenario that should be compared to the scenario presented in Sec.~\ref{The_Phase_Transition_Scenario} ($g_q=37$, $g_h=3$, $\tau_0=1\;$fm, $T_0=250\;$MeV, $T_c=170\;$MeV, $T_f=150\;$MeV) must have the same thermalization time~$\tau_0$, and in the case of an ideal massless pion gas, an initial temperature of 578~MeV. This extremely high value is a consequence of the small number of effective degrees of freedom present in the ideal massless pion gas ($g_h = 3$) and exhibits that the ideal pion gas is definitively not the best model for the description of the purely hadronic scenario. However, since the emphasis in this thesis is on bremsstrahlung processes in the QGP, we now ignore this fact.

\begin{figure}
        \centerline{\psfig{figure=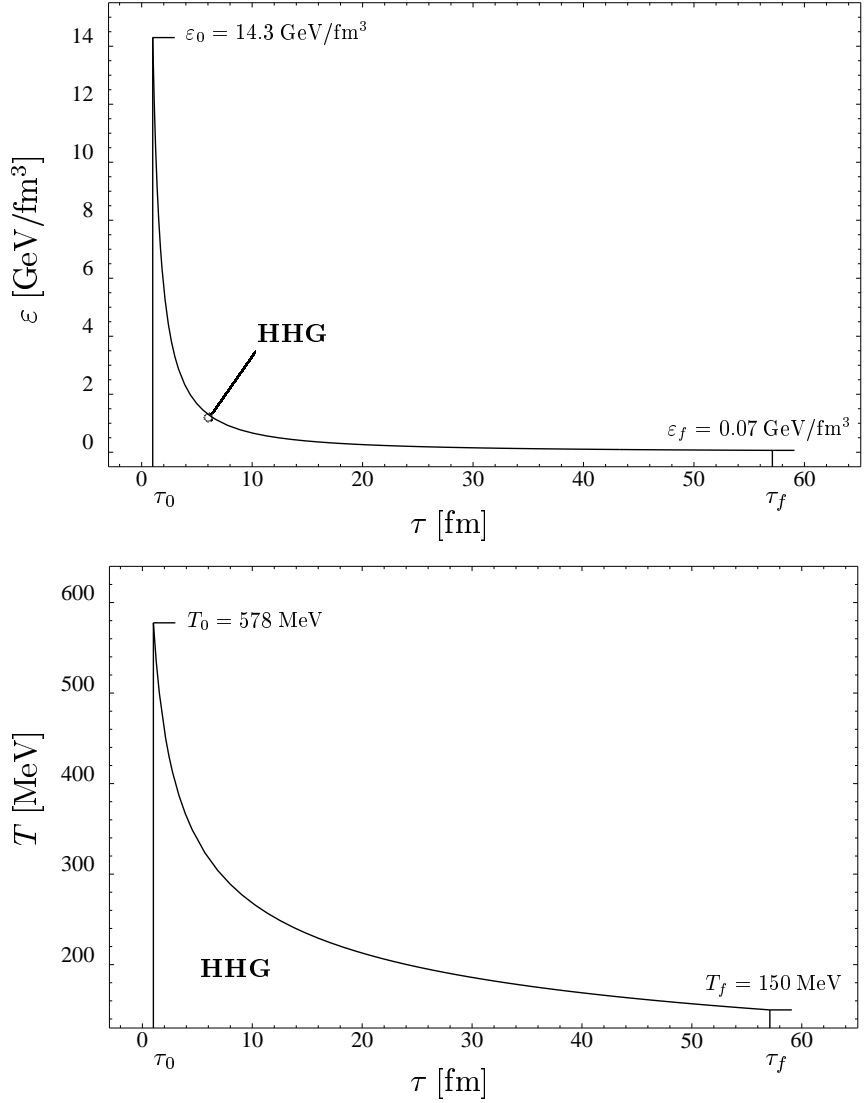,clip=,width=4.5in}}
\caption[Energy Density and Temperature Evolution in the No Phase Transition Scenario]{Energy Density and Temperature Evolution in the No Phase Transition Scenario. The model parameters were set as follows, $g_h=3$, $\tau_0=1\;$fm, $T_0=578\;$MeV, and $T_f=150\;$MeV. With these settings a direct comparison with the sample scenario presented in Sec.~\ref{The_Phase_Transition_Scenario} ($g_q=37$, $g_h=3$, $\tau_0=1\;$fm, $T_0=250\;$MeV, $T_c=170\;$MeV, $T_f=150\;$MeV) becomes meaningful as is explained in the text.}
\label{Fig_NTS_Energy_Density_Temperature_Evolution}
\end{figure}

\pagebreak
Considering the no phase transition scenario with the following parameters,
\begin{eqnarray}
           g_h & = & 3\;\;\;\mbox{(ideal massless pion gas)},
\nonumber\\
        \tau_0 & = & 1\; \mbox{fm},
\nonumber\\
           T_0 & = & 578\; \mbox{MeV},
\nonumber\\
           T_f & = & 150\; \mbox{MeV},
\label{nts_model_parameters}
\end{eqnarray}
we find the evolution of entropy density, energy density and temperature as illustrated in Figs.~\ref{Fig_Entropy_Density_Evolution} and~\ref{Fig_NTS_Energy_Density_Temperature_Evolution}, respectively. While the entropy density evolution in this purely hadronic scenario is identical to the one in the comparable phase transition scenario, the energy density and temperature evolution show, of course, a different behavior.
%The corresponding energy density and temperature evolution can be seen in Fig.~\ref{Fig_NTS_Energy_Density_Temperature_Evolution} and, of course, they look different than those obtained in the comparable phase transition scenario.
%
%
%
%----------------------------------------------------------
\subsection*{Lifetime of the HHG Phase}
%----------------------------------------------------------
%
%
In the no phase transition scenario, the lifetime of the HHG phase coincides with the period in which the fireball is in local thermal equilibrium. This state is reached at the initial time~$\tau_0$ that marks the starting point of the HHG phase. The ending point is just given by the freeze-out temperature~$T_f$, at which the hadrons cease to interact thermally. With Eq.~(\ref{T_hh}), we derive
\be
        \tau_f = \left( \frac{T_0}{T_f}\right)^3 \tau_0
\label{tau_f_hh}
\ee
and get for the lifetime of the HHG phase in the no phase transition scenario
\be
  \Delta \tau_{h} = \tau_{0} \left[ \left( \frac{T_{c}}{T_{f}} \right) ^{3} - 1 \right].
\label{lifetime_hh}
\ee
%The~$\tau_0$, $T_0$, and $T_f$ dependence of these expressions will also be plotted in Chap.~\ref{Systematic_Investigation:_Thermal_Photons}.

\cleardoublepage
%
%
%----------------------------------------------------------
\chapter{Photons}
\label{Photons}
%----------------------------------------------------------
%
The central element of the research for this thesis were hard thermal photon rates. In fact, the advent of new, extraordinary photon rates describing bremsstrahlung processes in QGP triggered the investigation presented in this thesis. This chapter will review the present-day rates for hard thermal photons from QGP and HHG. 

As mentioned in the introduction, photons are interesting probes of the fireball. Because of their purely electromagnetic nature, they have mean free paths much larger than the transverse size of the fireball, $\lambda_{mfp}\gg\sigma$. Thus, most photons produced in the fireball do not suffer any final state interaction. They reach the experimental detector directly from their origin in space-time. As a consequence, photon distributions stay far from equilibrium~\cite{KAPUSTA_1991} and provide undisturbed footprints from the various stages of nuclear matter in which they have been produced. Photons seem in this sense promising for the confirmation of QGP possibly produced in ultra-relativistic heavy ion collisions.

High-energy nucleus-nucleus collisions exhibit many (potential) photon sources. There are hard scattering processes of the initial partons during the very early stages of the collision. Among these processes that bring the system locally into thermal equilibrium are also photon producing reactions. Photons from these processes are referred to as {\em prompt photons}. When local thermal equilibrium is reached, there will be thermal photon production. Because the electrically charged constituents that emit photons are different in QGP and HHG, it is sensible to consider thermal photon emission from QGP and HHG separately. After freeze-out, hadronic decay processes as the Dalitz decays, $\pi^0 \rightarrow \gamma\gamma$ and $\eta \rightarrow \gamma\gamma$, contribute significantly in the measured photon spectra. 

In this thesis, we concentrate on thermal photon production, so prompt and decay photons are considered background. Because experiments provide direct photon yields, decay contributions are already subtracted as will be explained in Chap.~\ref{Analysis_of_Experimental_Data_on_Direct_Photon_Production}. The situation looks different for prompt photons that are part of the direct photon spectrum. It is the task of theorists to calculate and subtract the prompt photon yields in order to get the thermal photon spectrum. However, since the photon production in pre-equilibrium scatterings is mainly important for the very high momentum region of the considered spectra~\cite{ALAM_1996}, we only discuss prompt photon production qualitatively. Without the quantitative treatment, one must keep in mind that there are photon emitting processes besides thermal photon production when we inspect the direct photon yields at very high momenta.
%We only discuss photon production in pre-equilibrium scatterings qualitatively since they are mainly important for the region at the very high momentum upper edge of the direct photon spectrum~\cite{ALAM_1996}
%However, the photon production in pre-equilibrium scatterings is mainly important for the very high momentum region of the direct photon spectra~\cite{ALAM_1996}, which justifies our only qualitative discussion. 
%Since we do not account quantitatively for these reactions, one must keep in mind that there are additional photon emitting processes besides thermal photon production when we inspect direct photon yields.

This chapter is organized as follows. We start with a brief discussion of prompt photons. Next, a detailed review of the most recent thermal photons rates for QGP and HHG is presented. Decay photons are not covered because they do not appear in the experimental direct photon data. A description of how we get the photon spectrum from the production rates concludes this part of the thesis.
\section{Prompt Photons}
\label{Prompt_Photons}
%----------------------------------------------------------
%
%
In the very early pre-equilibrium stage of an ultra-relativistic heavy reaction, many hard scattering processes between the partons of the initial colliding nuclei steer the system towards local thermal equilibrium. Among these hard processes, there are also photon producing reactions, such as Compton scattering ($qg \rightarrow q\gamma$, $\bar{q}g \rightarrow \bar{q}\gamma$), $q\bar{q}$-annihilation ($q\bar{q} \rightarrow g\gamma$) and higher-order bremsstrahlung processes ($qg \rightarrow qg\gamma$, $\bar{q}g \rightarrow \bar{q}g\gamma$, etc.). Because the momentum transfer in these processes is large, the elementary rates on the parton level, $(dN/d^3p_{\gamma})_{ab\rightarrow \gamma c}$, can be calculated reliably within perturbative (zero-temperature) QCD. These rates must then be folded with the structure functions $F$ of the colliding nuclei A and B,
\be
        \left(E_{\gamma}\,\frac{d\sigma}{d^3p_{\gamma}}\right)_{AB\rightarrow \gamma C} = \int \frac{dx_a}{x_a}\frac{dx_b}{x_b}\,F_{a,A}(x_a)\,F_{b,B}(x_b)\,\left(E_{\gamma}\,\frac{d\sigma}{d^3p_{\gamma}}\right)_{ab\rightarrow \gamma c},
\ee
as is derived in~\cite{ALAM_1996}. In the above expression, $C$~stands for the final state particle(s) on the hadronic level produced besides the photon, $c$~denotes the same quantity on the partonic level, and $F_{j,J}(x_j)dx_j$ is the probability of having a parton $j$ inside the nucleus $J$, where the parton carries a momentum fraction between $x_j$ and $x_j + dx_j$ of the nucleus momentum. By summing the rates of all processes $AB\rightarrow \gamma C$ up to some specified order in the coupling constants incoherently, one obtains the total prompt photon yield $E_{\gamma}\,dN/d^3p_{\gamma}$ in the specified order.

The uncertainties in the extraction of the prompt photon yields are dominated by two aspects connected to the nuclear structure functions~$F_{j,J}(x_j)$. First, even for relatively hard photons ($E_{\gamma} \approx 4$~GeV), one is in the low-$x$ region, where the form of the structure functions is a matter of current research~\cite{DUKE_1984,GLUECK_1995_GLUECK_1998}. Second, the structure functions for the nuclei cannot be simply extrapolated from the nucleon structure functions: one has to take into account medium effects that are also the subject of ongoing discussions. Here, we do not go beyond the above qualitative remarks on prompt photons. For a quantitative treatment, the reader is referred to a recent investigation with emphasis on nuclear shadowing effects~\cite{HAMMON_1998}.
%
%
%
%----------------------------------------------------------
\section{Thermal Photons}
\label{Thermal_Photons}
%----------------------------------------------------------
%
%
The thermally moving constituents of the strongly interacting continuum, produced in an ultra-relativistic heavy ion reaction, undergo reactions and, in the case of hadrons, also decays. These processes are governed by the thermal distributions of the participating particles. Because many constituents of the continuum carry electrical charge, many reactions produce photons. These are the photons we refer to as {\em thermal photons}. 
%!!! excellent thermometer for ultra-relativistic heavy ion collisions
%
%If the thermal photon rates for QGP and HHG do not differ, the yields can serve as an excellent thermometer for ultra-relativistic heavy ion collisions~\cite{KAPUSTA_1991}.
%
%
%
%----------------------------------------------------------
\subsection{Thermal Photons from Quark-Gluon Plasma}
\label{Thermal_Photons_from_Quark-Gluon_Plasma}
\begin{figure}
        \centerline{\psfig{figure=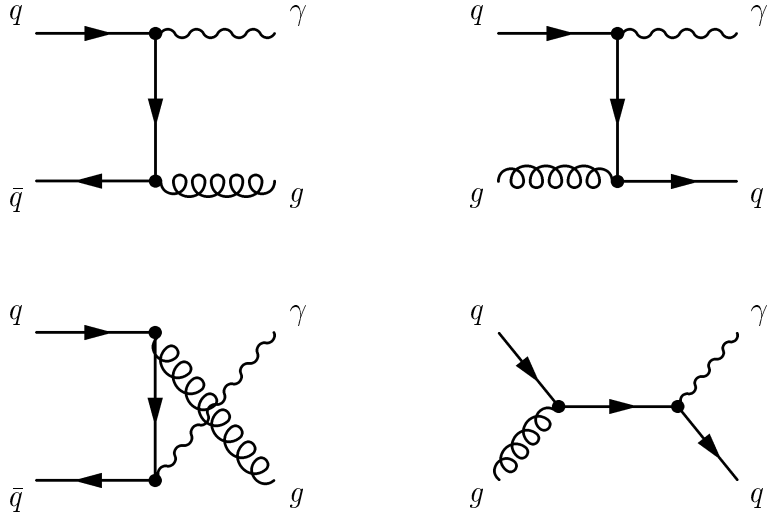,clip=,width=4.in}}
\caption[$q\bar{q}$-Annihilation and Compton Scattering with an Initial Gluon]{$q\bar{q}$-Annihilation and Compton Scattering with an Initial Gluon. The Feynman diagrams display the lowest order reactions in which a real photon is produced: $q\bar{q}$-annihilation (left hand side) and Compton scattering with an initial gluon (right hand side). Only the Compton scattering diagrams for a quark are shown because the same topology with reverted particle flow holds for the Compton scattering of an anti-quark. For massless quarks and soft momentum transfers, each process exhibits the well known logarithmically divergent behavior. In QGP, however, medium effects provide the quarks with a finite thermal mass that regulates all infrared divergences. A consistent inclusion of the medium effects is achieved within the Braaten-Pisarski method, where the soft quark propagator is replaced by an effective propagator that emerges from the resummation of hard thermal loops.}
\label{Fig_1-loop_processes}
\end{figure}

%----------------------------------------------------------
%
%
Thermal photon production from QGP is examined in finite temperature QCD~\cite{KAPUSTA_1989}. The lowest order photon emitting reactions in the QGP are $q\bar{q}$-annihilation,
\bea 
        q\bar{q} & \rightarrow & g\gamma,
\eea
and Compton scattering with an initial gluon,
\bea
        qg & \rightarrow & q\gamma, \\
        \bar{q}g & \rightarrow & \bar{q}\gamma,
\eea
which are shown as Feynman diagrams in Fig.~\ref{Fig_1-loop_processes}. By considering these processes for massless quarks in QCD, a logarithmically divergent behavior of the production rates is encountered for soft momentum transfers. This infrared divergence is well known since it appears in every QED and QCD tree diagram process in which a soft, massless particle is exchanged. However, QGP medium effects result in a non-vanishing thermal quark mass that serves as an infrared cutoff and renders the production rate being finite. A systematic treatment of the QGP medium effects that does not spoil gauge invariance became available with the {\em Braaten-Pisarski method} known as {\em hard thermal loop} (HTL) resummation technique~\cite{BRAATEN_1990}. Using this technique, the production rates of hard ($E_{\gamma} \gg T$) thermal photons from the above lowest order reactions have been calculated for thermal and chemical equilibrium~\cite{KAPUSTA_1991,BAIER_1991} and also for chemical {\em non}-equilibrium~\cite{TRAXLER_1996}, where deviations from the thermal distributions were described by fugacities. A close inspection of photon spectra from a fireball in chemical {\em non}-equilibrium has been found as being important but it is beyond the scope of this study. We concentrate only on photon emission from a system in thermal and chemical equilibrium at zero baryon density. The case of finite baryon density, which has also been investigated~\cite{TRAXLER_1995}, is here of no relevance.

\begin{figure}
        \centerline{\psfig{figure=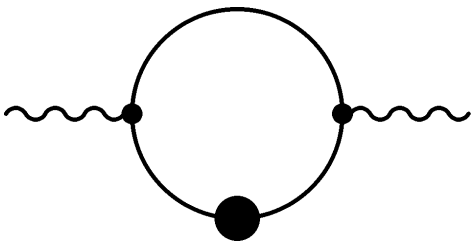,clip=,width=2.5in}}
\caption[Effective One-Loop Contribution to the Photon Self-Energy]{Effective One-Loop Contribution to the Photon Self-Energy. The hard thermal photon production rates for the processes shown in Fig.~\ref{Fig_1-loop_processes} arise from the imaginary part of this diagram by applying thermal cutting rules. One quark in the loop can be soft which means that its bare propagator must be dressed to avoid an infrared singularity. The dressing is indicated in the black blob on the lower quark propagator.}
\label{Fig_1-loop_contribution}
\end{figure}

How does one proceed (more explicitly) in computing the production rate of hard thermal photons from $q\bar{q}$-annihilation ($q\bar{q} \rightarrow g\gamma$) and Compton scattering with an initial gluon ($qg \rightarrow q\gamma$, $\bar{q}g \rightarrow \bar{q}\gamma$)? Following the {\em Braaten-Yuan prescription}~\cite{BRAATEN_1991}, a parameter~$\Lambda$ is introduced that separates high momentum transfers of the order~$T$ from low momentum transfers of the order $gT$, where the strong coupling constant is assumed much smaller than one, $g \ll 1$. This decomposes the rate into a hard and a soft part. The hard part has the infrared cutoff~$\Lambda$ and is calculated in relativistic kinetic theory with {\em bare} propagators and {\em bare} vertices which means that no medium effects are taken into account. The thermal distributions enter the derivation of the hard part in the following way
\bea
        E_{\gamma}\,\frac{dN}{d^4x\,d^3p_{\gamma}} = \int \prod_{i=1}^3 \frac{d^3p_i}{2E_i(2\pi)^3}
                & & \!\!\!\! \inv{2(2\pi)^3} \,f_1(p_1)\,f_2(p_2)\,[1\pm f_3(p_3)] \nonumber\\
                & & \;\;\;\; \times \sum |{\cal M}|^2\,(2\pi)^4\delta^4(p_1+p_2-p_3-p_{\gamma}), 
\eea
where $\sum |{\cal M}|^2$ is the squared scattering amplitude of the considered process summed over the initial and final parton states. 
%where ${\cal N}$ is a factor taking into account spin, color, and flavor of the participating partons and ${\cal M}$ is the scattering amplitude of the considered process. 
The particles in the initial state are labeled by $1$ and $2$, while those in final state are labeled by $3$ and $\gamma$. The $f$'s are the Fermi-Dirac or Bose-Einstein distribution functions depending on the spin of the corresponding particle. A fermion in the final state is Pauli suppressed while a boson in the final state is Bose enhanced. For the soft part with the ultraviolet cutoff~$\Lambda$, medium effects are crucial and any soft bare quark propagator is replaced by the HTL resummed propagator that contains the finite thermal quark mass. The soft quark propagators in other words become dressed. The soft part of the thermal emission rate is calculated from the retarded photon self-energy according to the relation~\cite{WELDON_1983,GALE_1991} 
\be
        E_{\gamma}\,\frac{dN}{d^4x\,d^3p_{\gamma}} = - \inv{(2\pi)^3}\,f_B(E_{\gamma})\,\mbox{Im}\,\Pi_{\mu}^{R,\mu} (p_{\gamma})
\ee
where $\Pi_{\mu}^{R,\mu}$ is the retarded polarization tensor of the photon and $f_B$ is the Bose-Einstein distribution function. $q\bar{q}$-annihilation and Compton scattering are contained in the one-loop contribution to the photon self-energy with one quark propagator dressed\footnote{The one-loop contribution to the photon self-energy with both quark propagators dressed does not enter into the computation of the hard thermal photon rates because the photon cannot be hard if both quarks in the loop are soft.}. This can be seen by applying the thermal cutting rules~\cite{KOBES_1985,KOBES_1986} on the diagram shown in Fig.~\ref{Fig_1-loop_contribution}. 
%The same diagram with both quark propagators dressed does not enter the computation of the hard thermal photon rates, which is clear because the photon cannot be hard if both quarks in the loop carry soft momenta. 
% shows the dressed one-loop contribution to the photon self-energy and the cut necessary to infer the corresponding physical processes described. 
%
By summing the hard and the soft part, both of which depend on~$\Lambda$, a finite result is obtained, which is independent of the separation-parameter~$\Lambda$. 

\begin{figure}
        \centerline{\psfig{figure=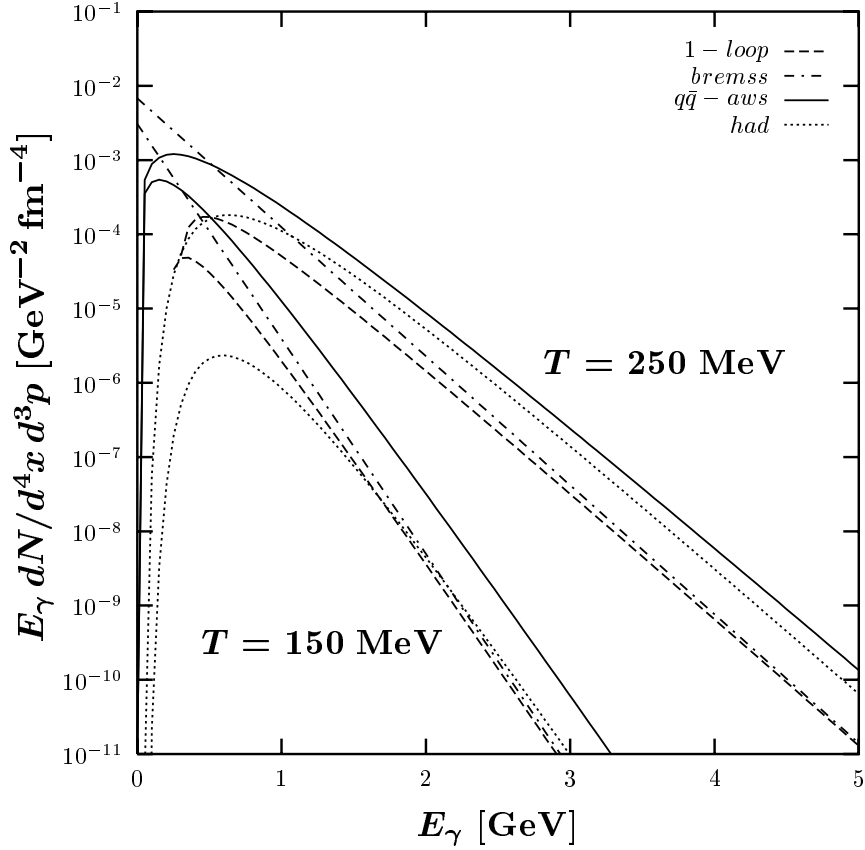,clip=,width=4.5in}}
\caption[Thermal Production Rates of Hard Photons from QGP and HHG]
{Thermal Production Rates of Hard Photons from QGP and HHG. The $E_{\gamma}$~dependence of the thermal photon production is presented for two temperatures, $T = 150\;\MeV$ and $T = 250\;\MeV$, where the dashed, dot-dashed, and solid lines represent the contributions from $1-loop$, $bremss$, and $q\bar{q}-aws$ processes in a two-flavored QGP
%\footnote{A two-flavored QGP ($N_f = 2$) was assumed in the illustration of the thermal photon production rates that depend on the number of flavors~$N_f$.}
, respectively, and the dotted line represents the contribution from $\pi\rho \rightarrow \pi\gamma$ and $\pi\rho \rightarrow a_1 \rightarrow \pi\gamma$ processes in HHG. Because all rates have been computed in the limit of hard photons, $E_{\gamma} \gg T$, the shape of the curves for $E_{\gamma} < 1\;\GeV$ should be ignored. It is important to notice that already the static thermal photon spectra exhibit the $q\bar{q}-aws$ process as the dominating one for $E_{\gamma} > 1\;\GeV$. This is a remarkable result since the $q\bar{q}-aws$ process is of higher order than the $1-loop$ processes.}
\label{Fig_thermal_photon_rates}
\end{figure}

\newpage
The net rate in the limit of hard photons, $E_{\gamma} \gg T$, has the following form 
\be
        \left. E_{\gamma}\,\frac{dN}{d^4x\,d^3p_{\gamma}} \,\right|_{1-loop} = 
        \frac{N_c C_F}{8\pi^2} \Big( \sum_f e_f^2 \Big)
        \alpha \alpha_s \ln\;\left(\frac{c\,E_{\gamma}}{\alpha_s\,T}\right)\,T^2\,e^{-E_{\gamma}/T},
\label{1-loop}
\ee
where $c = 0.23$ is a constant, $e_f$ is the electric charge of the quark with flavor $f$ in units of the electron charge $e$, and the sum runs over all flavors assumed in the QGP, e.g.,
\be
        \sum_f e_f^2 \; = \; \left( \frac{2}{3} \right)^2 + \left( \frac{1}{3} \right)^2 \; = \; \frac{5}{9},
\ee
for a two-flavored QGP. We 
%will assume a two-flavored QGP and 
use expression~(\ref{1-loop}) with the standard value $N_c = 3$ and $C_F = 4/3$, which is the corresponding Casimir operator of the fundamental representation of color SU(3). Further, we insert for the QED coupling constant $\alpha = 1/137$ and for the QCD coupling constant the lattice QCD data parameterization~\cite{KARSCH_1988}
\be
        \alpha_s(T) = \frac{6\,\pi}{(33-2\,N_f)\,\ln(8\,T/T_c)}.
\label{alpha_s}
\ee
The rate~(\ref{1-loop}) is labeled with the subscript $1-loop$ to remind that the soft part was computed from the one-loop contribution to the photon self-energy. Its dependence on the photon energy $E_{\gamma}$ for two fixed temperatures, $T = 150\;\MeV$ and $T = 250\;\MeV$, is presented with the dashed lines in Fig.~\ref{Fig_thermal_photon_rates}, where a two-flavored QGP has been assumed.

\begin{figure}
        \centerline{\psfig{figure=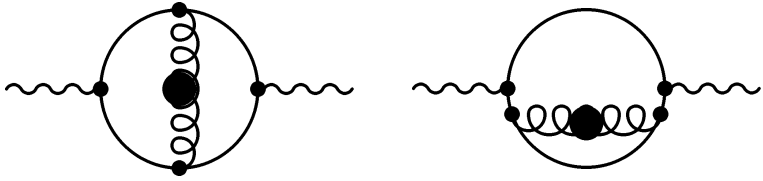,clip=,width=4.in}}
\caption[Effective Two-Loop Contributions to the Photon Self-Energy]{Effective Two-Loop Contribution to the Photon Self-Energy. In thermal QCD, bremsstrahlung processes have been investigated on the two-loop contributions to the photon self-energy within the framework of the Braaten-Pisarski method. By applying thermal cutting rules on the diagrams, the bremsstrahlung processes that are illustrated in Fig.~\ref{Fig_2-loop_processes} emerge. The black blob indicates an effective gluon propagator which is necessary since the gluon can be soft. It is basically this effective gluon propagator which causes the fact that bremsstrahlung processes arise in the same order of the coupling constants as the one-loop processes.}
\label{Fig_2-loop_contributions}
\end{figure}

\begin{figure}
        \centerline{\psfig{figure=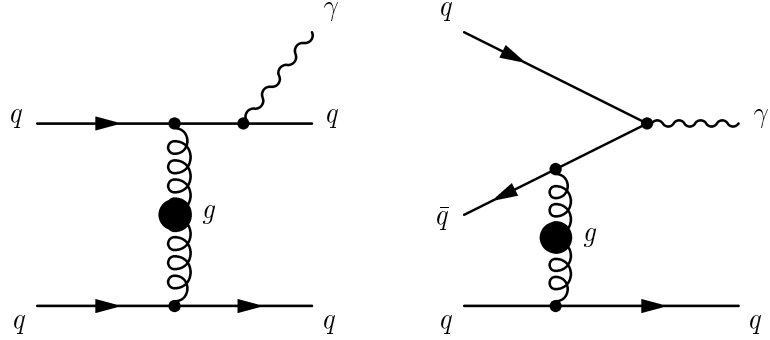,clip=,width=4.in}}
\caption[Feynman Diagrams of the Bremsstrahlung Processes]{Feynman Diagrams of the Bremsstrahlung Processes. By applying thermal cutting rules on the effective two-loop contributions to the photon-self energy, one finds the physical bremsstrahlung processes with a real photon in the final state. The diagram on the left hand side is just the ordinary bremsstrahlung process involving two quarks. The dressing on the gluon propagator indicated by the black blob is necessary to avoid an quadratical infrared divergence for soft momentum transfers. There are similar processes which are not shown but also described in the production rate $bremss$: $q\bar{q} \rightarrow q\bar{q}\gamma$, $\bar{q}\bar{q} \rightarrow \bar{q}\bar{q}\gamma$, $qg \rightarrow qg\gamma$, and $\bar{q}g \rightarrow \bar{q}g\gamma$. The diagram on the right hand side is a reaction that can be described as $q\bar{q}$-annihilation with an additional scattering on a quark. Again, other similar processes obtained from the two-loop contributions to the photon-self energy are not illustrated but included in the production rate $q\bar{q}-aws$, such as $q\bar{q}\bar{q} \rightarrow \bar{q}\gamma$ and $q\bar{q}g \rightarrow g\gamma$. Reactions of the type shown on the right hand side, $q\bar{q}-aws$, are the ones that surprisingly dominate the thermal photon spectra from the QGP.}
\label{Fig_2-loop_processes}
\end{figure}

The $1-loop$~rate has been employed in many investigations on real photons as {\em the only} measure for thermal photon production in QGP~\cite{ARBEX_1995,ALAM_1996,CLEYMANS_1997,SRIVASTAVA_1994,HAMMON_1998}. This {\em must} be modified because bremsstrahlung processes considered in thermal QCD turned out to contribute at the same order in the coupling constants as the one-loop processes~\cite{AURENCHE_1998}. The thermal rates for the bremsstrahlung processes have been calculated from the {\em two-loop} contributions to the photon self-energy that are pictured in Fig.~\ref{Fig_2-loop_contributions}. The thick black dot on the gluon propagator indicates an HTL resummed propagator which is necessary since the gluon can be soft. 
By applying the thermal cutting rules on the two-loop contribution to the photon self-energy, physical scattering processes such as the ones shown in Fig.~\ref{Fig_2-loop_processes} are obtained. It can be seen that besides ``ordinary'' bremsstrahlung, Fig.~\ref{Fig_2-loop_processes} (a), another process of $q\bar{q}$-annihilation with an additional scattering in the medium, Fig.~\ref{Fig_2-loop_processes} (b), is contained in the two-loop contributions to the photon self-energy. This process has in fact been found as the dominating source for hard photons in the QGP.

\newpage
The technical derivation of Aurenche et al. leads in the limit $E_{\gamma} \gg T$ to the production rate for ordinary bremsstrahlung photons
\be
        \left. E_{\gamma}\,\frac{dN}{d^4x\,d^3p_{\gamma}} \,\right|_{bremss} = 
        \frac{16}{\pi^3} \ln(2) (J_T-J_L) \frac{N_c C_F}{8\pi^2} \Big( \sum_f e_f^2 \Big) 
        \alpha \alpha_s\,T^2\,e^{-E_{\gamma}/T}
\label{bremss}
\ee
and to the production rate for photons from  $q\bar{q}$-annihilation with an additional scattering in the medium
\be
        \left. E_{\gamma}\,\frac{dN}{d^4x\,d^3p_{\gamma}} \,\right|_{q\bar{q}-aws} =  
        \frac{16}{3\pi^3}(J_T-J_L)\frac{N_c C_F}{8\pi^2} \Big( \sum_f e_f^2 \Big)
        \alpha \alpha_s\,E_{\gamma}\,T\,e^{-E_{\gamma}/T},
\label{qqbar-aws}
\ee
where the factor $(J_T-J_L)$ in the above expressions stands for the difference of two integrals that depend only on the number of colors $N_c$ and flavors $N_f$ assumed in the QGP. With the standard value $N_c = 3$, one gets for a two-flavored QGP, 
%which we will assume,
\be
        (J_T-J_L) = 8.71.
\ee
The dependence of~(\ref{bremss}) and~(\ref{qqbar-aws}) on the photon energy $E_{\gamma}$ for two fixed temperatures, $T = 150\;\MeV$ and $T = 250\;\MeV$, under the assumption of a two-flavored QGP is shown graphically in Fig.~\ref{Fig_thermal_photon_rates} with the dot-dashed lines representing the $bremss$~contribution and the solid lines representing the $q\bar{q}-aws$~contribution. The dominance of the $q\bar{q}$-annihilation process with an additional scattering in the medium can be seen clearly. It can be traced back to the factor $E_{\gamma}T\exp(-E_{\gamma}/T)$ that favors the production of hard thermal photons. In Chap.~\ref{Systematic_Investigation:_Thermal_Photons}, one will realize also on the thermal photon spectra that the bremsstrahlung processes, $bremss$ and $q\bar{q}-aws$, do not only give small corrections but lead to a significant enhancement in the thermal photon yield from QGP.

There are several uncertainties in the thermal photon production rates which have been derived in finite temperature QCD by applying the Braaten-Pisarski method to account for long range medium effects in the QGP. The rates listed explicitly have been obtained for thermal and chemical equilibrium. As is shown in~\cite{STRICKLAND_1994,TRAXLER_1996,SRIVASTAVA_1997}, abandoning the assumption of a chemically equilibrated QGP significantly alters the one-loop result. Thus, an inspection of bremsstrahlung processes from a system in chemical {\em non}-equilibrium would be very interesting and, in fact, it should be the subject of a future systematic investigation. Another uncertainty is connected to processes contained in the three or more loop contributions to the photon self-energy. Photon self-energy terms with a higher number of loops could again contribute in the same order of the coupling constants as the one- and two-loop contributions. A confirmation of this speculation would definitively raise fundamental questions concerning the validity of the effective perturbative expansion based on the HTL resummation technique. Finally, the HTL resummation technique is based on the assumption $g \ll 1$ which does not hold even in the hot stage of an ultra-relativistic heavy ion collision: for a two-flavored QGP at a high temperature, e.g., $T = 3\,T_c$, Eq.~(\ref{alpha_s}) gives $\alpha_s = 0.2$ which corresponds to $g = \sqrt{4 \pi \alpha_s} = 1.6$. However, an effective field theory cleaner than the one achieved from the HTL resummation technique and also suitable to calculate thermal photon rates for QGP is not available.
%
%
%----------------------------------------------------------
\subsection{Thermal Photons from Hot Hadronic Gas}
\label{Thermal_Photons_from_Hot_Hadronic_Gas}
%----------------------------------------------------------
%
%
In our simple dynamical model of nuclear collisions, we treat the HHG as an ideal massless pion gas. This is, of course, only an approximation in which thermodynamic quantities are obtained analytically. Pions are not massless and the HHG phase will also exhibit other mesons, such as $\rho$, $\eta$, and $\omega$ mesons. We do not account for these facts in the consideration of the fireball evolution as already explained in Chap.~\ref{A_Simple_Model_for_Ultra-Relativistic_Heavy_Ion_Collisions}. However, we implement thermal photon production rates from reactions of mesons more massive than pions. This approach is to some extend inconsistent, but sufficiently precise for the purpose of this investigation that centers on the effects of the bremsstrahlung processes in QGP.

The thermal photon rates for reactions of the thermalized secondary hadrons have been derived in an effective field theory that is renormalizable and provides gauge invariant results. Scattering processes of $\pi$, $\rho$, and $\eta$ mesons,
\bea
        \pi\rho & \rightarrow & \pi\gamma,
\label{pirho-pigamma}\\
        \pi\pi  & \rightarrow & \rho\gamma,
\label{pipi-rhogamma}\\ 
        \pi\pi  & \rightarrow & \eta\gamma,
\label{pipi-etagamma}\\ 
        \pi\eta & \rightarrow & \pi\gamma,
\label{pieta-pigamma}\\
        \pi\pi  & \rightarrow & \gamma\gamma,
\label{pipi-gammagamma}
\eea
 and also vector-meson decays,
\bea
        \rho^0 & \rightarrow & \pi^+ \pi^- \gamma,
\label{rho-pipigamma}\\ 
        \omega & \rightarrow & \pi^0\gamma,
\label{omega-pigamma} 
\eea
have been examined and the Compton scattering of a pion on an initial $\rho$ meson, $\pi\rho \rightarrow \pi\gamma$, has been explored as the dominating photon source in HHG~\cite{KAPUSTA_1991}. A later investigation of the $\pi\rho \rightarrow \pi\gamma$ scattering but this time through the $a_1$-resonance, 
\be
        \pi\rho \; \rightarrow \; a_1 \; \rightarrow \; \pi\gamma,
\label{pirho-a1-pigamma}
\ee
discovered a significant enhancement of the $\pi\rho \rightarrow \pi\gamma$ contribution~\cite{XIONG_1992}. Concentrating on hard thermal photons, we consider the dominating processes~(\ref{pirho-pigamma}) and~(\ref{pirho-a1-pigamma}) as the only photon sources in the HHG and apply the parametrization from~\cite{XIONG_1992} divided by a factor of two, which was suggested by Kevin Haglin~\cite{XIONG_1992_HAGLIN} due to an encountered isospin miscounting in~\cite{XIONG_1992},
\be
        \left. E_{\gamma}\,\frac{dN}{d^4x\,d^3p_{\gamma}} \,\right|_{had} = 
        1.2\,T^{2.15}\,e^{-1/(1.35\,T\,E_{\gamma})^{0.77}}\,e^{-E_{\gamma}/T},
\label{had}
\ee
where $E_{\gamma}$ and $T$ should be given in units of GeV. The dotted lines in Fig.~\ref{Fig_thermal_photon_rates} illustrate the dependence of this rate on the photon energy $E_{\gamma}$ for two fixed temperatures, $T = 150\;\MeV$ and $T = 250\;\MeV$. Other decay processes besides the ones listed above can be neglected: either they are too slow, such as the Dalitz decays, or they involve more massive and consequently stronger Boltzmann-suppressed hadrons. 
%for a significant contribution to the thermal photon yield.

Uncertainties arise mainly from the assumption of a thermally and chemically equilibrated HHG phase and from potential hadronic in-medium effects, such as decreasing masses and increasing widths, that have been ignored in the derivation of the above parameterization. In addition, there is no fundamental theory for the strong force on the hadron level which means that the coupling constants of the effective theory must be inferred from various experiments that do not display the environment produced in an ultra-relativistic heavy ion collision.
%
%
%----------------------------------------------------------
\section{Photon Spectra}
%----------------------------------------------------------
%
%
It has been mentioned before that photon rates must be convoluted with the space-time evolution of the fireball to obtain the photon spectra comparable with experimental data. This section details on the explicit form of the integration~(\ref{integration_over_space-time_evolution}) in the fireball picture described.
% in Chap.~\ref{A_Simple_Model_for_Ultra-Relativistic_Heavy_Ion_Collisions}. 

Because we concentrate on central collisions with vanishing impact parameter and consider an only longitudinally expanding tube of strongly interacting matter, the integrations in the directions transverse to the beam can be performed independently of the production rate
\be
        \int d^4x = \pi\,R_A^2 \; \int dt\;dz,
\label{int_transverse}  
\ee
where expression~(\ref{R_A}) is used. In the remaining two-dimensional integration, we substitute the center-of-mass coordinates $t$ and $z$ of the emitting fluid cell by proper time $\tau$ and rapidity fluid cell rapidity $y'$,
\bea
        t & = & \tau \sinh y', \\
        z & = & \tau \cosh y',
\eea
since in the latter variables one takes advantage of the symmetries contained in the Bjorken model. The fluid cell rapidity has been renamed from $y$ to $y'$ and in the following $y$ will be referred to as the photon rapidity in the center-of-mass frame. The substitution results in the form
\be
        \int dt\;dz = \int_{\tau_1}^{\tau_2} d\tau\,\tau \; \int_{-y_{nucl}}^{+y_{nucl}} dy',
\label{int_t_z}
\ee
where $\tau_1$ and $\tau_2$ denote the starting and ending point of the considered collision phase, respectively. If any nuclear stopping is neglected, $y_{nucl}$ is the center-of-mass rapidity of the projectiles and the identity~\cite{WONG_1994}
\be
        y_{nucl} = \mbox{arcosh} \left( \frac{\sqrt{s}}{2 \cdot A \cdot \GeV} \right)
\label{y_nucl}
\ee
holds, which means $\sqrt{s}$ limits the range of the fluid cell rapidities. Next, we rewrite
\be
        E\, \frac{1}{d^3p} = \frac{1}{d^2p_{\perp}\,dy},
\label{E/d^3p}
\ee
where $E$, $p$, $p_{\perp}$, and $y$ represent energy, momentum, transverse momentum, and rapidity of the thermal photons, respectively. 
%$E$ is the energy, $p$ is the momentum, $p_{\perp}$ is the transverse momentum, and $y$ is the rapidity referred to the photon. 
The index $\gamma$, formerly used to label a quantity that refers to the emitted thermal photon, has been suppressed. By inserting Eqs.~(\ref{int_transverse}), (\ref{int_t_z}), and (\ref{E/d^3p}) into the integration~(\ref{integration_over_space-time_evolution}), one arrives finally at the expression that is evaluated numerically to obtain the thermal photon spectra in the center-of-mass $(CM)$ frame 
\be
        \left( \frac{dN}{d^2p_{\perp}\,dy}\right)_{(CM)} = 
        \pi\,R_A^2 \; \int_{\tau_1}^{\tau_2} d\tau\,\tau \; \int_{-y_{nucl}}^{+y_{nucl}} dy'\, 
        \left( E\, \frac{dN}{d^4x\, d^3p}\right)_{(LR)}.
\label{photon_spectrum}
\ee
The integrand of the above two-dimensional integral is the production rate in the considered collision phase (pure QGP, mixed, or pure HHG phase). Because production rate expressions have been computed for the local rest frame (LR) of the emitting fluid cell, the following form of photon energy must be used in the integrand,
\be
        E_{(LR)} = p_{\perp} \, \cosh (y'-y).
\ee
During the {\em mixed phase}, the QGP volume fraction~$\lambda$ enters the integrand,
\be
        \left( E\, \frac{dN}{d^4x\, d^3p}\right)_{(LR)} 
        \; = \; \lambda \; \left(E \,\frac{dN}{d^4x\,d^3p} \,\right)_{(LR)}^{QGP}
        \; + \; [1-\lambda] \; \left(E ,\frac{dN}{d^4x\,d^3p} \,\right)_{(LR)}^{HHG}
\ee
with the rates labeled by the superscripts $QGP$ and $HHG$ referring to the QGP and the HHG state of matter, respectively. 
%If different photon emitting processes represented in different production rates are studied, which will be the case for the QGP state of matter, one gets the total photon yield from these processes simply by adding the yields from the separate processes. 
%

\cleardoublepage
%
%
%
%----------------------------------------------------------
\chapter{Systematic Investigation}
\label{Systematic_Investigation:_Thermal_Photons}
%----------------------------------------------------------
%
%
After the introduction on ultra-relativistic heavy ion physics, the description of the simple model for heavy ion reactions in the transparent region, and the discussion of the thermal photon production rates, we now present the results of the systematic investigation on the thermal photon spectra. The starting point are the spectra obtained in the two sample scenarios that have been introduced in Chap.~\ref{A_Simple_Model_for_Ultra-Relativistic_Heavy_Ion_Collisions}, one with phase transition and the other without. Then, we concentrate only on the phase transition scenario, where the one presented in Chap.~\ref{A_Simple_Model_for_Ultra-Relativistic_Heavy_Ion_Collisions} serves as the basis for the systematic investigation, in which the dependence of the spectra is carefully examined on the {\em mass number of the projectiles}~$A$, the {\em projectile rapidity}~$y_{nucl}$, the {\em thermalization time}~$\tau_0$, the {\em initial temperature}~$T_0$, the {\em transition temperature}~$T_c$, and the {\em freeze-out temperature}~$T_f$.
%
%
%----------------------------------------------------------
\section{Thermal Photon Spectra in the Sample Scenarios}
%----------------------------------------------------------
%
%
The fireball evolution provided by the simple, well understood model explicitly described in Chap.~\ref{A_Simple_Model_for_Ultra-Relativistic_Heavy_Ion_Collisions} was illustrated on two sample scenarios, a {\em phase transition scenario} and a {\em no phase transition scenario}. The model parameters were specified for the phase transition sample scenario in~(\ref{pts_model_parameters}) and for the no phase transition sample scenario in~(\ref{nts_model_parameters}), where $T_0$ was adjusted according to Eq.~(\ref{T_0_relation}) to allow a direct comparison of both scenarios. We now extract the thermal photon spectra for both sample scenarios by applying Eq.~(\ref{photon_spectrum}) with the necessary additional parameters set to
\be
        y_{nucl} = 8.6,
\label{y_nucl_model_parameter}
\ee
corresponding to the pursued LHC center-of-mass collision energy of $\sqrt{s} = 5500\;A\cdot\GeV$, and
\be
        A = 208,
\label{A_model_parameter}
\ee
corresponding to the mass number of the lead nuclei that will be the projectiles in the LHC heavy ion physics program. To study photons from the central, low-baryon density region which is best described by the employed model, all photon spectra are presented for {\em photons at mid-rapidity},
\be
        y = 0.
\label{mid-rapidity}
\ee
%
%
%----------------------------------------------------------
\subsection{Thermal Photon Spectra in the Phase Transition Scenario}
%----------------------------------------------------------
%
%
In the phase transition scenario, several combinations of the spectra give insights into the importance of the different rates, the two states of matter (QGP and HHG), and the three collision phases (pure QGP phase, MP, and pure HHG phase), where we are primarily interested in the effects of the QGP bremsstrahlung processes. The upper diagram in Fig.~\ref{Fig_pts_new_old_qgp_tot} shows therefore the total\footnote{total = pure QGP phase contribution + MP contribution + pure HHG phase contribution} thermal photon yields with bremsstrahlung processes (solid line) referred to as {\em New} and without bremsstrahlung processes (dashed line) referred to as {\em Old}. The effect of the QGP bremsstrahlung processes is clearly most pronounced for high-$p_{\perp}$ photons. This behavior is due to the fact that the high-$p_{\perp}$ region is mainly populated through processes at high temperatures. In the phase transition scenario, these are the processes in the pure QGP phase and the MP to which bremsstrahlung processes (if included) belong. The lower diagram in Fig.~\ref{Fig_pts_new_old_qgp_tot} illustrates the decomposed contribution of these processes from the QGP state of matter, where the dashed, dot-short-dashed, and dot-long-dashed lines indicate the contributions from the $1-loop$, $bremss$, and $q\bar{q}-aws$ production rates, respectively, and the solid line indicates the sum of the contributions from the three QGP rates. This diagram exhibits most clearly the astonishing importance of the bremsstrahlung processes: {\em the inclusion of the bremsstrahlung processes enhances the total thermal photon yield from the QGP state of matter by about one order of magnitude over the complete considered $p_{\perp}$-range}, and already the $q\bar{q}-aws$ processes alone outshine the $1-loop$ processes that give the total yield from the QGP if bremsstrahlung processes are neglected.
\befig
        \centerline{\psfig{figure=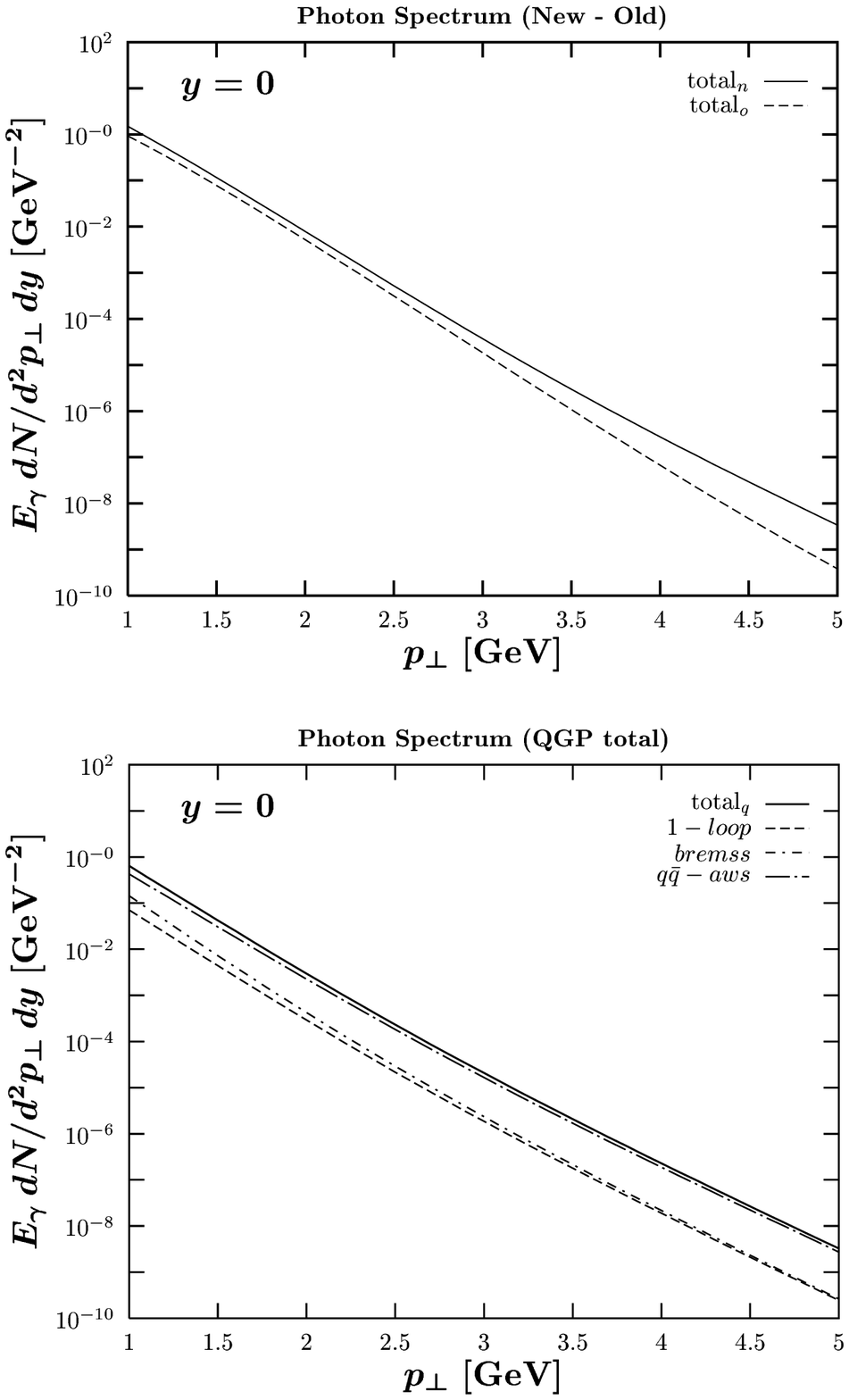,clip=,width=4.in}}
\caption[Total Thermal Photon Spectra and Thermal Photon Spectra from the QGP State of Matter in the Phase Transition Sample Scenario]{Total Thermal Photon Spectra and Thermal Photon Spectra from the QGP State of Matter in the Phase Transition Sample Scenario. The total thermal photon yields are shown in the upper plot, where the solid and dashed lines are obtained with and without QGP bremsstrahlung processes, respectively. In the lower plot, the thermal photon yields from the QGP state of matter are presented, where the dashed, dot-short-dashed, and dot-long-dashed lines indicate the spectra from the $1-loop$, $bremss$, and $q\bar{q}-aws$ rates, respectively. The total QGP contribution is given in the solid line if bremsstrahlung processes are included.}
% and in the dashed line if bremsstrahlung processes are neglected.}
\label{Fig_pts_new_old_qgp_tot}
\efig

In Fig.~\ref{Fig_pts_qgp_hhg}, the yields from the two states of matter, QGP and HHG, can be compared with (New) and without (Old) the QGP two-loop contributions taken into account. The spectra from QGP and HHG are presented in the dashed and dotted lines, respectively, and the sum is presented in the solid line. Without bremsstrahlung processes in the QGP, HHG controls the total thermal photon spectrum almost over the complete considered transverse momentum range, $1\;\GeV < p_{\perp} < 4.5\;\GeV$, while with these processes, QGP becomes the dominant photon source for $p_{\perp} > 3\;\GeV$.
\befig
        \centerline{\psfig{figure=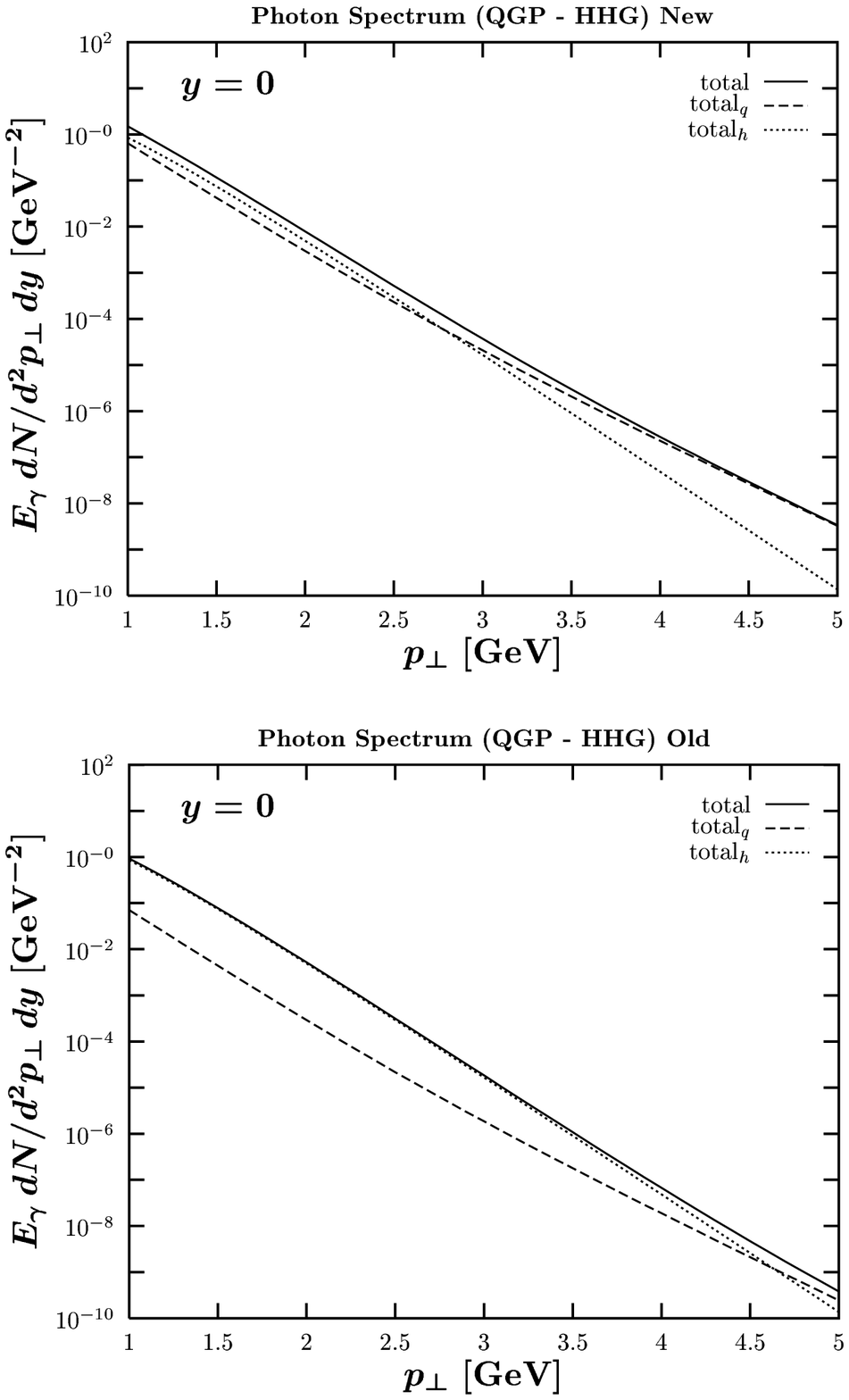,clip=,width=4.in}}
\caption[Thermal Photon Spectra from the QGP and the HHG State of Matter in the Phase Transition Sample Scenario]{Thermal Photon Spectra from the QGP and the HHG State of Matter in the Phase Transition Sample Scenario. The thermal photon yields from the QGP and the HHG state of matter are represented in the dashed and the dotted lines, respectively, while the sum of both is the total thermal photon yield represented in the solid line. QGP bremsstrahlung processes are included in the upper plot and neglected in the lower plot.}
\label{Fig_pts_qgp_hhg}
\efig

It is also interesting to inspect the photon yields from the different stages of the collision, the pure QGP phase, the MP, and the pure HHG phase, that are shown in Fig.~\ref{Fig_pts_qgp_mp_hhg}. The upper plot, where the $bremss$ and the $q\bar{q}-aws$ rates were included, exhibits the pure QGP phase (dashed line) as the dominant thermal photon source in the high-$p_{\perp}$ region. 
%over the MP (dot-dashed line) and the pure HHG phase (dotted line). 
Further, in the upper and the lower plot, where for the QGP state of matter only the $1-loop$ rate was implemented, one can clearly identify different slopes corresponding to different temperature ranges of the specific collision stage. Due to the Boltzmann factor, $\exp(-E/T)$, that is contained in each of the applied rates as the essential factor for the shape of the spectrum, the relation between slope and temperature can be expressed approximately in the form
\be
        \ln \left( \frac{dN}{d^2p_{\perp}\,dy}\right) \; \propto \; -\,\inv{T},
\ee
which means a flat spectrum indicates a high temperature.
\befig
        \centerline{\psfig{figure=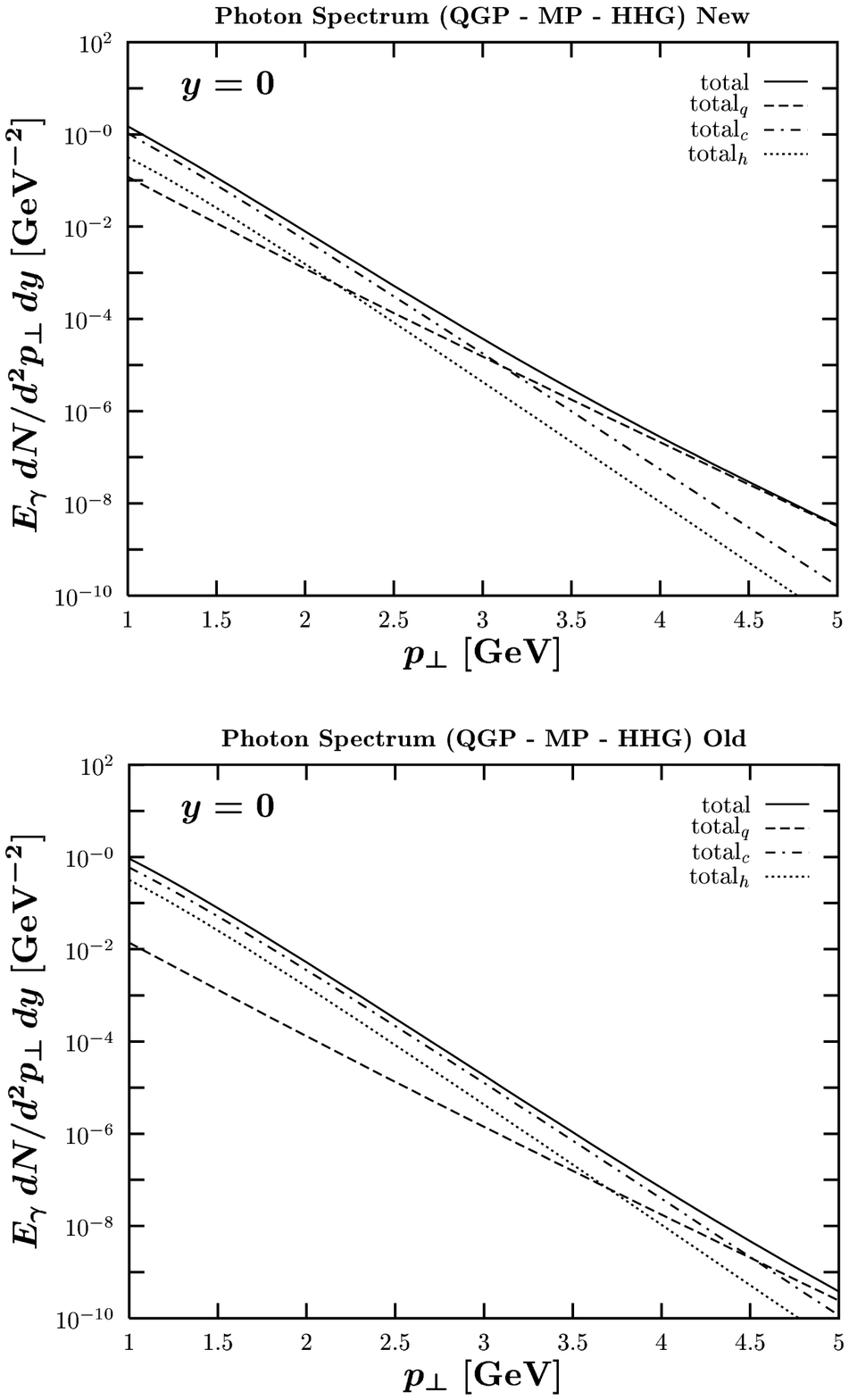,clip=,width=4.in}}
\caption[Thermal Photon Spectra from the Different Collision Stages in the Phase Transition Sample Scenario]{Thermal Photon Spectra from the Different Collision Stages in the Phase Transition Sample Scenario. The dashed, dot-dashed, and dotted lines correspond to the thermal photon yields from the pure QGP, the MP, and the pure HHG stages of the collision, respectively, and the solid line corresponds to the total thermal photon yield, where QGP two-loop processes 
%, $bremss$ and $q\bar{q}-aws$, 
are included in the upper diagram and neglected in the lower diagram.}
\label{Fig_pts_qgp_mp_hhg}
\efig
%
%
%----------------------------------------------------------
\newpage
\subsection{Thermal Photon Spectra in the No Phase Transition Scenario}
\label{Thermal_Photon_Spectra_in_the_No_Phase_Transition_Scenario}
%----------------------------------------------------------
%
%
While in the phase transition scenario it was interesting to analyze the total photon spectrum in several ways, the basic message from the no phase transition scenario can be illustrated in just one line which is presented in Fig.~\ref{Fig_nts}. This line, which indicates the thermal photon yield from the HHG state of matter, is obtained with the production rate given in Eq.~(\ref{had}). 
\befig
        \centerline{\psfig{figure=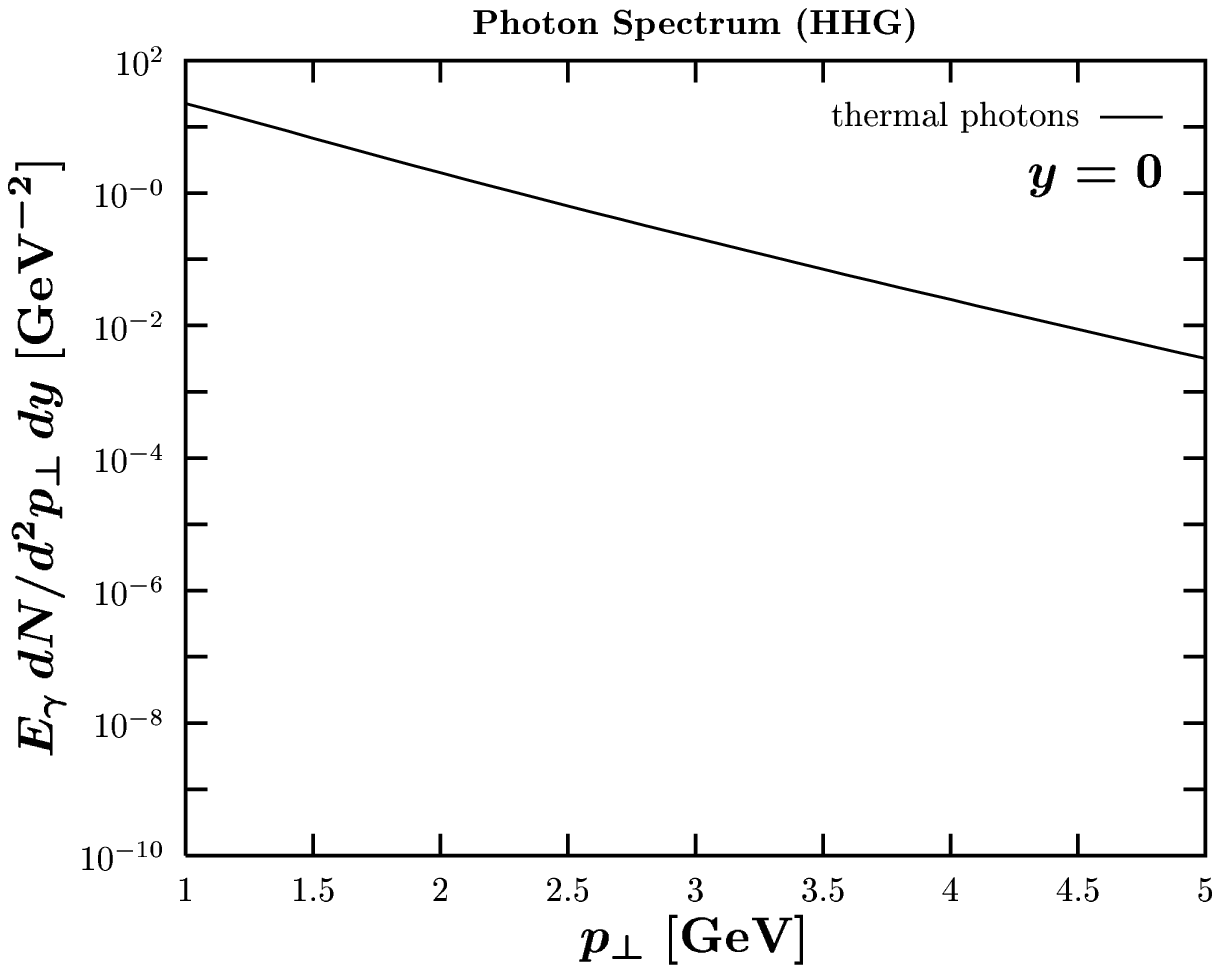,clip=,width=4.in}}
\caption[Thermal Photon Spectrum in the No Phase Transition Sample Scenario]{Thermal Photon Spectrum in the No Phase Transition Sample Scenario.}
\label{Fig_nts}
\efig
It is trivial that the scenario without a phase transition, in which no QGP state of matter is assumed, does not allow to study the effects of QGP bremsstrahlung processes. Nevertheless, the comparison of the total thermal photon spectrum from the no phase transition scenario with the one from the phase transition scenario is very important because the quality of thermal photons as a potential signature for the QGP formation can be tested in this way. Here, the comparison of the total thermal photon spectra exhibits a significant abundance in the no phase transition scenario that increases from one order of magnitude for photons with $p_{\perp} = 1\;\GeV$ up to about six orders of magnitude for photons with $p_{\perp} = 5\;\GeV$. This behavior can be traced back to the temperature evolution shown in the lower half of Fig.~\ref{Fig_NTS_Energy_Density_Temperature_Evolution}. The high initial temperature of $T_0 = 578\;\MeV$ in the no phase transition scenario, which was chosen to have the same initial entropy density as in the phase transition scenario, leads to a very high mean temperature that benefits thermal photon production especially in the high-$p_{\perp}$ range. In fact, the extremely high value for $T_0$ was a consequence of the massless pion gas used to model the HHG state of matter and its small number of effective degrees of freedom ($g_h = 3$) that entered the derivation of the initial temperature from the initial entropy density. A higher number of effective degrees of freedom, which is present in a more realistic description of the HHG including finite hadron masses and a higher number of hadrons~\cite{CLEYMANS_1997}, would result in a lower initial temperature and consequently in a lower thermal photon yield. 
%Here is the problem! While for the HHG temperatures attained in the phase transition scenario, $T_f < T < T_c$, the massless pion gas might be a reasonable approximation, it is only a crude approximation for the higher temperatures reached in the no phase transition scenario. 
For an adequate treatment of the purely hadronic scenario, one must therefore implement such a more realistic description of the HHG that includes finite hadron masses and a higher number of hadrons. This will not be presented here and the important careful comparison of the thermal photon spectra from the phase transition scenario with those from the no phase transition scenario is postponed for future work, which means that we will concentrate from now on only on the scenario with a phase transition.

The remaining part of this chapter is devoted to the dependence of the thermal photon spectra on the model parameters in the phase transition scenario. We take the phase transition sample scenario with the parameter settings~(\ref{pts_model_parameters}), (\ref{y_nucl_model_parameter}), and (\ref{A_model_parameter}) as the basis. Then, sticking to the two-flavored ideal massless parton gas ($g_q = 37$) and to the ideal massless pion gas ($g_h =3$), we vary every other parameter separately over an instructive range and inspect the corresponding influence on the thermal photon spectra.
%
%
%----------------------------------------------------------
\section{\boldmath $A$ - Mass Number of Projectile}
%----------------------------------------------------------
%
%
The mass number~$A$ of the projectile governs the transverse size of the fireball tube. Since we concentrate on central collisions with zero impact parameter, the transverse size of the only longitudinally expanding fireball equals the transverse size of the projectile. The thermal photon spectra are therefore proportional to the geometrical cross section of the projectile,
\be
        \frac{dN}{d^2p_{\perp}\,dy} \; \propto \; \pi\,R_A^2,
\ee
which can be seen from Eq.~(\ref{photon_spectrum}). This proportionality together with the phenomenological formula~(\ref{R_A}) for the nuclear radius leads directly to the $A$-dependence of the thermal photon spectra,
\be
        \frac{dN}{d^2p_{\perp}\,dy} \; \propto \; A^{2/3}.
\label{A_dependence}
\ee
In Fig.~\ref{Fig_A_dependence}, the total photon spectrum is presented for sulfur ($A = 32$) and gold ($A = 197$) projectiles with (solid line) and without (dashed line) bremsstrahlung processes included. This diagram illustrates clearly the dependence of the thermal photon spectra on the mass number~$A$ of the projectile. With the above expression~(\ref{A_dependence}), it is found that the displayed spectra differ by a factor of about 3.5. For lead ($A = 208$) projectiles, the total thermal photon spectra have already been shown in the upper half of Fig.~\ref{Fig_pts_new_old_qgp_tot}, and by comparing these spectra with the ones for the gold projectiles, only a marginal difference is found.
\befig[t]
        \centerline{\psfig{figure=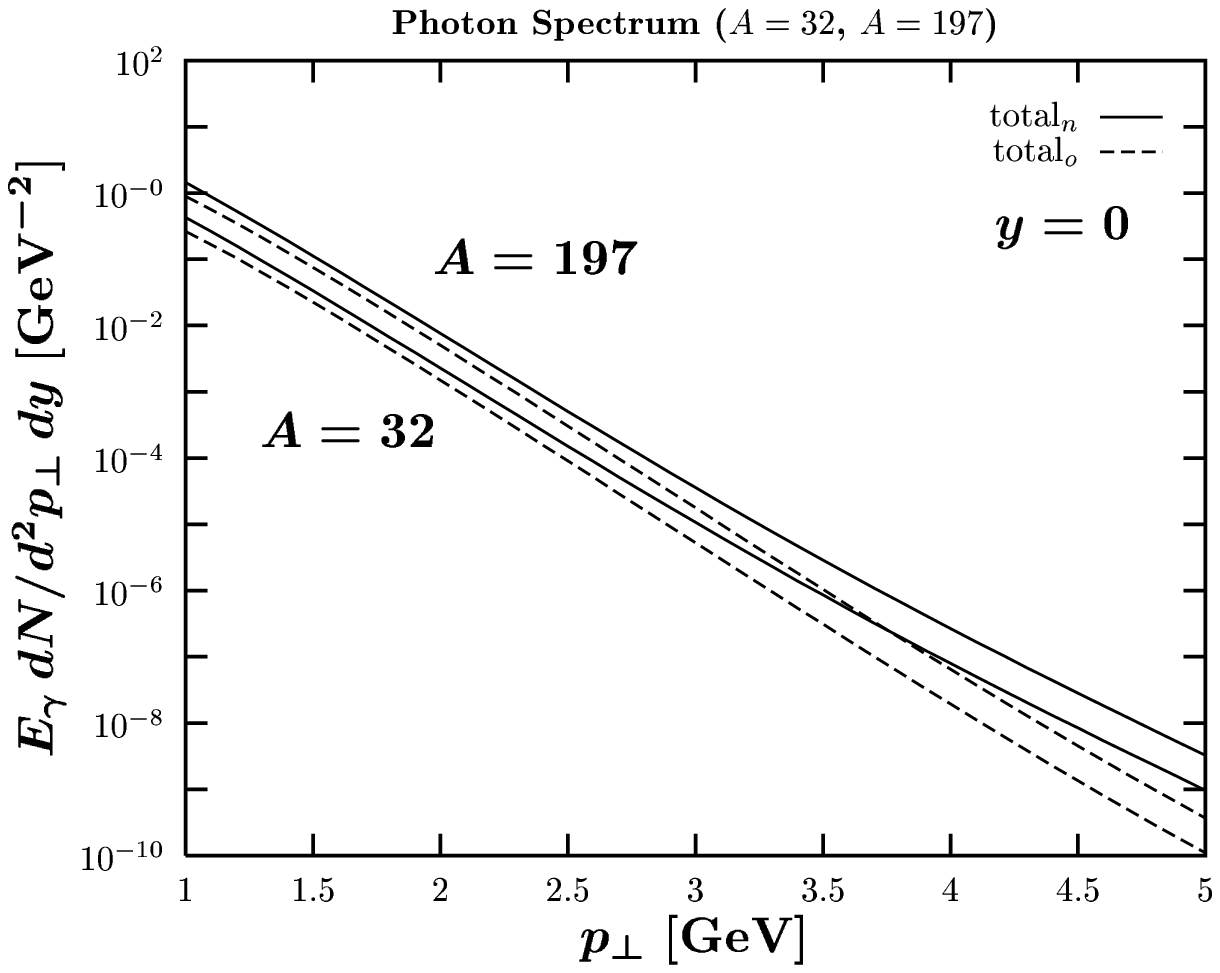,clip=,width=4.in}}
\caption[Total Thermal Photon Spectra for $A = 32$ and $A = 197$ in the Phase Transition Scenario]{Total Thermal Photon Spectra for $A = 32$ (Upper Curves) and $A = 197$ (Lower Curves) in the Phase Transition Scenario. As in the upper diagram of  Fig.~\ref{Fig_pts_new_old_qgp_tot}, the solid lines are obtained if QGP bremsstrahlung processes are included and the dashed lines are obtained if QGP bremsstrahlung processes are neglected.}
\label{Fig_A_dependence}
\efig

Different accelerators provide different projectiles depending on the employed heavy ion injector. Table~\ref{Tab_Accelerator_Properties} summarizes the projectiles that have been applied at the CERN SPS and that will be applied at the BNL RHIC and the CERN LHC. Collisions of gold and lead nuclei support the highest thermal photon yields but unfortunately also the highest background yields. However, due to the progressive photon spectrometer LEDA used in the WA98 experiment, the direct photon data from WA98 is expected to provide interesting information soon on the thermal photon production in ultra-relativistic heavy ion reactions.
\begin{table}
\centering
\begin{tabular}{|c|c|c|c|c|c|c|c|}
\hline
Accelerator / & Beam-Target /              & Experiment & $A$   & $\sqrt{s}$    & $y_{nucl}$ & $dN/dy$ & $T_0$ \\
Collider      & Beam-Beam                  &            &       & [A$\cdot$GeV] &            &         & [MeV] \\ 
\hline\hline
SPS           & $^{32}S - ^{197}\!\! Au$   & WA80       & $32$  & 20            & 3.0        & 200     & 185   \\ 
              & $^{208}Pb - ^{208}\!\! Pb$ & WA98       & $208$ & 17            & 2.8        & 800     & 195   \\ 
\hline\hline
RHIC          & $^{197}Au - ^{197}\!\! Au$ & PHENIX     & $197$ & 200           & 5.3 &\textgreater 1200 &\textgreater 225
\\
LHC           & $^{208}Pb - ^{208}\!\! Pb$ & ALICE      & $208$ & 5500          & 8.6 &\textgreater 2500 &\textgreater 280
\\
\hline
\end{tabular}
\caption[Accelerator and Experiment Specific Quantities Relevant for the Extraction of the Thermal Photon Spectra]{Accelerator- and Experiment Specific Quantities Relevant for the Extraction of the Thermal Photon Spectra. The entries for the projectile rapidity $y_{nucl}$ were obtained with Eq.~(\ref{y_nucl}) from the corresponding center-of-mass energy~$\sqrt{s}$. For the multiplicity distribution~$dN/dy$, we specified values in accordance with the literature~\cite{BJORKEN_1983,SATZ_1992,ALAM_1996,NA49_1996}. With these values, the $T_0$ entries were computed using Eq.~(\ref{initial_conditions}) and the canonical value for the thermalization time $\tau_0 = 1\;\fm$.}
\label{Tab_Accelerator_Properties}
\end{table}
%
%
%----------------------------------------------------------
\section{\boldmath $y_{nucl}$ - Projectile Rapidity}
%----------------------------------------------------------
%
%
The projectile rapidity~$y_{nucl}$ limits the rapidity range of the photon emitting fluid cells if any stopping of the projectiles during the collision is neglected. It is calculated from the center-of-mass energy according to Eq.~(\ref{y_nucl}), which has also been used for the computation of the accelerator specific values of~$y_{nucl}$ listed in Tab.~\ref{Tab_Accelerator_Properties}. The assumption of no nuclear stopping in an ultra-relativistic heavy ion collision is, of course, arguable. However, the dependence of the thermal photon spectra at mid-rapidity on the fluid cell rapidity limits, or~$y_{nucl}$, is anyway extremely weak. Figure~\ref{Fig_y_prime_dependence} illustrates this fact on the emission of mid-rapidity photons, $y = 0$, with two transverse momenta, $p_{\perp} = 1\;\GeV$ and $p_{\perp} = 5\;\GeV$, from fluid cells with different rapidities~$y'$ at the temperature $T = 250\;\MeV$. It can be seen that the contribution to the mid-rapidity photon yield decreases rapidly for both transverse momenta with increasing fluid cell rapidity~$y'$. This behavior is easily understood as a consequence of the Boltzmann factor, which is an essential part of every thermal photon production rate considered. For example, a fluid cell at the rapidity $y' = 4$ must emit a photon having an energy of $E_{(LR)} = 1\;\GeV\cdot\cosh 4 \approx 27\;\GeV$ for it to appear at mid-rapidity with a transverse momentum of $p_{\perp} = 1\;\GeV$. Due to the extremely high energy, the Boltzmann factor makes this event an extremely rare one. In conclusion, we can neglect the weak dependence on~$y_{nucl}$ 
%and also any nuclear stopping power 
as we are concentrating on the photon spectra at mid-rapidity. 
\befig[t]
        \centerline{\psfig{figure=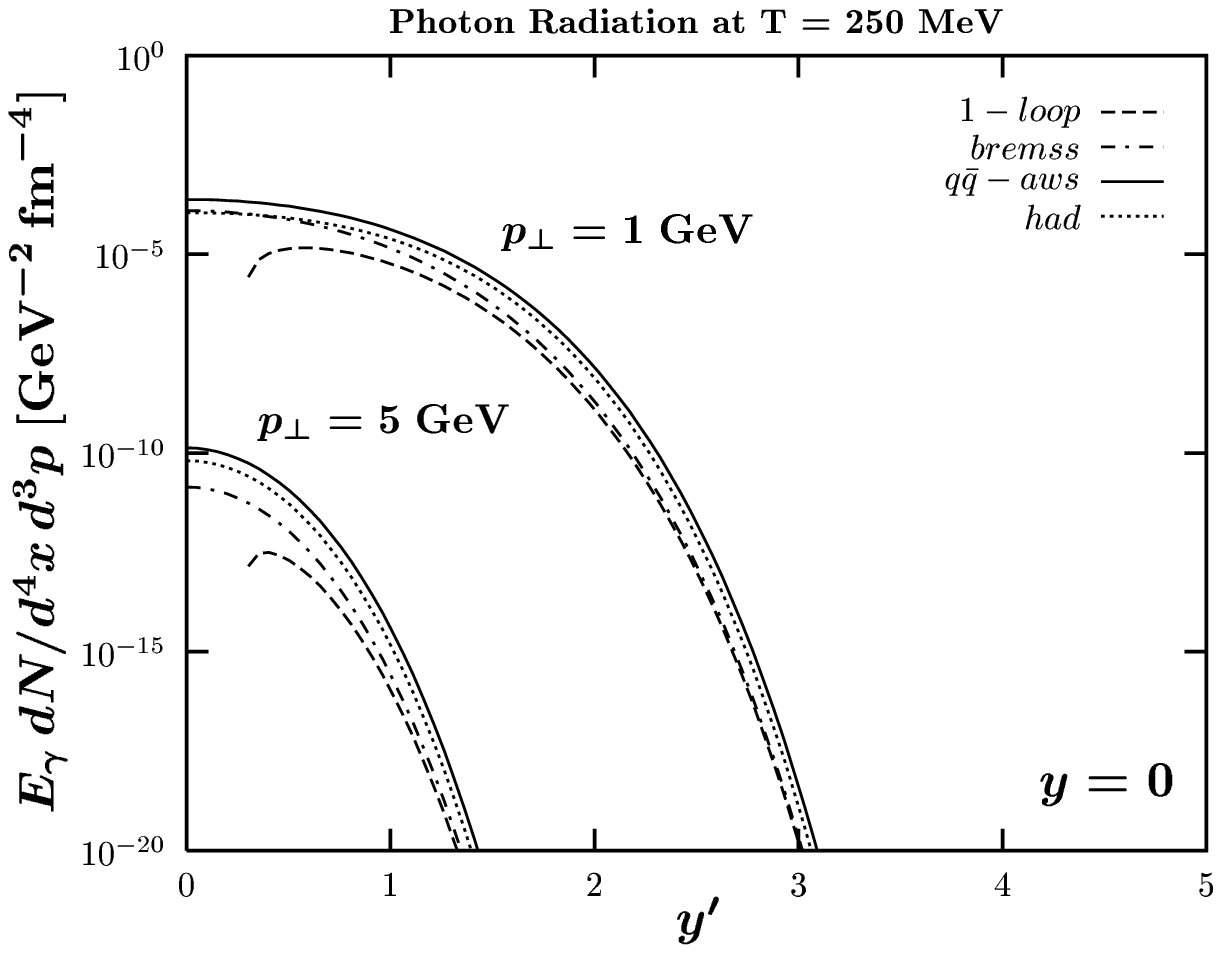,clip=,width=4.in}}
\caption[Thermal Emission of Mid-Rapidity Photons with $p_{\perp} = 1\;\GeV$ and $p_{\perp} = 5\;\GeV$ as a Function of the Fluid Cell Rapidity~$y'$ at the Temperature $T = 250\;\MeV$]{Thermal Emission of Mid-Rapidity Photons with $p_{\perp} = 1\;\GeV$ and $p_{\perp} = 5\;\GeV$ as a Function of the Fluid Cell Rapidity~$y'$ at the Temperature $T = 250\;\MeV$. As in Fig.~\ref{Fig_thermal_photon_rates}, the dashed, dot-dashed, solid, and dotted lines indicate the thermal photon production rates $1-loop$, $bremss$, $q\bar{q}-aws$, and $had$, respectively.}
\label{Fig_y_prime_dependence}
\efig
%
%The fluid cell rapidity~$y'$ enters the thermal photon rates through the photon energy that has for mid-rapidity photons the form
%\be
%        E_{(LR)} = p_{\perp} \cosh y',
%\ee
%where $(LR)$ refers to the local rest frame of the considered fluid cell. 
%
%
%----------------------------------------------------------
\newpage
\section{Initial Conditions}
%----------------------------------------------------------
%
%
The initial conditions determine the complete evolution of the strongly interacting continuum and, therefore, their significant influence on the thermal photon spectra is clear in advance. Properties of the collision, as the mass number~$A$ of the projectile, the projectile rapidity~$y_{nucl}$, and the impact parameter~$b$, should unambiguously specify these important initial conditions, but there are uncertainties as the nuclear stopping power and the time scale for equilibration that complicate their assessment substantially. How can we then estimate the initial conditions? A first hint on the thermalization time is given by the strong interaction (proper) time scale~\cite{ALAM_1996},
\be
        \tau_0 \approx \inv{\Lambda_{QCD}} \approx 1\;\fm,
\label{canonical_value}
\ee
which coincides with the canonical value, $\tau_0 = 1\;\fm$, used in many investigations. In the Bjorken model, an additional expression can be derived that relates the thermalization time~$\tau_0$ and the initial temperature~$T_0$ to the multiplicity distribution~$dN/dy$~\cite{HWA_1985}
\be
        T_0^3\tau_0 = \frac{2\,\pi^4}{45\,\zeta(3)\,\pi R_A^2\,4 a_k} \, \frac{dN}{dy},
\label{initial_conditions}
\ee
where for the phase transition scenario with a two-flavored QGP in the initial state, $a_k = a_q = 37\pi^2/90$. While the multiplicity distribution~$dN/dy$ at the CERN SPS can be measured experimentally, it must be extrapolated from high energy $p$-$p$ and $p$-$\bar{p}$ collision data for RHIC and LHC. By going through the literature~\cite{BJORKEN_1983,SATZ_1992,ALAM_1996,NA49_1996}, we could specify the~$dN/dy$ values listed in Tab.~\ref{Tab_Accelerator_Properties}, where the RHIC and LHC entries present lower limits. The initial temperatures given in this table were calculated with Eq.~(\ref{initial_conditions}) for the canonical thermalization time, $\tau_0 = 1\;\fm$. In several works~\cite{ALAM_1996,CLEYMANS_1997,SRIVASTAVA_1994}, Eq.~(\ref{initial_conditions}) has been used together with another relation for~$\tau_0$ and~$T_0$~\cite{KAPUSTA_1992}
\be
        \tau_0 \approx \inv{3\,T_0}
\ee
and with measurements or estimations of~$dN/dy$ as the key to the initial conditions. A different approach in the framework of hydrodynamics is presented in~\cite{SOLLFRANK_1997}, where hadronic spectra are fitted to experimental data by fixing the initial conditions, which are then used to extract photon and dilepton yields. More sophisticated theoretical calculations beyond hydrodynamics are also available to determine the initial conditions since transport models can simulate the early non-equilibrium phase of the reaction from kinetic theory. In transport models on the partonic level, e.g., the parton cascade model (PCM)~\cite{GEIGER_1992_GEIGER_1995}, parton momentum distributions are considered locally and the corresponding cell is revealed as being in local thermal equilibrium if these distributions exhibit an exponential (thermal) and isotropic distribution. Here, no initial conditions will be derived from observables or transport models, we instead restrict ourselves to the variation of~$\tau_0$ and~$T_0$ within the realistic ranges that contain values from the methods discussed above.
%
%
%----------------------------------------------------------
\subsection{\boldmath $\tau_0$ - Thermalization Time}
%----------------------------------------------------------
%
%
The thermalization time~$\tau_0$ indicates the equilibration time scale, more specifically, it denotes the proper time point after the maximum overlap of the colliding nuclei in which the system reaches local thermal equilibrium. In the Bjorken model, local thermal equilibrium sets in at a {\em constant} proper time~$\tau_0$ that describes a hyperbola in the Minkowski diagram as shown in Fig.~\ref{Fig_Space-Time_Evolution}. The dependence of the thermal photon spectra on~$\tau_0$ has the following simple form
\be
        \frac{dN}{d^2p_{\perp}\,dy} \; \propto \; \tau_0^2,
\label{tau_0_dependence}
\ee
which is now derived analytically. Whenever the proper time~$\tau$ appeared in a model quantity, 
%in Chap.~\ref{A_Simple_Model_for_Ultra-Relativistic_Heavy_Ion_Collisions}
such as the temperature~$T$, it was scaled by a factor of~$1/\tau_0$. Thus, it is sensible to perform the substitution
\be
         \tau = \tau_0\; w
%\nonumber\\        d\tau & = & \tau_0\,dw
\label{substitution}
\ee
in the computation of the spectra according to Eq.~(\ref{photon_spectrum}). By rewriting Eq.~(\ref{photon_spectrum}) as
\be
        \frac{dN}{d^2p_{\perp}\,dy} \; = \; \pi\,R_A^2 \; 
                                   \int_{\tau_1}^{\tau_2} d\tau\,\tau \; f\left( T(\tau/\tau_0),\;...\right)
\ee
with the integrand~$f$ denoting the production rate already integrated over the fluid cell rapidities~$y'$, the result of this substitution that leads to
\be
        \frac{dN}{d^2p_{\perp}\,dy} \; = \; \pi\,R_A^2 \; 
                        \tau_0^2\; \int_{w_1(=\tau_1/\tau_0)}^{w_2(=\tau_2/\tau_0)} dw\,w \; f\left( T(w),\;...\right)
\ee
can be seen very clearly as one can directly read off the $\tau_0$-dependence claimed in Eq.~(\ref{tau_0_dependence}). 

The quadratical dependence~(\ref{tau_0_dependence}) is also confirmed in Fig.~\ref{Fig_tau_0_dependence}, where the total thermal photon yields with (solid line) and without (dashed line) bremsstrahlung processes are displayed for three thermalization times, $\tau_0 = 0.1$~fm, 0.5~fm, and 1~fm. It is here instructive to note that each of these three values implies a different entropy, which can be realized by recalling that~$\tau_0$ determines the initial volume
\be
        V_0 = 2\,\pi\,R_A^2\,\tau_0
\ee
and, therefore, the entropy
\be
        S = s_0 V_0
\ee
if all other parameters remain unchanged.
\befig
        \centerline{\psfig{figure=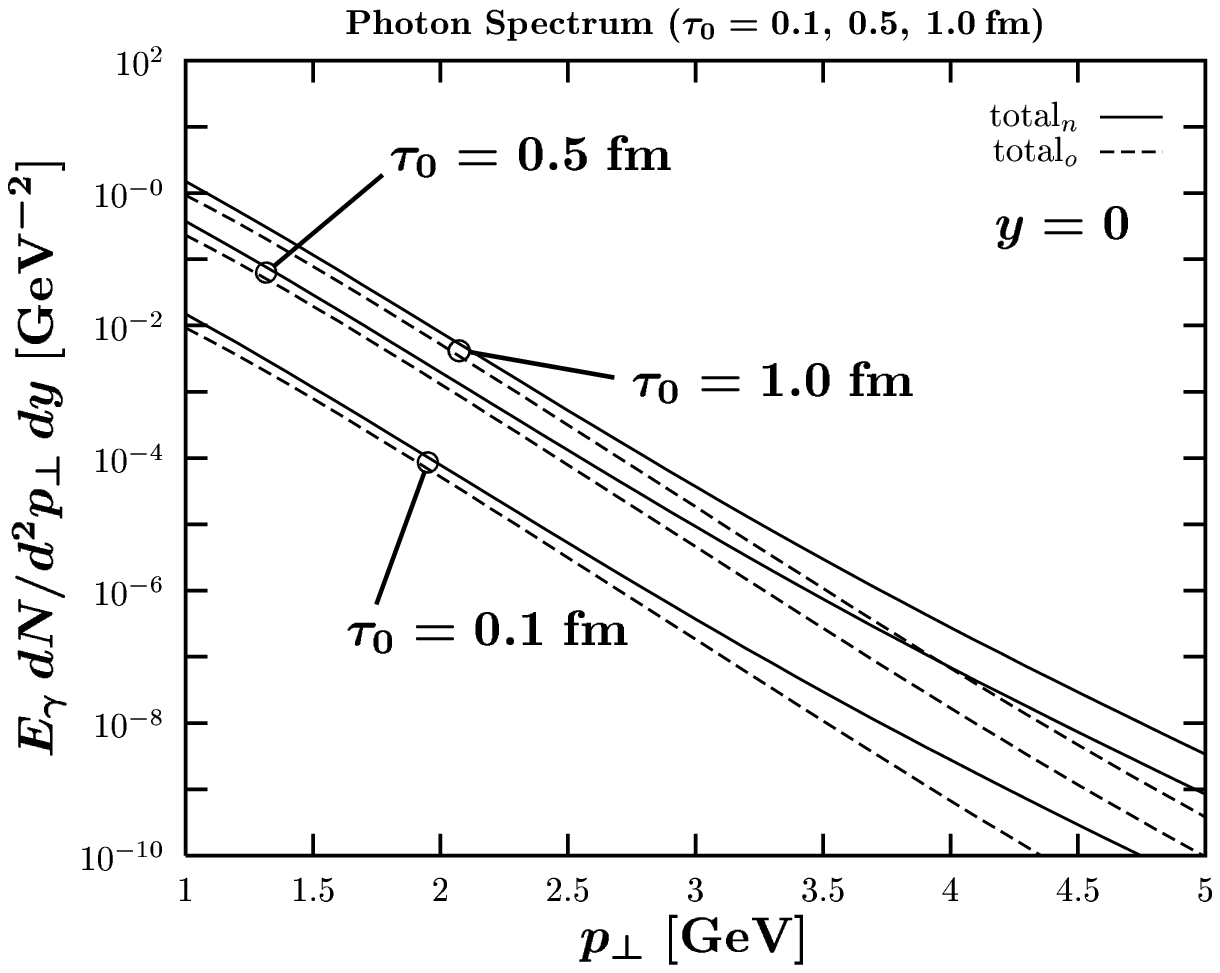,clip=,width=4.in}}
\caption[Total Thermal Photon Spectra for $\tau_0 = 0.1$~fm, 0.5~fm, and 1~fm in the Phase Transition Scenario]{Total Thermal Photon Spectra for $\tau_0 = 0.1$~fm (Lower Curves), 0.5~fm (Middle Curves), and 1~fm (Upper Curves) in the Phase Transition Scenario. As in the upper diagram of Fig.~\ref{Fig_pts_new_old_qgp_tot}, the solid lines are obtained if QGP bremsstrahlung processes are included and the dashed lines are obtained if QGP bremsstrahlung processes are neglected.}
\label{Fig_tau_0_dependence}
\efig
%
%
%----------------------------------------------------------
\newpage
\subsection{\boldmath $T_0$ - Initial Temperature}
%----------------------------------------------------------
%
%
The initial temperature~$T_0$ is the temperature of the fluid cells at the thermalization time~$\tau_0$. Before this proper time point, a temperature cannot be defined because the system is in its non-equilibrium phase still heading towards local thermal equilibrium. For the dependence of the thermal photon spectra on the initial temperature, it is not possible to give a relation as simple as~(\ref{A_dependence}) or~(\ref{tau_0_dependence}), which means that the impact of the initial temperature must be studied on the extracted thermal photon spectra. 

The importance of the initial temperature is already anticipated by the $T$-dependence of the thermal production rates and by the $T_0$-dependence of the lifetimes of the collision stages. Because the evolution of the entropy density~(\ref{entropy_density_evolution}) provides the time scale for the complete hydrodynamic expansion, the cubic $T_0$-dependence of the initial entropy density,
\be
        s(\tau_0) \; = \; s_q(\tau_0) \; \propto \; T_0^3,
\ee
is transferred to the lifetimes of the collision stages as can be seen in Eqs.~(\ref{lifetime_q}), (\ref{lifetime_c}), and~(\ref{lifetime_h}), and in Fig.~\ref{Fig_T_0_lifetimes}.
\befig[t]
        \centerline{\psfig{figure=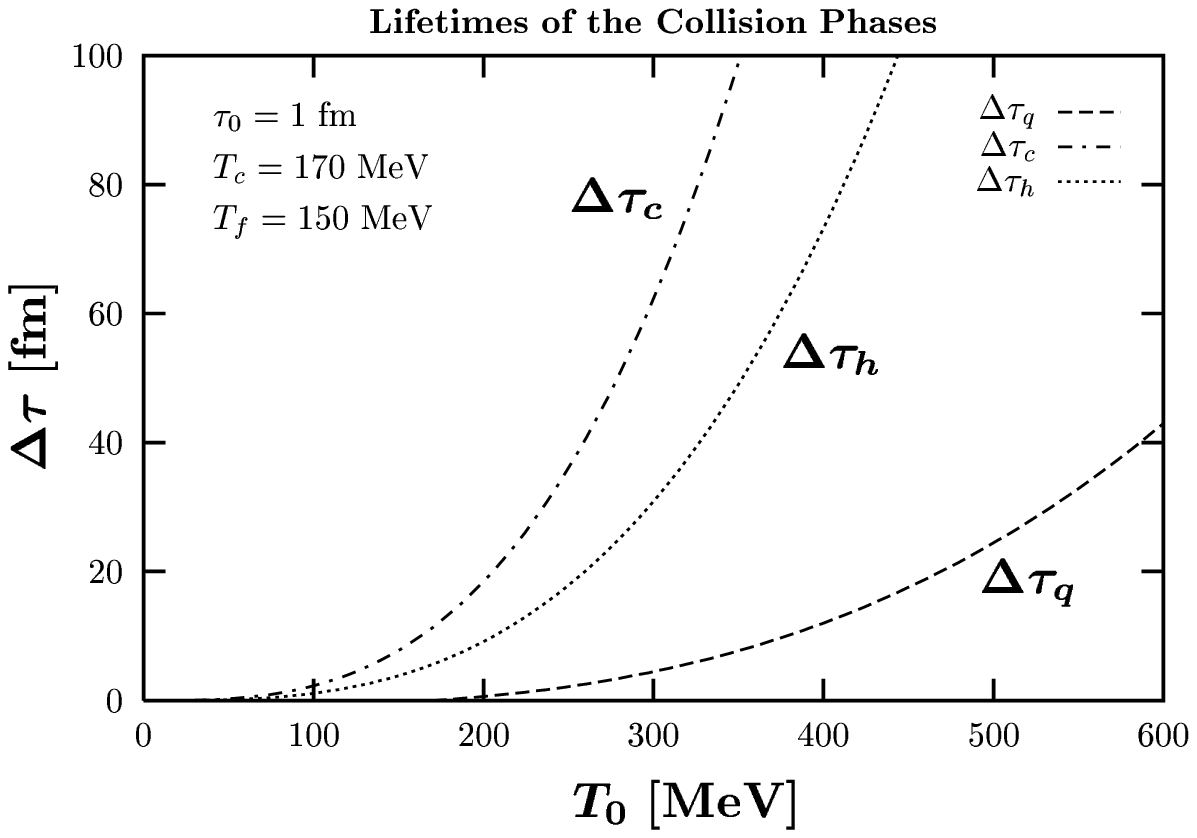,clip=,width=4.in}}
\caption[Lifetimes of the Collision Stages and their Dependence on the Initial Temperature~$T_0$]{Lifetimes of the Collision Stages and their Dependence on the Initial Temperature~$T_0$. The dashed, dot-dashed, and dotted lines illustrate the $T_0$-dependence of the lifetimes of the pure QGP, MP, and pure HHG collision stage, respectively.}
\label{Fig_T_0_lifetimes}
\efig
%

%\newpage
The above anticipation is confirmed in Figs.~\ref{Fig_T_0_200_new_old_qgp_tot} and~\ref{Fig_T_0_300_new_old_qgp_tot}, where the total photon spectra (upper diagrams) with and without bremsstrahlung processes and the photon spectra from the different processes in the QGP state of matter (lower diagrams) are illustrated for $T_0 = 200\;\MeV$ and $T_0 = 300\;\MeV$. In the upper diagrams of these figures, it can be seen that an increase of the initial temperature from $T_0 = 200\;\MeV$ (Fig.~\ref{Fig_T_0_200_new_old_qgp_tot}) up to $T_0 = 300\;\MeV$ (Fig.~\ref{Fig_T_0_300_new_old_qgp_tot}) leads to an increase in the total thermal photon yields by more than one order of magnitude and to an increase in the effect of the bremsstrahlung processes especially in the high-$p_{\perp}$ region. It is the significantly longer lifetime of the thermalized collision phase and the $T$-dependence of the thermal production rates that supports high thermal photon yields in the case of high initial temperatures. 
%and, in turn, it is the shorter QGP lifetime that weakens the bremsstrahlung effect for lower initial temperatures. 
The lower diagrams in Figs.~\ref{Fig_T_0_200_new_old_qgp_tot} and~\ref{Fig_T_0_300_new_old_qgp_tot} affirm the above observations concerning the contributions from the QGP state of matter. The QGP contribution for $T_0 = 300\;\MeV$ exceeds the one for $T_0 = 200\;\MeV$ by one order of magnitude at $p_{\perp} = 1\;\GeV$ and by more than three orders of magnitude at $p_{\perp} = 5\;\GeV$. A close look at the shapes of the QGP photon spectra is also interesting. While the $T_0 = 200\;\MeV$ spectra basically exhibit one slope that indicates a small variation in the temperature, the $T_0 = 300\;\MeV$ spectra exhibit two different slopes, where the steeper one results from the low temperature contribution of the MP and the flatter one corresponds to the high temperature contribution of the pure QGP phase. If one could really neglect transverse expansion, the occurrence of two different slopes in the thermal photon spectra would be a probable signature for a long lived MP, and thus, for the deconfinement phase transition. However, by taking into account also transverse expansion, the Doppler effect can mimic high temperatures and cause a flat slope~\cite{ALAM_1996,CLEYMANS_1997}.

It is also instructive to look at Figs.~\ref{Fig_T_0_200_qgp_hhg} and~\ref{Fig_T_0_300_qgp_hhg} that present for $T_0 = 200\;\MeV$ and $T_0 = 300\;\MeV$, respectively, the contributions from the two states of matter with and without bremsstrahlung processes taken into account. In the $T_0 = 200\;\MeV$ illustration (Fig.~\ref{Fig_T_0_200_qgp_hhg}), one sees a HHG contribution that completely dominates the thermal photon spectrum if bremsstrahlung processes in the QGP are neglected (lower diagram). This changes significantly by including the production rates from the two-loop calculations (upper diagram). A similar behavior is observed for $T_0 = 300\;\MeV$ (Fig.~\ref{Fig_T_0_300_qgp_hhg}), where the inclusion of the QGP bremsstrahlung processes extends the region, in which the QGP thermal photon contribution dominates, from $p_{\perp} > 4\;\GeV$ to $p_{\perp} > 2\;\GeV$.

The comparison of the photon spectra of the different collision phases for $T_0 = 200\;\MeV$ and $T_0 = 300\;\MeV$ manifests our above statement concerning the influence of the collision phase lifetime and the temperature range on the thermal photon spectra: a small lifetime of the pure QGP phase at relatively low temperatures results in small thermal photon yields from this phase (Fig.~\ref{Fig_T_0_200_qgp_mp_hhg}), while longer lifetimes at higher temperatures particularly benefit the population of the high-$p_{\perp}$ region (Fig.~\ref{Fig_T_0_300_qgp_mp_hhg}).
\befig
        \centerline{\psfig{figure=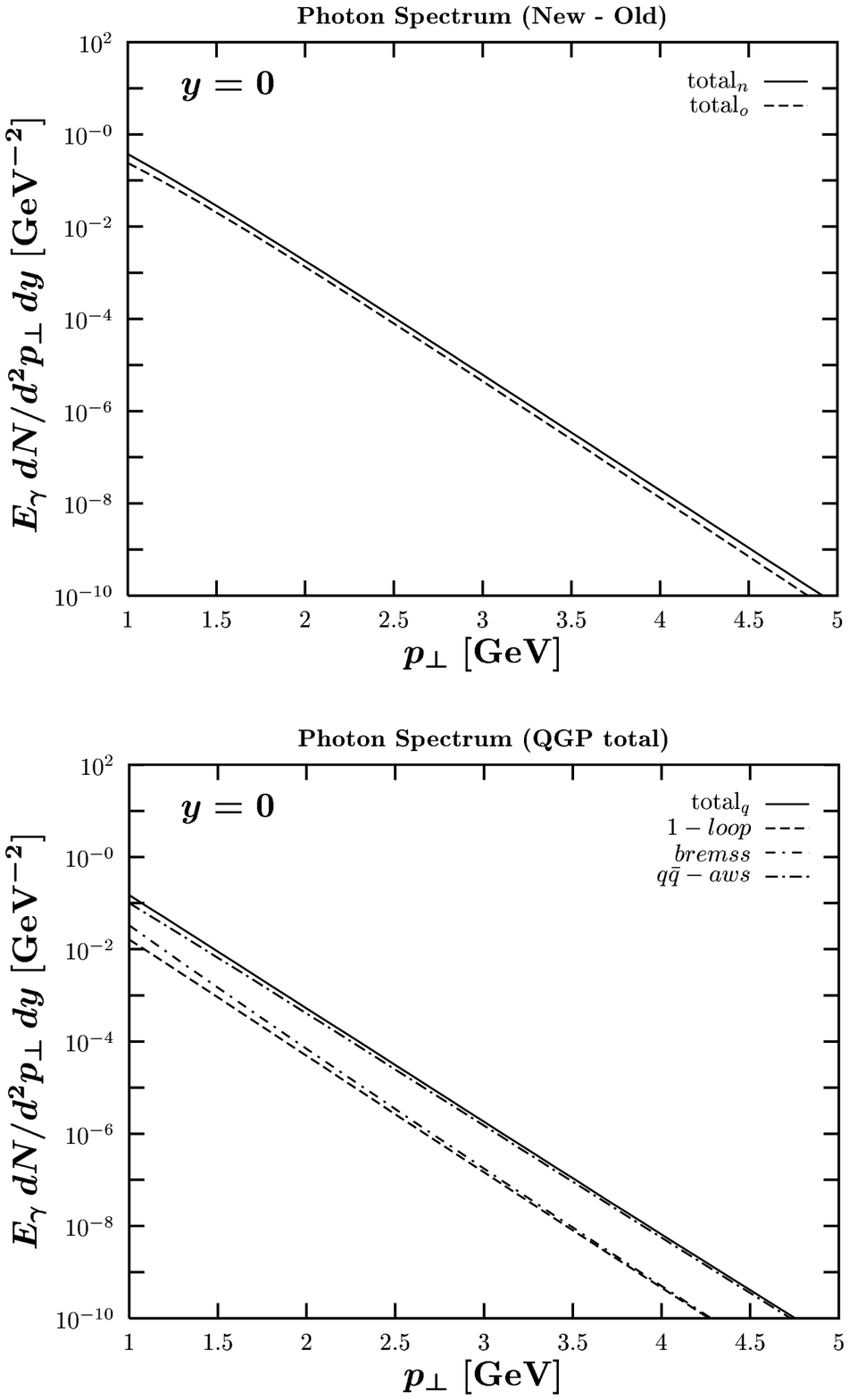,clip=,width=4.in}}
\caption[Total Thermal Photon Spectra and Thermal Photon Spectra from the QGP State of Matter in the Phase Transition Scenario with $T_0 = 200\;\MeV$]{Total Thermal Photon Spectra (Upper Diagram) and Thermal Photon Spectra from the QGP State of Matter (Lower Diagram) in the Phase Transition Scenario with $T_0 = 200\;\MeV$. Same as Fig.~\ref{Fig_pts_new_old_qgp_tot} but for $T_0 = 200\;\MeV$.}
\label{Fig_T_0_200_new_old_qgp_tot}
\efig
\befig
        \centerline{\psfig{figure=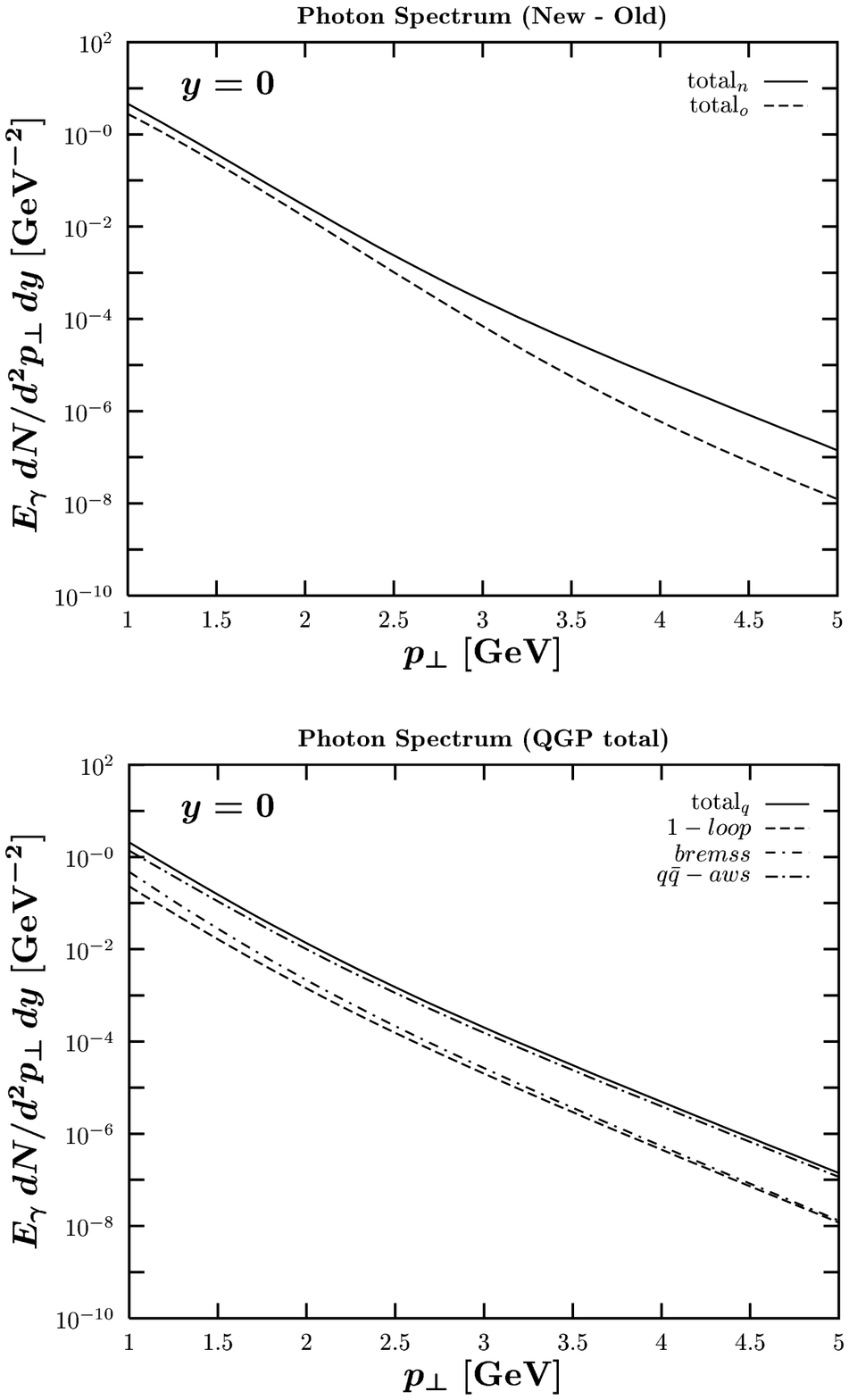,clip=,width=4.in}}
\caption[Total Thermal Photon Spectra and Thermal Photon Spectra from the QGP State of Matter in the Phase Transition Scenario with $T_0 = 300\;\MeV$]{Total Thermal Photon Spectra (Upper Diagram) and Thermal Photon Spectra from the QGP State of Matter (Lower Diagram) in the Phase Transition Scenario with $T_0 = 300\;\MeV$. Same as Fig.~\ref{Fig_pts_new_old_qgp_tot} but for $T_0 = 300\;\MeV$.}
\label{Fig_T_0_300_new_old_qgp_tot}
\efig
\befig
        \centerline{\psfig{figure=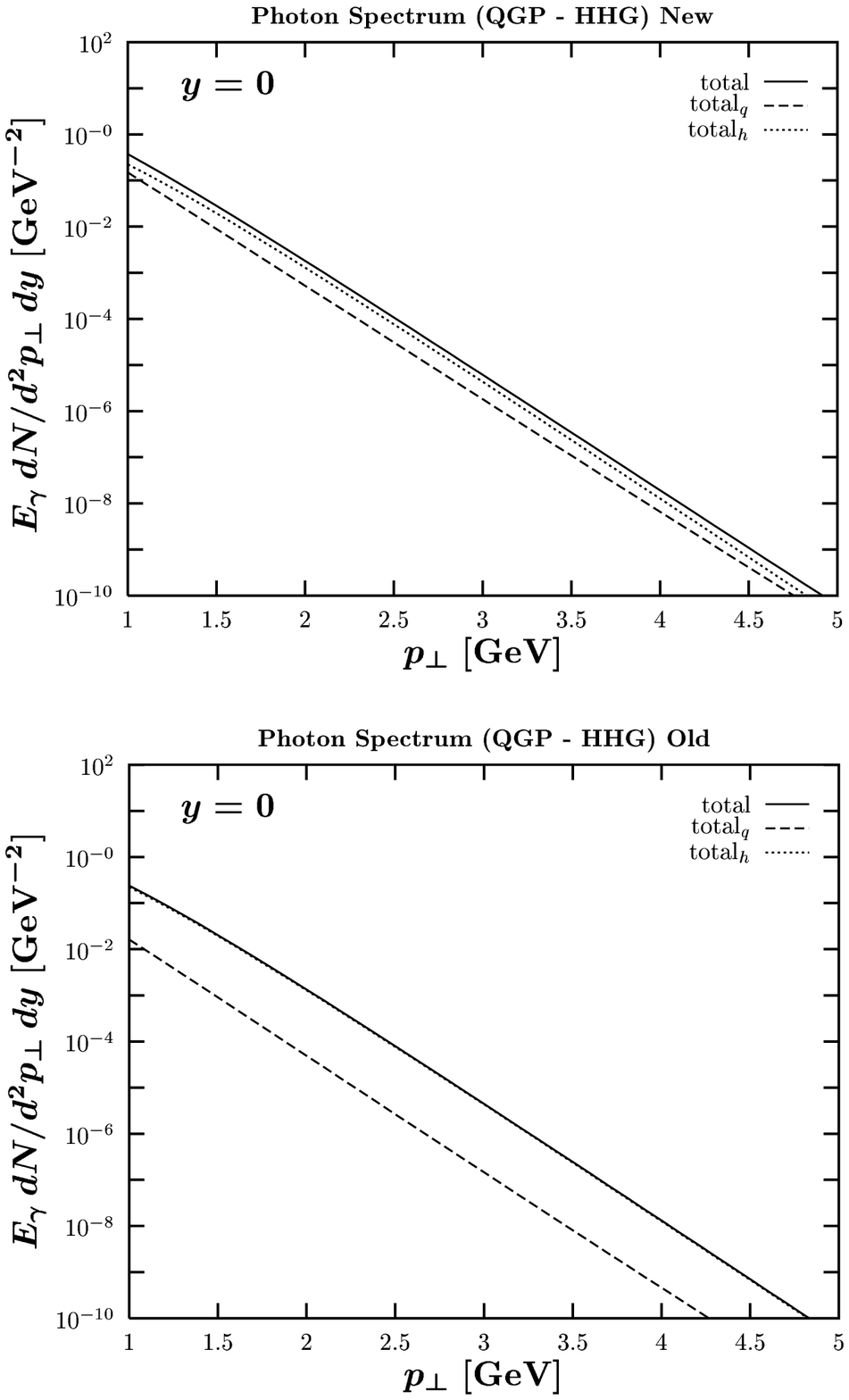,clip=,width=4.in}}
\caption[Thermal Photon Spectra from the QGP and the HHG State of Matter in the Phase Transition Scenario with $T_0 = 200\;\MeV$]{Thermal Photon Spectra from the QGP and the HHG State of Matter in the Phase Transition Scenario with $T_0 = 200\;\MeV$. Same as Fig.~\ref{Fig_pts_qgp_hhg} but for $T_0 = 200\;\MeV$.}
\label{Fig_T_0_200_qgp_hhg}
\efig
\befig
        \centerline{\psfig{figure=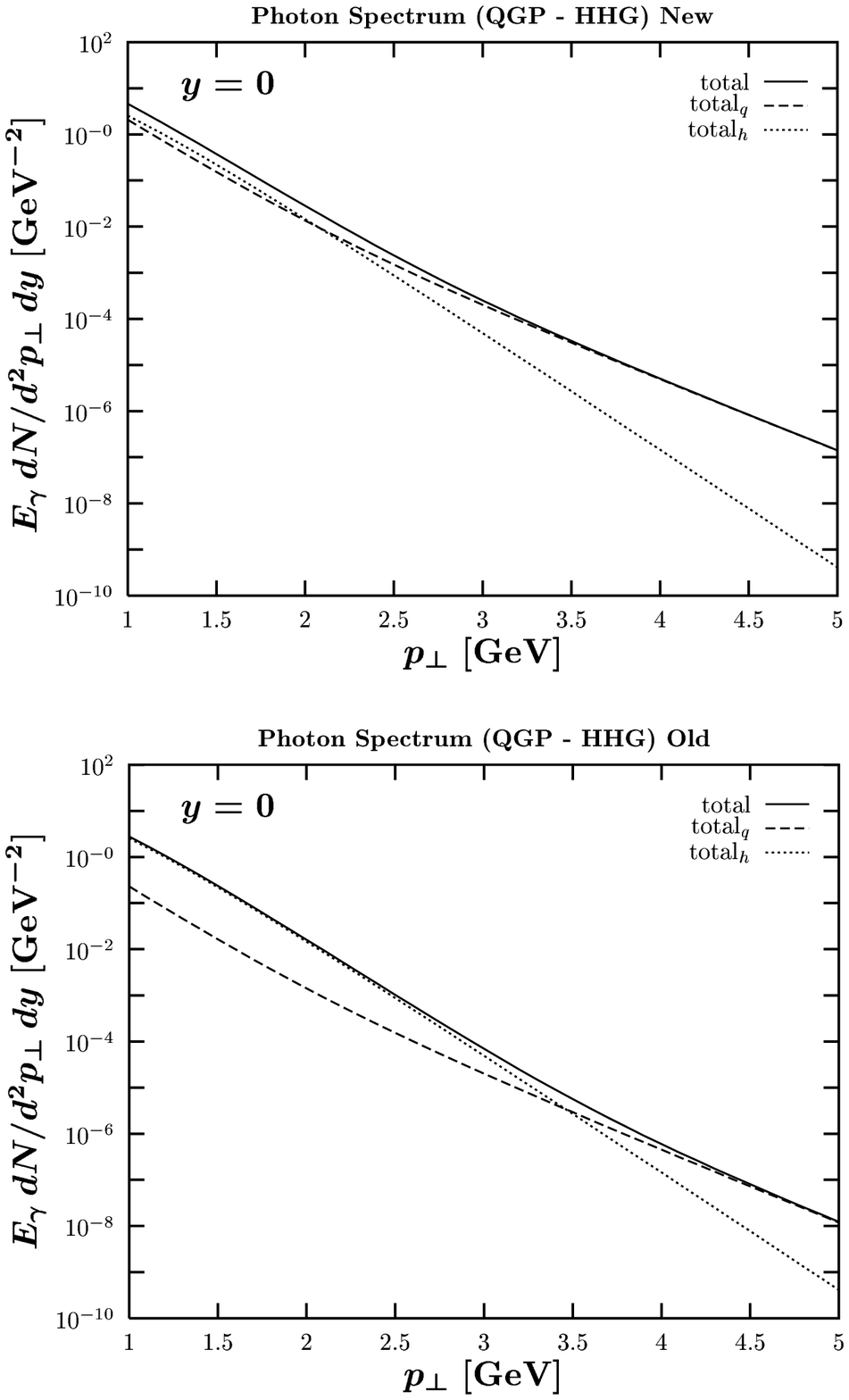,clip=,width=4.in}}
\caption[Thermal Photon Spectra from the QGP and the HHG State of Matter in the Phase Transition Scenario with $T_0 = 300\;\MeV$]{Thermal Photon Spectra from the QGP and the HHG State of Matter in the Phase Transition Scenario with $T_0 = 300\;\MeV$. Same as Fig.~\ref{Fig_pts_qgp_hhg} but for $T_0 = 300\;\MeV$.}
\label{Fig_T_0_300_qgp_hhg}
\efig
\befig
        \centerline{\psfig{figure=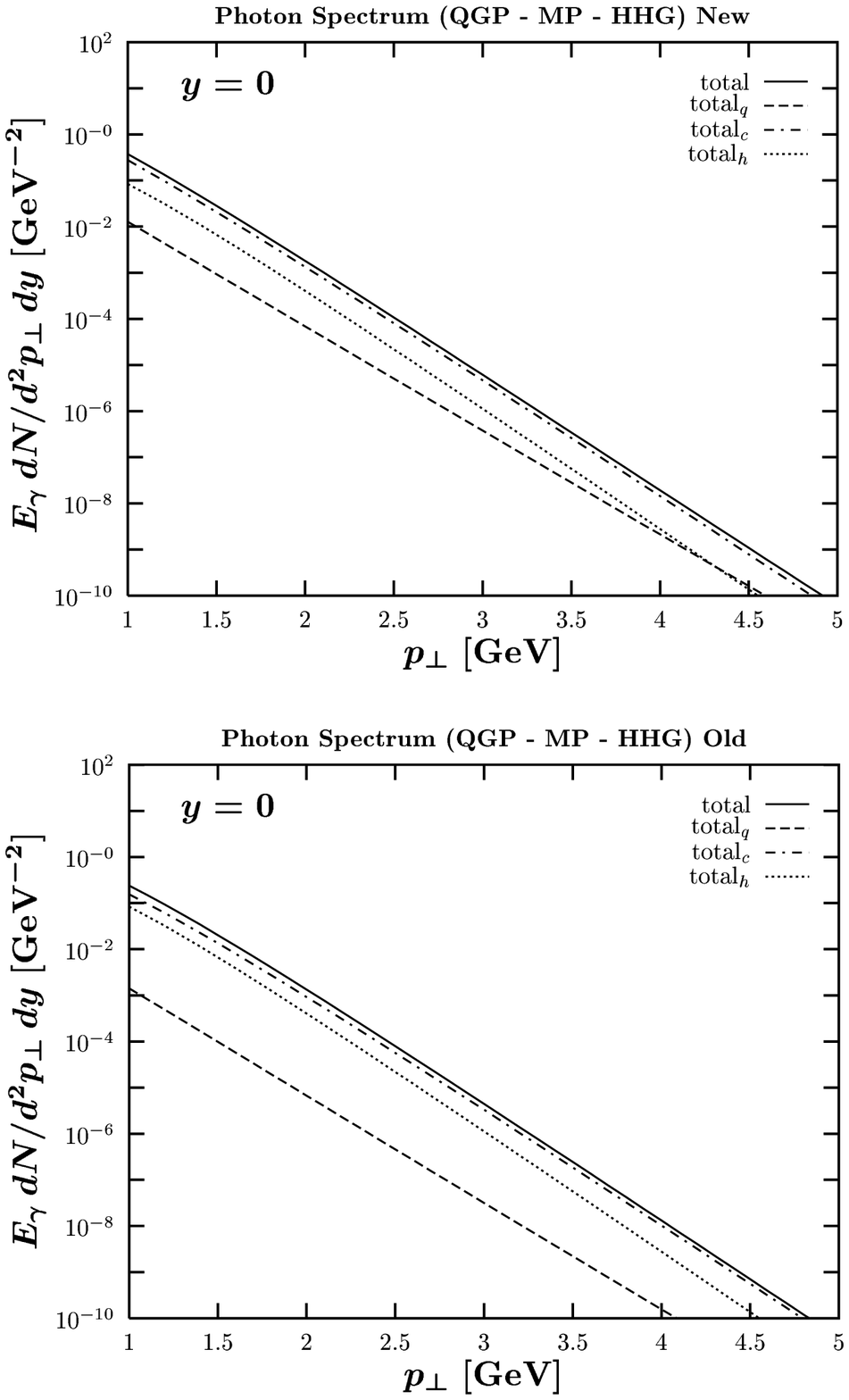,clip=,width=4.in}}
\caption[Thermal Photon Spectra from the Different Collision Stages in the Phase Transition Scenario with $T_0 = 200\;\MeV$]{Thermal Photon Spectra from the Different Collision Stages in the Phase Transition Scenario with $T_0 = 200\;\MeV$. Same as Fig.~\ref{Fig_pts_qgp_mp_hhg} but for $T_0 = 200\;\MeV$.}
\label{Fig_T_0_200_qgp_mp_hhg}
\efig
\befig
        \centerline{\psfig{figure=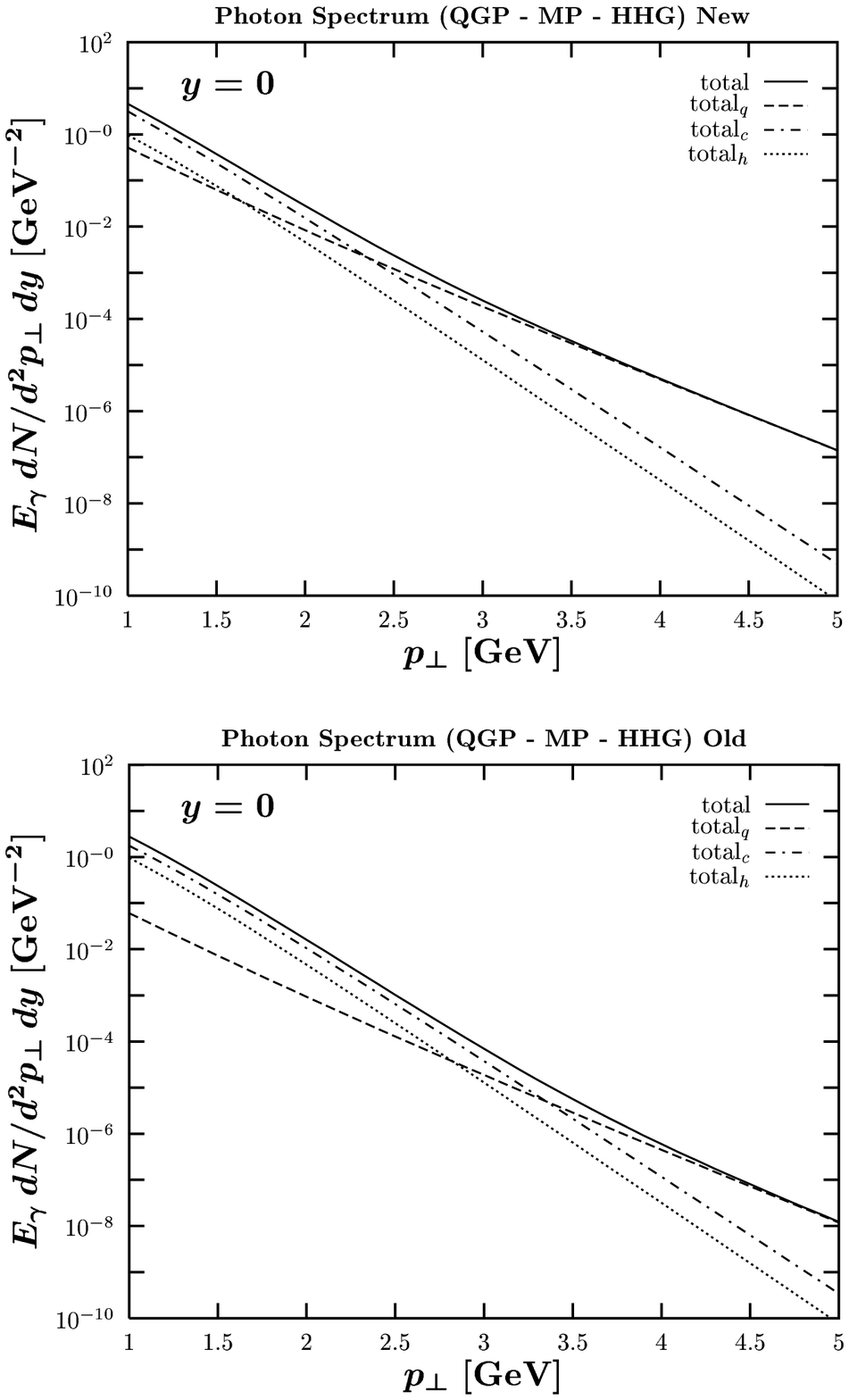,clip=,width=4.in}}
\caption[Thermal Photon Spectra from the Different Collision Stages in the Phase Transition Scenario with $T_0 = 300\;\MeV$]{Thermal Photon Spectra from the Different Collision Stages in the Phase Transition Scenario with $T_0 = 300\;\MeV$. Same as Fig.~\ref{Fig_pts_qgp_mp_hhg} but for $T_0 = 300\;\MeV$.}
\label{Fig_T_0_300_qgp_mp_hhg}
\efig
%
%
%----------------------------------------------------------
\newpage
\section{\boldmath $T_c$ - Transition Temperature}
\label{T_c-Transition_Temperature}
%----------------------------------------------------------
%
%
The transition temperature~$T_c$ characterizes the critical point in which quarks and gluons become confined, or, vice versa, in which hadrons become deconfined. It should depend only on the nature of the strong interaction. 
%, and for zero baryon density, it should not depend on the projectiles and the accelerator properties. Because the assumption of a vanishing baryon density is just an approximation that will be most valid for the central region in RHIC and LHC collisions, experiments at SPS and RHIC/LHC could uncover different values for~$T_c$. 
Experimentally, the search for the transition temperature in ultra-relativistic heavy ion experiments is the search for discontinuities indicating the existence of the QGP. According to the simple model discussed in Chap.~\ref{A_Simple_Model_for_Ultra-Relativistic_Heavy_Ion_Collisions} and Eq.~(\ref{T_c}), particularly, a different approach in the spirit of the bag model is also available, where values of the bag constant~$B$ accessible through hadron spectroscopy can be used to infer~$T_c$. The most sophisticated theoretical tool for the investigation of the transition temperature is lattice QCD, which has been applied to pure gluon theory ($N_f = 0$) and also to theories including dynamical quarks ($N_f = 4, 2, 2+1$). Depending upon the number of flavors embedded in the lattice QCD calculations, different values for the transition temperature have been found~\cite{WONG_1994,CSERNAI_1994} and we restrict ourselves to the transition temperature range
\be
        150\;\MeV\;\> {\buildrel <\over \sim}\> \;T_c\;\> {\buildrel <\over \sim}\> \;200\;\MeV
\ee
that contains most of the predicted values.

For the dependence of the thermal photon spectra on the transition temperature~$T_c$, again no simple formula can be given and the thermal photon spectra extracted within the simple model must be examined. Before proceeding in this way, it is instructive to investigate the $T_c$-dependence of the collision phase lifetimes $\Delta \tau_{q}$, $\Delta \tau_{c}$, and $\Delta \tau_{h}$. There is no influence of $T_c$ on the total lifetime 
\be
        \Delta \tau = \Delta \tau_{q} + \Delta \tau_{c} + \Delta \tau_{h}
\ee
of the local thermal equilibrium phase since the freeze-out time given in Eq.~(\ref{tau_f}) is independent of~$T_c$. Instead, the transition temperature toggles only the distribution of the total lifetime on the different collision stages, pure QGP phase, MP, and pure HHG phase, as is illustrated in Fig.~\ref{Fig_T_c_lifetimes}.
\befig[t]
        \centerline{\psfig{figure=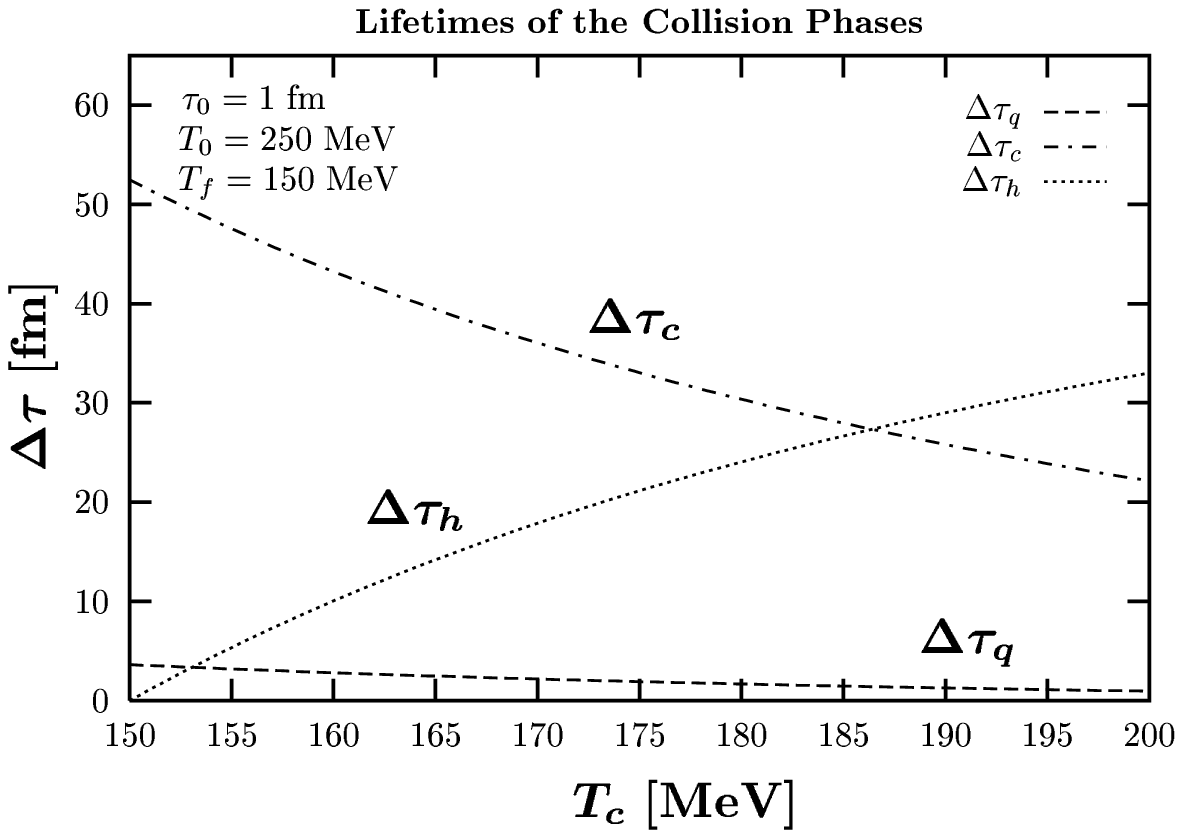,clip=,width=4.in}}
\caption[Lifetimes of the Collision Stages and their Dependence on the Transition Temperature~$T_c$]{Lifetimes of the Collision Stages and their Dependence on the Transition Temperature~$T_c$. The dashed, dot-dashed, and dotted lines illustrate the $T_c$-dependence of the lifetimes of the pure QGP, MP, and pure HHG collision stage, respectively. By summing all three lifetimes in thought, it can be seen that the total collision lifetime is independent of~$T_c$.}
\label{Fig_T_c_lifetimes}
\efig
For low values of the transition temperature, e.g., $T_c = 160\;\MeV$, the lifetimes of the pure QGP phase and the MP are relatively large and the lifetime of the pure HHG phase is relatively short. This changes by going to high transition temperatures, e.g., $T_c = 200\;\MeV$, where the lifetimes of the pure QGP phase and the MP are relatively short and the lifetime of the pure HHG phase is relatively long. This behavior of the lifetimes predicts the behavior of the thermal photon spectra under the variation of~$T_c$: contributions from the QGP state of matter will be higher for low values of~$T_c$, while contributions from the HHG state of matter will be higher for high values of~$T_c$. The thermal photon spectra shown in Figs.~\ref{Fig_T_c_qgp_hhg_new} and~\ref{Fig_T_c_new_old} affirm the above expectation. In Fig.~\ref{Fig_T_c_qgp_hhg_new}, one can compare the spectra from the two states of matter for $T_c = 160\;\MeV$ (upper diagram) with those for $T_c = 200\;\MeV$ (lower diagram) and see that the contribution from the QGP state of matter is dominant for $T_c = 160\;\MeV$, while that from the HHG state of matter is dominant for $T_c = 200\;\MeV$. This results in the QGP bremsstrahlung effect being more pronounced for $T_c = 160\;\MeV$ than for $T_c = 200\;\MeV$ as is shown in Fig.~\ref{Fig_T_c_new_old}. 
%This results in the effect of the QGP bremsstrahlung processes being more pronounced for $T_c = 160\;\MeV$ than for $T_c = 200\;\MeV$ as is exhibited in Fig.~\ref{Fig_T_c_new_old}. 
The total thermal photon yield displayed in both figures is also affected by going from  $T_c = 160\;\MeV$ to $T_c = 200\;\MeV$. It increases by about a factor of five due to a higher mean temperature in the scenario with $T_c = 200\;\MeV$.
\befig
        \centerline{\psfig{figure=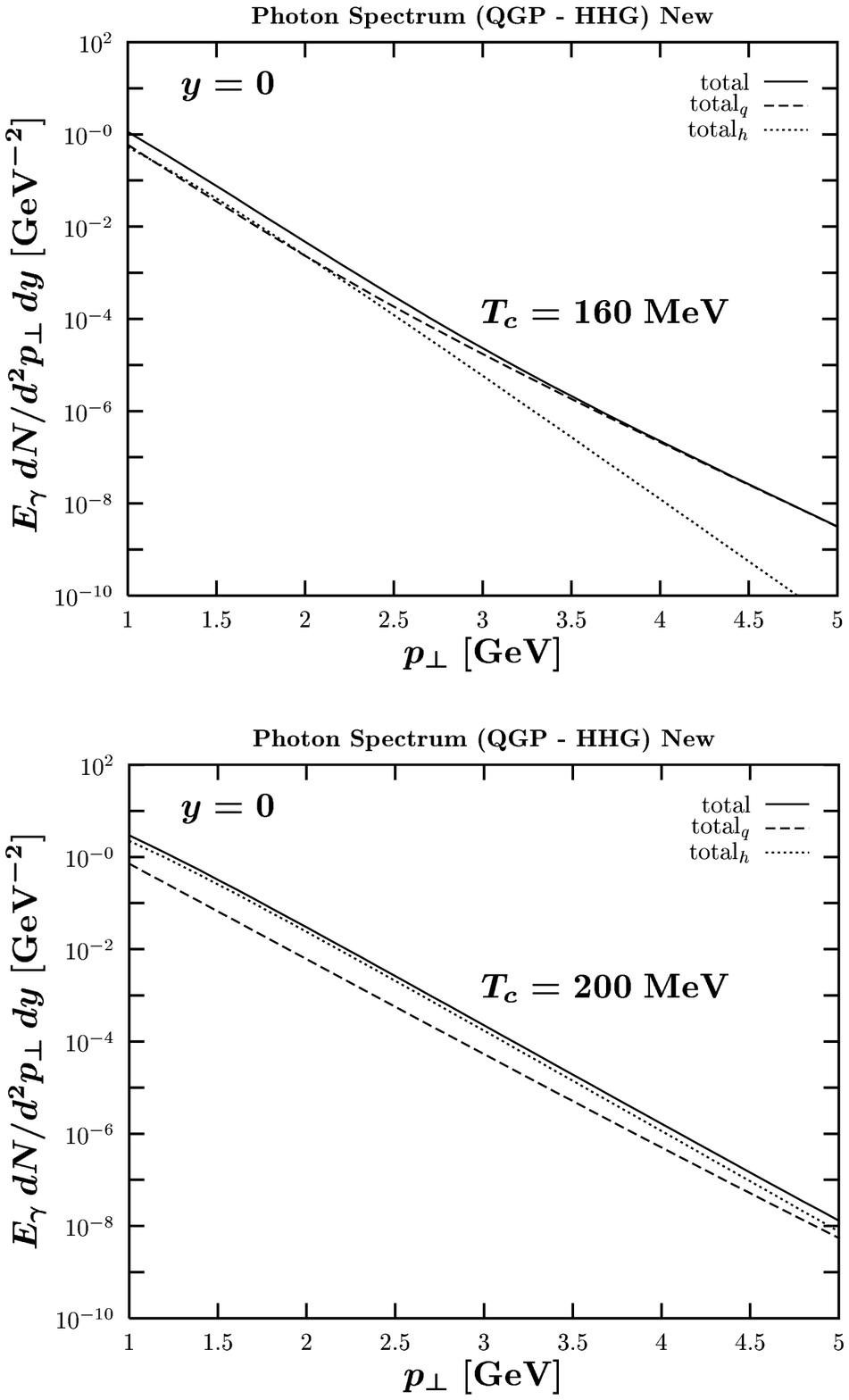,clip=,width=4.in}}
\caption[Thermal Photon Spectra from the QGP and the HHG State of Matter in the Phase Transition Scenario for $T_c = 160\;\MeV$ and $T_c = 200\;\MeV$]{Thermal Photon Spectra from the QGP and the HHG State of Matter in the Phase Transition Scenario for $T_c = 160\;\MeV$ (Upper Diagram) and $T_c = 200\;\MeV$ (Lower Diagram). Same as Fig.~\ref{Fig_pts_qgp_hhg} but for $T_c = 160\;\MeV$ and $T_c = 200\;\MeV$ with QGP bremsstrahlung processes included in both diagrams.}
\label{Fig_T_c_qgp_hhg_new}
\efig
\befig
        \centerline{\psfig{figure=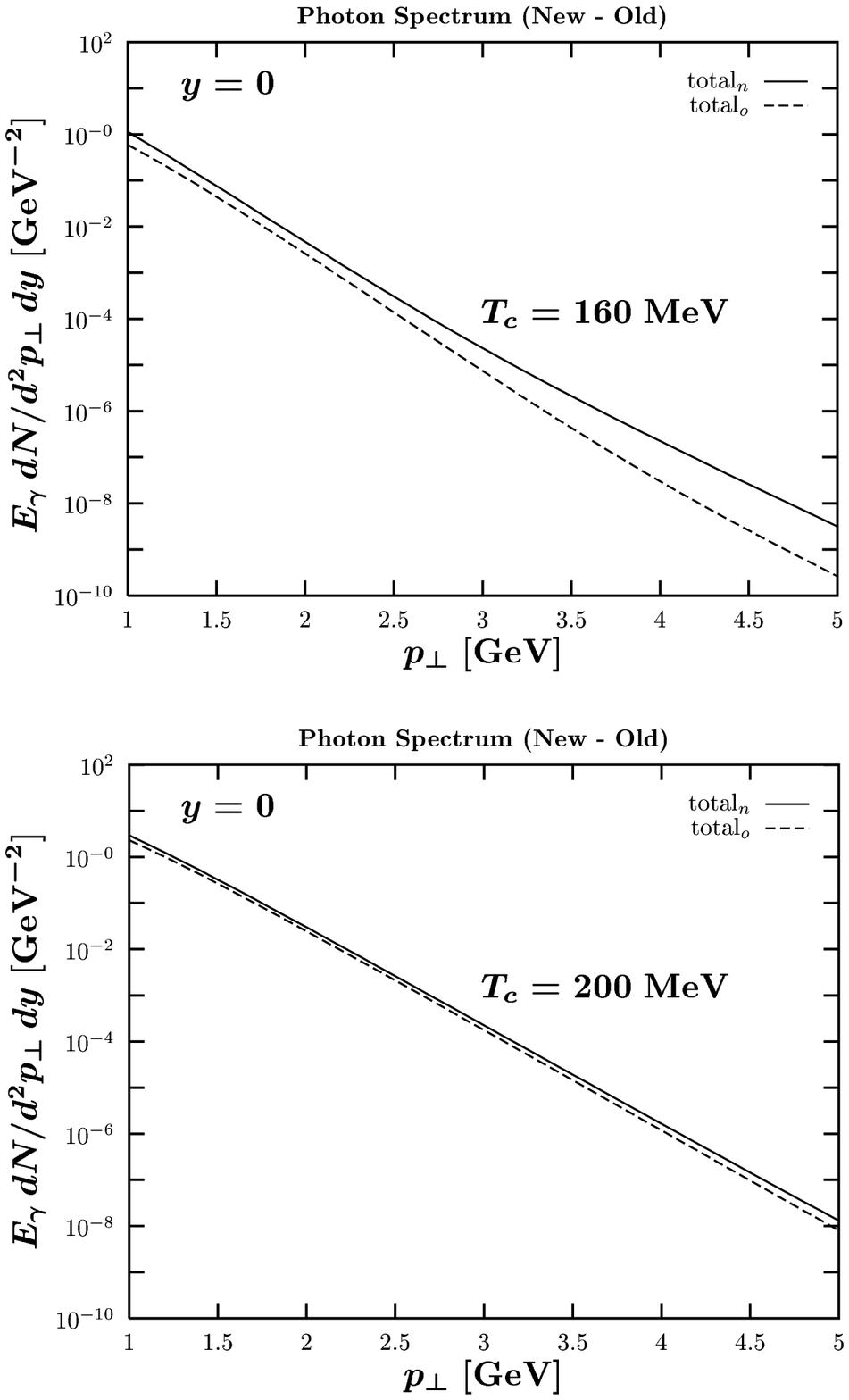,clip=,width=4.in}}
\caption[Total Thermal Photon Spectra in the Phase Transition Scenario for $T_c = 160\;\MeV$ and $T_c = 200\;\MeV$]{Total Thermal Photon Spectra in the Phase Transition Scenario for $T_c = 160\;\MeV$ (Upper Diagram) and $T_c = 200\;\MeV$ (Lower Diagram). Same as the upper diagram in Fig.~\ref{Fig_pts_new_old_qgp_tot} but for $T_c = 160\;\MeV$ and $T_c = 200\;\MeV$.}
\label{Fig_T_c_new_old}
\efig
%
%
%----------------------------------------------------------
\section{\boldmath $T_f$ - Freeze-Out Temperature}
%----------------------------------------------------------
%
%
The freeze-out temperature~$T_f$ defines the proper time point at which the system becomes dilute. At later times, the strongly interacting matter cannot be considered fluid-like, which means a hydrodynamical description is not justified. Instead, the picture of hadrons heading as free particles towards the detector is a more appropriate one for the phase beyond freeze-out. The definition of the freeze-out temperature is a matter of taste since there is no rigid criterion~\cite{ALAM_1996}. A common approach is however Landau's {\em mean free path consideration}: when the mean free path of the hadrons~$\lambda_{mfp}^{h}$ is of the order of the transverse size of the fireball,
\be
        \lambda_{mfp}^{h} = \inv{n_{h}\sigma} \approx R_A
\label{mfp_consideration}
\ee
with the number density~$n_h$ and the total cross section~$\sigma$, the system cannot be treated as fluid-like anymore. By applying this criterion, a freeze-out (hyper-)surface for each hadron species present in the strongly interacting continuum can be defined. An alternative, more involved way to determine~$T_f$ is to combine experimental hadron spectra with theoretical flow investigations. With the above means, $T_f$~values can be found in the range~\cite{BRAUN-MUNZINGER_1998}
\be
        100\;\MeV\;\> {\buildrel <\over \sim}\> \;T_f\;\> {\buildrel <\over \sim}\> \;160\;\MeV,
\ee
which is also the region over which we vary the freeze-out temperature~$T_f$.

As in the preceding sections, the look on the lifetimes of the collision stages gives a first insight into the influence of the investigated parameter. For the freeze-out temperature~$T_f$, only the lifetime of the pure HHG phase expressed in Eq.~(\ref{lifetime_h}) shows a dependence in a way that it increases with decreasing~$T_f$. A lower value of~$T_f$ consequently leads to a higher thermal photon yield from the pure HHG collision phase. However, because the temperature at the end of the pure HHG collision phase is relatively low, the enhancement from the longer lived pure HHG collision phase should be relatively small. Figure~\ref{Fig_T_f_new_old} confirms this expectation on the thermal photon yields obtained for~$T_f = 100\;\MeV$ (upper diagram) and~$T_f = 160\;\MeV$ (lower diagram).
\befig
        \centerline{\psfig{figure=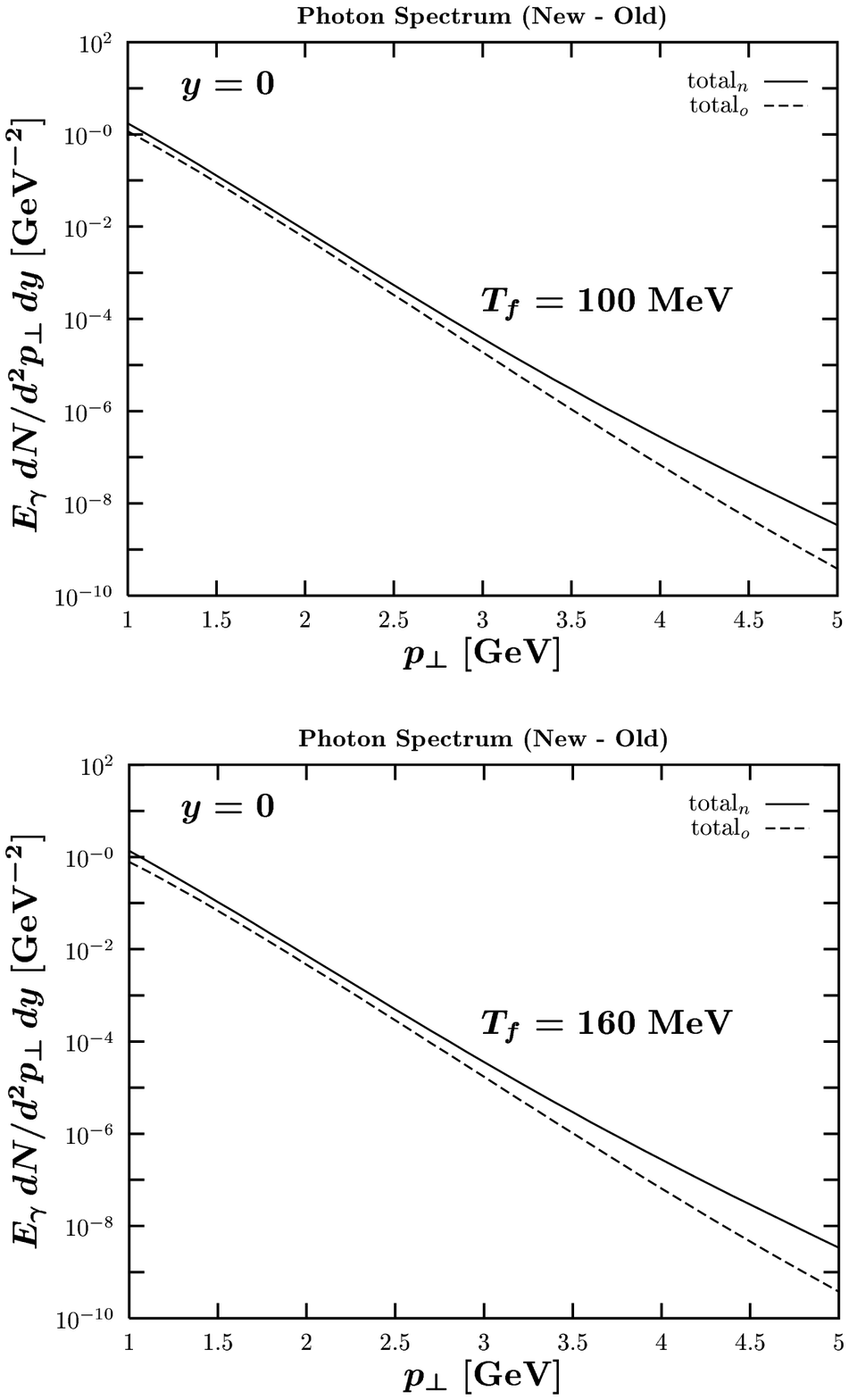,clip=,width=4.in}}
\caption[Total Thermal Photon Spectra in the Phase Transition Scenario for $T_f = 100\;\MeV$ and $T_f = 160\;\MeV$]{Total Thermal Photon Spectra in the Phase Transition Scenario for $T_f = 100\;\MeV$ (Upper Diagram) and $T_f = 160\;\MeV$ (Lower Diagram). Same as the upper diagram in Fig.~\ref{Fig_pts_new_old_qgp_tot} but for $T_f = 100\;\MeV$ and $T_f = 160\;\MeV$.}
\label{Fig_T_f_new_old}
\efig
The marginal difference in the yields is only slightly visible in the low-$p_{\perp}$ region. Thus, for the thermal photon spectra in the considered transverse momentum range, $1\;\GeV < p_{\perp} < 5\;\GeV$, the influence of the parameter~$T_f$ is weak and negligible. This fact will change if transverse expansion is included~\cite{ALAM_1996,CLEYMANS_1997}, which should be done systematically in a future extension of this investigation.

\cleardoublepage
%
%
%
%----------------------------------------------------------
\addtocontents{toc}{\protect\newpage}
\chapter{Experimental Data and Comparison with other Works}
\label{Analysis_of_Experimental_Data_on_Direct_Photon_Production}
\begin{figure}[p]
        \centerline{\psfig{figure=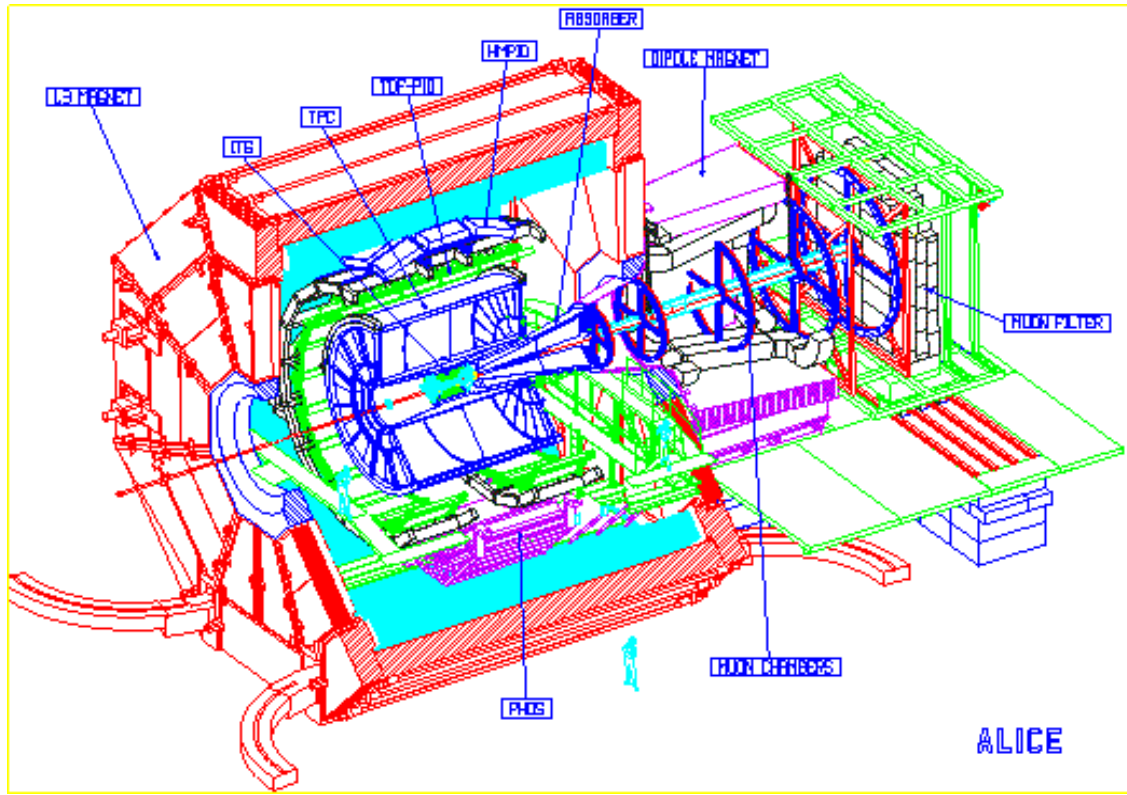,clip=,width=5.9in,angle=270}}
        \caption[The ALICE Detector]{The ALICE Detector. For the measurement of the direct photon spectrum, the photon spectrometer PHOS will be used. It is planned as a state-of-the-art electromagnetic calorimeter with very high granularity. The extreme segmentation is necessary to handle the extreme multiplicities ($dN/dy > 2500$) expected due to the prospected position in the mid-rapidity region.}
\label{Fig_ALICE_detector}
\end{figure}

%----------------------------------------------------------
%
%
As mentioned in the introduction, the examination of experimental data is the most important element of a systematic theoretical investigation. Unfortunately, the only reliable data on direct photon production in ultra-relativistic heavy ion reactions are the upper limits from fixed target $200\;A\cdot\GeV$ $S + Au$ collisions at the CERN SPS presented by the WA80 collaboration~\cite{WA80_1996,CERES_1996_KAMPERT_1997}. From the successor experiment WA98, only very preliminary data has been presented which showed direct photon production in fixed target $158\;A\cdot\GeV$ $Pb + Pb$ collisions at the CERN SPS. The final WA98 results are eagerly awaited and expected to be published in the near future. In the meanwhile, the WA98 detector has been disassembled and the electromagnetic calorimeter LEDA (LEadglass Detector Array) has been shipped to the BNL RHIC, where it will be used for the measurement of the direct photon spectrum in the PHENIX experiment~\cite{PHENIX_1998}. The higher temperatures and the large fireball volumes that will be achieved at RHIC (due to the higher center-of-mass energy of $\sqrt{s} = 200\;A\cdot\GeV$ and the employed $^{197}\!Au$~projectiles) will support higher multiplicities and higher thermal photon yields. RHIC is presently in its testing phase with first data to be taken in fall 1999. This means final direct photon data from RHIC will be available in about five years. Even higher energies ($\sqrt{s} = 5500\;A\cdot\GeV$) will be reached in collisions of $^{208}\!Pb$~nuclei at the CERN LHC that is in its construction phase and will be completed in the year 2005. For this machine, the dedicated heavy ion experiment ALICE (A Large Ion Collider Experiment)~\cite{ALICE_1995_96} is planned. As illustrated in Fig.~\ref{Fig_ALICE_detector}, it will be equipped with the finely segmented photon spectrometer PHOS which will record the ALICE direct photon data. This data might be the decisive one, however, it will take about a decade from now for it to be analyzed and published.

Under the above circumstances, it is clear that presently only the WA80 data can be inspected, while patience is required for the WA98, PHENIX, and ALICE data. We compare our results not only with the WA80 upper limits, but also with other theoretical works. There are many studies within one-fluid and even three-fluid hydrodynamical models that simulate also transverse expansion and employ more realistic EOS's for the HHG. In parallel to this thesis, the QGP bremsstrahlung processes have been considered in such a one-fluid 2+1 hydrodynamical model~\cite{SRIVASTAVA_1999,SRIVASTAVA_SINHA_1999} and we will exclusively confront our results with the ones obtained in~\cite{SRIVASTAVA_1999,SRIVASTAVA_SINHA_1999}. Other investigations that do not account for QGP bremsstrahlung processes are also briefly reviewed. They provide hints on interesting aspects that should be examined in a future extension of this investigation.
%
%
%----------------------------------------------------------
\section{WA80 Upper Limits}
%----------------------------------------------------------
%
%
%The main challenge in the experimental extraction of direct photon yields is the separation of the large background due to final state $\pi^0$ and $\eta$ decays. These decay photon yields must be removed from the data in order to get the direct photon yields. Further sources for background could be the production of high-energy photons within jets.
The measurement of the direct photon yields in $200\;A\cdot\GeV$ $S + Au$ collisions at the CERN SPS was one of the main goals in the experiment WA80. Facing a high multiplicity environment ($dN/dy \approx 200$) and the difficulty of large backgrounds from the {\em Dalitz decays} $\pi^0 \rightarrow \gamma\gamma$ and $\eta \rightarrow \gamma\gamma$, the WA80 collaboration performed the analysis not event-by-event but on a statistical basis. In principle, the direct photon yield $\gamma^{dir}$ is obtained by identifying the background photon yield $\gamma^{bkgd}$ and subtracting it from the total observed photon yield $\gamma^{obs}$,
\be
        \gamma^{dir} \; = \; \gamma^{obs} - \; \gamma^{bkgd}.
\ee
However, the WA80 analysis followed an alternative approach less sensitive to systematic error in which the photon yields were normalized with the $\pi^0$ yield
\be
        \left(\frac{\gamma}{\pi^0}\right)^{dir} \; = \; 
        \left(\frac{\gamma}{\pi^0}\right)^{obs} - \left(\frac{\gamma}{\pi^0}\right)^{bkgd}.
\ee
This is reasonable since the majority of photons detected at a given $p_{\perp}$ originates from the decay of $\pi^0$'s at nearly the same $p_{\perp}$. A different measure for the direct photon yield is also the fraction $(\gamma/\pi^0)^{obs}/(\gamma/\pi^0)^{bkgd}$ that indicates direct photons if larger than one. 

\begin{figure}[p]
        \centerline{\psfig{figure=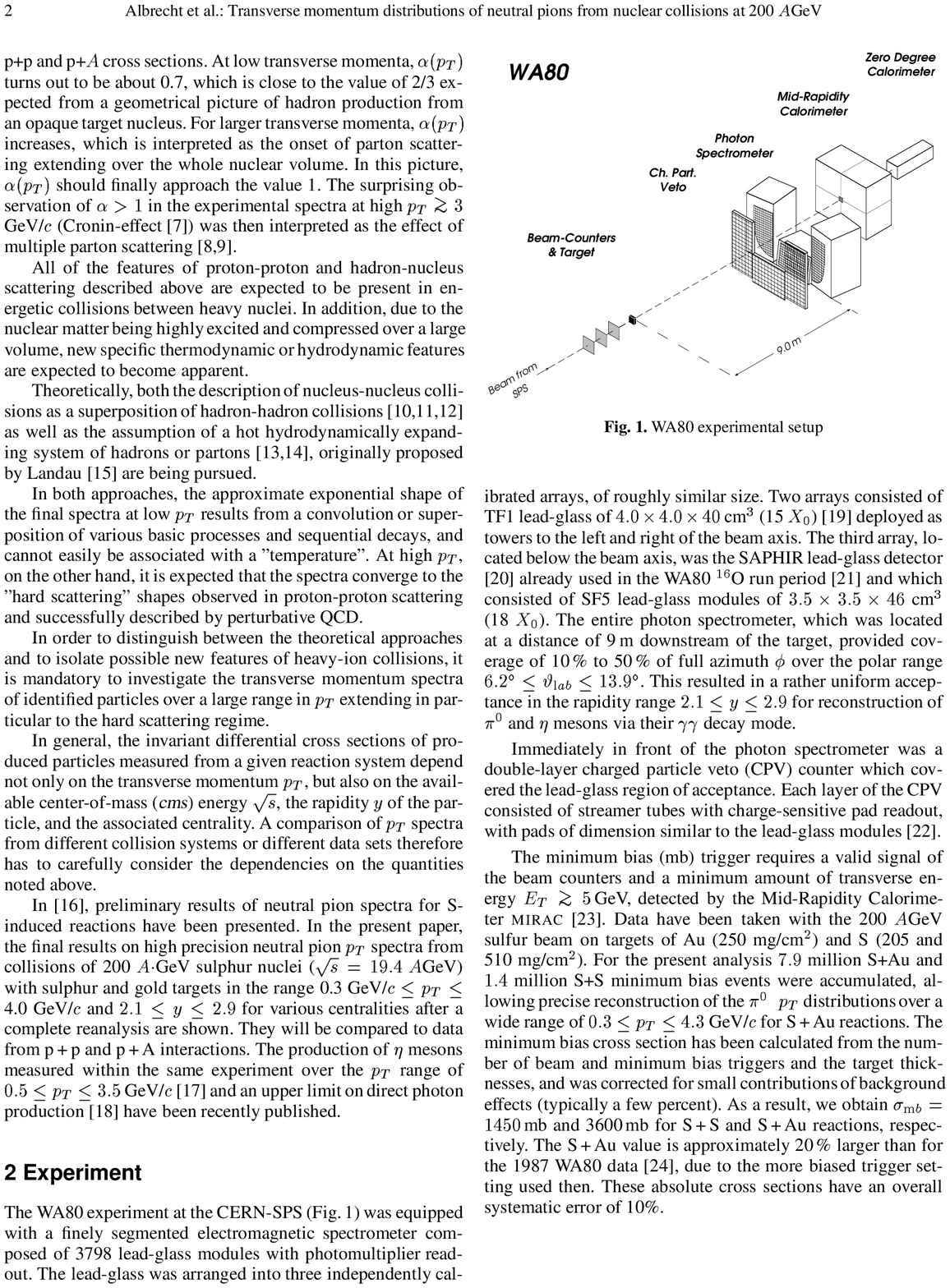,clip=,width=5.in}}
        \caption[The WA80 Detector]{The WA80 Detector. Most important in the analysis of the direct photon production has been the photon spectrometer. It was a lead glass calorimeter of fine granularity and good energy resolution that covered the laboratory frame pseudorapidity range $2.1 < \eta < 2.9$ (corresponding to the mid-rapidity range in the center-of-mass frame).}
\label{Fig_WA80_detector}
\end{figure}

The above expression exhibits the identification of the background photons as the crucial element of the direct photon analysis. Because 98\% of the background photons are originating from the Dalitz decays $\pi^0 \rightarrow \gamma\gamma$ and $\eta \rightarrow \gamma\gamma$, it is clear that the identification of background photons coincides with the reconstruction of $\pi^0$ and $\eta$ mesons. Equipped with a lead glass photon spectrometer of fine granularity and good energy resolution, 
%which covered the laboratory frame pseudorapidity range $2.1 < \eta < 2.9$ (corresponding to the mid-rapidity range in the center-of-mass frame), 
the WA80 detector shown in Fig.~\ref{Fig_WA80_detector} was well suited for this task. Using the {\em invariant mass method}
\be
        M_{\gamma\gamma}^2 = (E_{\gamma_1} + E_{\gamma_2})^2 + (\vec{p}_{\gamma_1} + \vec{p}_{\gamma_2})^2,
\ee
$\pi^0$ and $\eta$ mesons were reconstructed up to transverse momenta of several GeV. The main uncertainties in this method, which in fact control the systematic errors, arise from shower overlap due to the high photon multiplicity and from combinatorial background. The combinatorial background, however, could be determined with the {\em event mixing method} in which the invariant mass method is applied on photon pairs from artificial uncorrelated events.

The final result from the WA80 analysis of {\em central collisions} is an average direct photon excess over background sources of $5.0\% \pm 0.8\% (\mbox{stat.}) \pm 5.8\% (\mbox{syst.})$ over the transverse momentum range $0.5\;\GeV \le p_{\perp} \le 2.5\;\GeV$. With the measured direct photon yields and the statistical and systematical errors obtained for central collisions, also upper limits at the 90\% confidence level have been calculated for the direct photon yield in each $p_{\perp}$-bin up to about 3.5 GeV~\cite{WA80_1996}. These upper limits are the most interesting experimental results with regard to this investigation and we examine them within our approach to the thermal photon yields.

In Chap.~\ref{Systematic_Investigation:_Thermal_Photons}, the initial temperature~$T_0$ has been found as a model-parameter with a very strong effect on the thermal photon spectra. We therefore use the WA80 upper limits to extract an upper limit for the initial temperature~$T_0$ in the phase transition scenario with the remaining parameters set as follows
\begin{eqnarray}
        y_{nucl} & = & 3.0      \;\;\;\;(\sqrt{s} = 20\;A\cdot\GeV),
\nonumber\\
             A & = & 32         \;\;\;\;(^{32}S - ^{197}\!\! Au),
\nonumber\\
           g_q & = & 37         \;\;\;\;\mbox{(two-flavored QGP)},
\nonumber\\
           g_h & = & 3          \;\;\;\;\;\mbox{(ideal massless pion gas)},
\nonumber\\
        \tau_0 & = & 1\; \mbox{fm},
\nonumber\\
           T_c & = & 170\; \mbox{MeV},
\nonumber\\
           T_f & = & 150\; \mbox{MeV}.
\label{WA80_comparison}
\end{eqnarray}
Because the focus of this thesis is on the effects of the QGP bremsstrahlung processes, two maximum values for the initial temperature~$T_0^{max}$ are obtained such that the computed thermal photon yield does not exceed the WA80 upper limits. In Fig.~\ref{Fig_WA80_T_0_max_T_c_170}, one sees that with QGP bremsstrahlung processes (solid line) a maximum initial temperature of~$T_0^{max} = 185\;\MeV$ (upper diagram) is found, while without QGP bremsstrahlung processes (dashed line) a maximum initial temperature of~$T_0^{max} = 200\;\MeV$ (lower diagram) is found. 
%In the upper diagram of Fig.~\ref{Fig_WA80_T_0_max_T_c_170}, one sees that the upper limits (data points) allow a maximum initial temperature of~$T_0^{max} = 185\;\MeV$ if QGP bremsstrahlung processes are included (solid line). Without QGP bremsstrahlung contributions (dashed line),  
By lowering the transition temperature~$T_c$, these maxima can be extended to higher values since a lower transition temperature reduces the total thermal photon yield. (See Sec.~\ref{T_c-Transition_Temperature}.) We illustrate the effect of a $10\;\MeV$ lower transition temperature of
\bea
        T_c & = & 160\; \mbox{MeV},
\eea
in Fig.~\ref{Fig_WA80_T_0_max_T_c_160}, where the maximum initial temperatures~$T_0^{max} = 195\;\MeV$ (upper diagram) and~$T_0^{max} = 215\;\MeV$ (lower diagram) are identified with (solid line) and without (dashed line) QGP bremsstrahlung contributions, respectively. The difference between the maxima with and without QGP bremsstrahlung processes has grown because the QGP phase becomes more important with lower values of~$T_c$ and higher values of~$T_0$. In summary, the inclusion of the QGP bremsstrahlung processes reduces the possible range of initial temperatures with regard to the WA80 upper limits down to lower values by about $15\;\MeV$ for high $T_c$'s and about $20\;\MeV$ for low $T_c$'s. There is basically no effect of the QGP bremsstrahlung processes on the shape of the thermal photon spectrum which is a consequence of a relatively small QGP contribution in the above scenarios. For higher initial temperatures as expected at RHIC and LHC, the difference in the maximum initial temperatures and also in the shape of the spectrum due to the bremsstrahlung effects will be more pronounced.
\befig
        \centerline{\psfig{figure=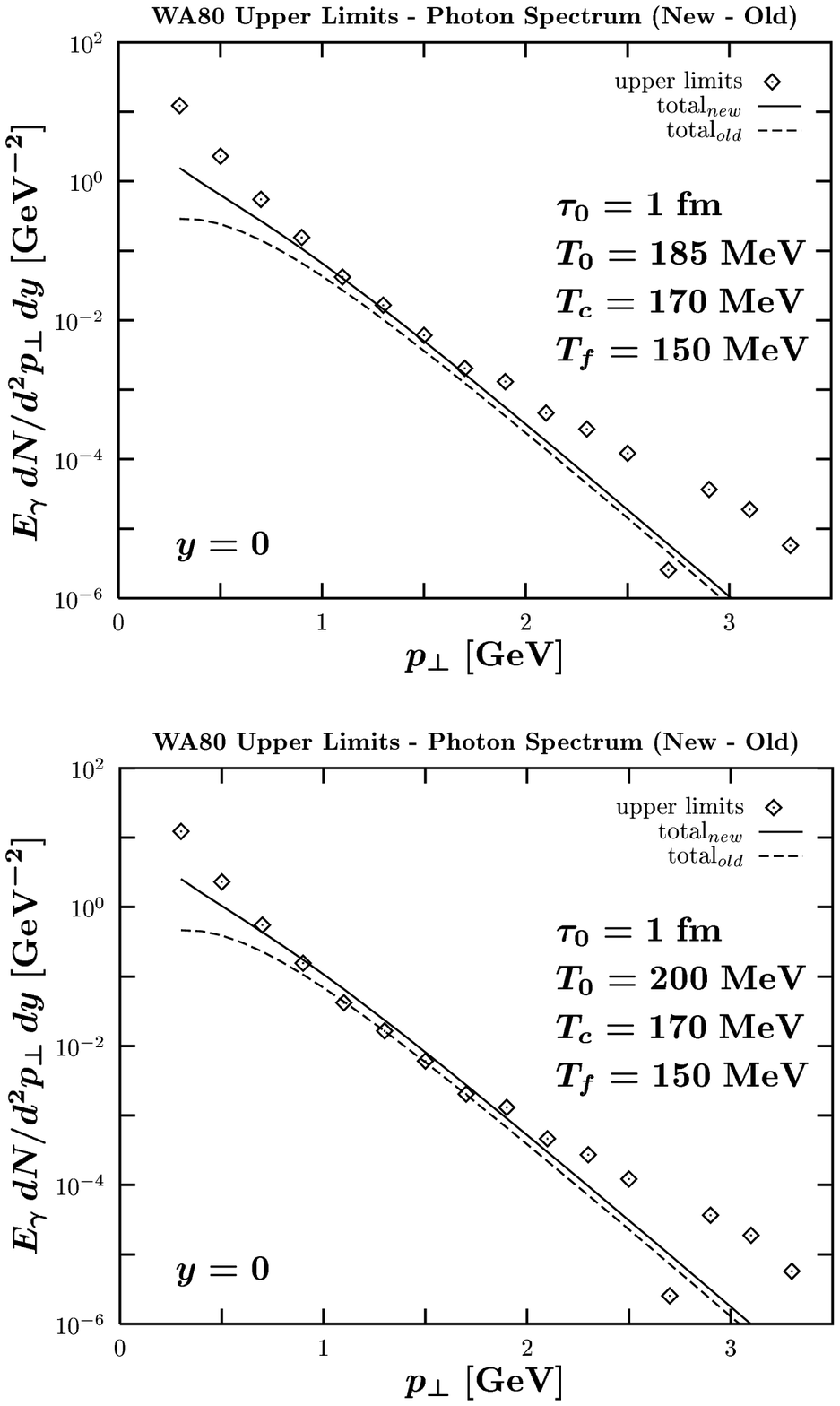,clip=,width=4.in}}
\caption[WA80 Upper Limits and the Total Thermal Photon Spectra in the Phase Transition Scenarios ($T_c = 170\;\MeV$) with $T_0 = 185\;\MeV$ and $T_0 = 200\;\MeV$]{WA80 Upper Limits and the Total Thermal Photon Spectra in the Phase Transition Scenarios ($T_c = 170\;\MeV$) with $T_0 = 185\;\MeV$ (Upper Diagram) and $T_0 = 200\;\MeV$ (Lower Diagram). The WA80 upper limits at the 90\% confidence level for the direct photon yield are represented in the data points, while the solid and dashed lines indicate the computed total thermal photon yields with and without QGP bremsstrahlung contributions, respectively.}
\label{Fig_WA80_T_0_max_T_c_170}
\efig
\befig
        \centerline{\psfig{figure=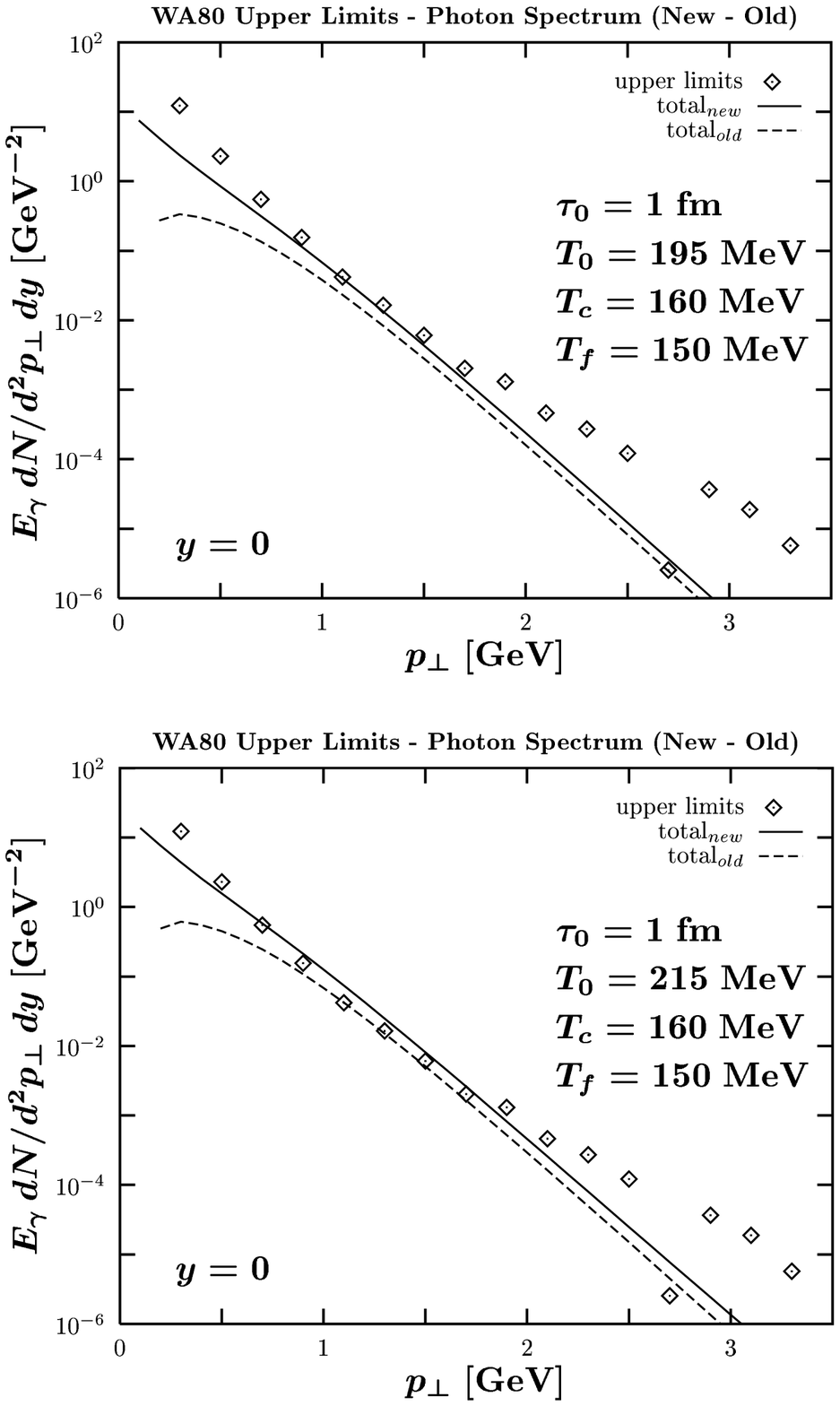,clip=,width=4.in}}
\caption[WA80 Upper Limits and the Total Thermal Photon Spectra in the Phase Transition Scenarios ($T_c = 160\;\MeV$) with $T_0 = 195\;\MeV$ and $T_0 = 215\;\MeV$]{WA80 Upper Limits and the Total Thermal Photon Spectra in the Phase Transition Scenarios ($T_c = 160\;\MeV$) with $T_0 = 195\;\MeV$ (Upper Diagram) and $T_0 = 215\;\MeV$ (Lower Diagram). As in Fig.~\ref{Fig_WA80_T_0_max_T_c_170}, the data points represent the WA80 upper limits and the solid and dashed lines represent the computed total thermal photon yields with and without QGP bremsstrahlung contributions, respectively.}
\label{Fig_WA80_T_0_max_T_c_160}
\efig

Additional remarks should be made on the comparison of the WA80 upper limits with the computed thermal photon spectra. Since the WA80 upper limits have been obtained from the measured direct photon yield and the corresponding statistical and systematical errors, not too much attention should be paid to the shape indicated by the upper limits. For example, the values of the upper limits in the region $p_{\perp} \ge 2\;\GeV$, which are high in comparison with the extracted thermal photon spectra, can be attributed to large statistical (less multiplicity) and large systematical (shower overlap) errors. One can, of course, also suspect a flattening of the direct photon spectrum due to prompt photons and transverse expansion, but we think that such statements inferred from the WA80 upper limits cannot be made without having a favor for speculation. On the other hand, it is interesting that the WA80 upper limits provide space for such phenomena. Another deviation of the shape of the thermal photon spectrum from the shape suggested by the upper limits is observed for $p_{\perp} \le 1\;\GeV$. Although a lowering of the freeze-out temperature~$T_f$ would slightly reduce the gap, one should remember that the implemented rates have been calculated in the asymptotic regime $E_{\gamma} \gg T$, which means that only the thermal photon spectra for $p_{\perp} \ge 1\;\GeV$ should be seriously considered.
%
%
%----------------------------------------------------------
\newpage
\section{Comparison with other Works}
%----------------------------------------------------------
%
%
In parallel to this thesis, an older investigation of thermal photon production in $200\;A\cdot\GeV$ $S + Au$ collisions~\cite{SRIVASTAVA_1994} has been modified~\cite{SRIVASTAVA_SINHA_1999} by implementing the QGP bremsstrahlung processes and a HHG EOS that included every hadron listed in the particle data table~\cite{PDB_1996} up to a mass of 2.5 GeV. The employed model describes besides a longitudinal also a cylindrically symmetric transverse expansion, where the SHASTA algorithm~\cite{SHASTA_1986} was used to solve the 2+1 hydrodynamic equations with the EOS. This EOS has been constructed as is outlined in Chap.~\ref{A_Simple_Model_for_Ultra-Relativistic_Heavy_Ion_Collisions}, but with the mentioned much richer HHG EOS in the modified version. The transition and the freeze-out temperatures were set to $T_c = 160\;\MeV$ and $T_f = 120\;\MeV$ and with Eq.~(\ref{initial_conditions})\footnote{For the projectile radius, the slightly different phenomenological expression $R_A = 1.2 \mbox{ fm } A^{1/3}$ was used.} and $dN/dy = 225$, an initial temperature of $T_0 = 203\;\MeV$ was obtained by assuming $\tau_0 = 1\;\fm$. For the thermal photon production from the QGP state of matter, the rates were implemented as in this thesis, while a different approach was followed for the photon emission from the HHG state of matter: the results of Kapusta et al.~\cite{KAPUSTA_1991} and Xiong et al.~\cite{XIONG_1992} were applied directly instead of using a parameterization.
%besides rates for the dominant processes, $\pi\rho \rightarrow \pi\gamma$ and $\pi\rho \rightarrow a_1 \rightarrow \pi\gamma$, also the results of Kapusta et al.~\cite{KAPUSTA_1991} for the other, not so dominant processes, (\ref{pipi-rhogamma}) - (\ref{omega-pigamma}), were taken into account.

We now extract thermal photon spectra in the phase transition scenario with the following parameters
\begin{eqnarray}
        y_{nucl} & = & 3.0      \;\;\;\;(\sqrt{s} = 20\;A\cdot\GeV),
\nonumber\\
             A & = & 32         \;\;\;\;(^{32}S - ^{197}\!\! Au),
\nonumber\\
           g_q & = & 37         \;\;\;\;\mbox{(two-flavored QGP)},
\nonumber\\
           g_h & = & 3          \;\;\;\;\;\mbox{(ideal massless pion gas)},
\nonumber\\
        \tau_0 & = & 1\; \mbox{fm},
\nonumber\\
           T_0 & = & 203\; \mbox{MeV},
\nonumber\\
           T_c & = & 160\; \mbox{MeV},
\nonumber\\
           T_f & = & 120\; \mbox{MeV}.
\label{Srivastava_Sinha_Comparison}
\end{eqnarray}
and compare the obtained results shown in Fig.~\ref{Fig_Srivastava_Sinha_Comparison} with the results of~\cite{SRIVASTAVA_SINHA_1999} shown in Fig.~\ref{Fig_Srivastava_Sinha_1999}. Concentrating on the region in which the implemented rates are valid, $p_{\perp} \ge 1\;\GeV$, one finds basically identical thermal photon yields from the QGP state of matter (dashed lines in  Fig.~\ref{Fig_Srivastava_Sinha_Comparison} and dot-dashed lines in Fig.~\ref{Fig_Srivastava_Sinha_1999}) and slightly different thermal photon yields from the HHG state of matter (dotted lines in  Fig.~\ref{Fig_Srivastava_Sinha_Comparison} and dashed lines in Fig.~\ref{Fig_Srivastava_Sinha_1999}). The slight differences in the yields from the HHG state of matter can be attributed to the different implemented production rates, the different HHG EOS, and/or the transverse behavior of the fireball. However, one sees clearly no strong effects due to the transverse expansion and the richer HHG EOS, which confirms the validity of the simple, well understood model in the considered phase transition scenario.
\befig
        \centerline{\psfig{figure=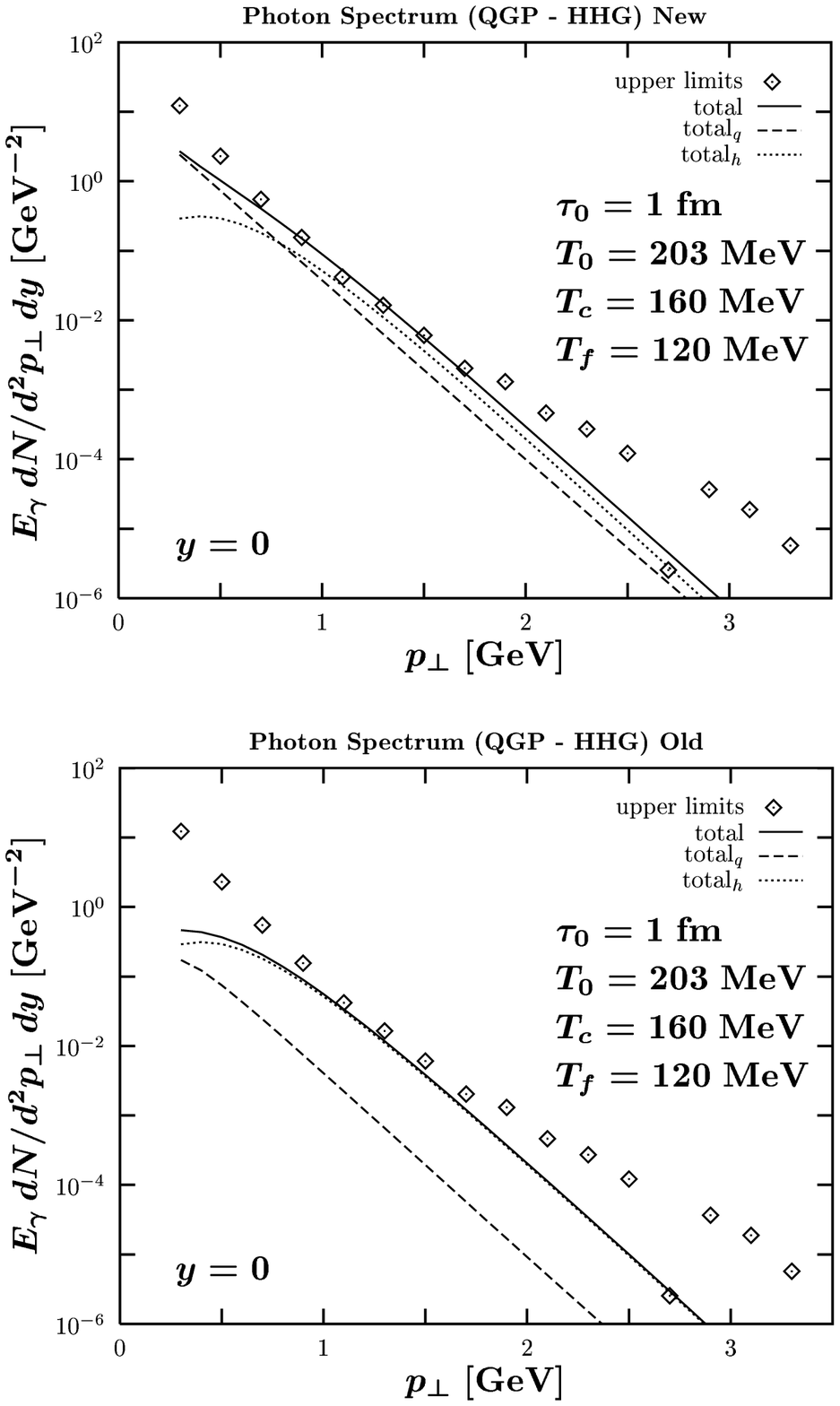,clip=,width=4.in}}
\caption[WA80 Upper Limits and Thermal Photon Spectra from the QGP and the HHG State of Matter]{WA80 Upper Limits and the Thermal Photon Spectra from the QGP and the HHG State of Matter. The thermal photon yields from the QGP and the HHG state of matter are represented in the dashed and the dotted lines, respectively, while the sum of both is the total thermal photon yield represented in the solid line. QGP bremsstrahlung processes are included in the upper plot and neglected in the lower plot. The WA80 upper limits are indicated in the data points.}
\label{Fig_Srivastava_Sinha_Comparison}
\efig
\befig
        \centerline{\psfig{figure=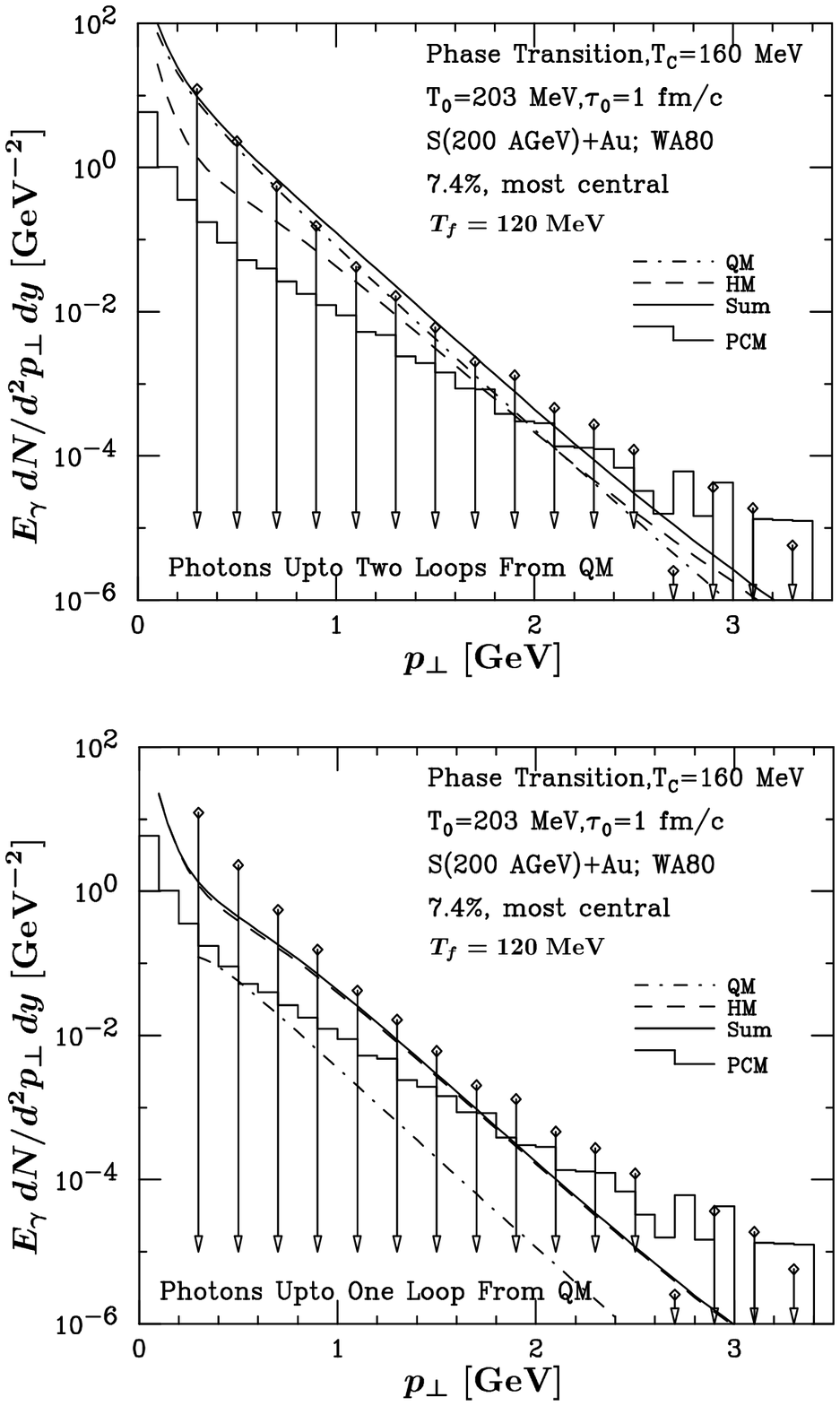,clip=,width=4.in}}
\caption[WA80 Upper Limits and Thermal Photon Spectra from the QGP and the HHG State of Matter Extracted by Srivastava and Sinha]
{WA80 Upper Limits and the Thermal Photon Spectra from the QGP and the HHG State of Matter Extracted by Srivastava and Sinha~\cite{SRIVASTAVA_SINHA_1999}. The thermal photon yields from the QGP and the HHG state of matter are represented in the dot-dashed (QM) and the dashed (HM) lines, respectively, while the sum of both is the total thermal photon yield represented in the solid (Sum) line. QGP bremsstrahlung processes are included in the upper plot and neglected in the lower plot. The WA80 upper limits are indicated in the data points with the arrows pointing downwards. A parton cascade model result for direct photon production is illustrated in the histogram (PCM).}
\label{Fig_Srivastava_Sinha_1999}
\efig

In~\cite{SRIVASTAVA_SINHA_1999}, also the no phase transition scenario is examined and in comparison to their earlier work~\cite{SRIVASTAVA_1994}, in which a different HHG EOS was employed, much lower thermal photon yields are obtained. As has already be seen in~\cite{CLEYMANS_1997}, this is a consequence of the much richer HHG EOS that alters the yields only slightly in the phase transition scenario but significantly in the no phase transition scenario. The underlying mechanism is here, as discussed in Sec.~\ref{Thermal_Photon_Spectra_in_the_No_Phase_Transition_Scenario}, the derivation of the initial temperature from the initial entropy density, in which also the EOS of the initial state of matter enters. For example, for the high number of effective degrees of freedom present in the rich HHG EOS, a lower $T_0$-value is obtained than for the small number of effective degrees of freedom present in the massless ideal pion gas. Thus, the choice of the HHG EOS determines the initial temperature in the no phase transition scenario, which has a significant influence on the thermal photon spectra.

The approach used in~\cite{SRIVASTAVA_SINHA_1999} was also applied in~\cite{SRIVASTAVA_1999}, where thermal photon spectra are examined for central $Pb + Pb$ collisions at SPS, RHIC, and LHC under the assumption of a {\em three}-flavored QGP state of matter ($N_f = 3$) and a lower freeze-out temperature of $T_f = 100\;\MeV$. Table~\ref{Tab_initial_conditions} shows the initial temperatures~$T_0$ that were obtained\footnote{The initial temperatures were again calculated from Eq.~(\ref{initial_conditions}) but this time with $a_k = a_q = 47.5\,\pi^2/90$ according to $N_f = 3$.} for the displayed assumptions of the thermalization time~$\tau_0$ and the estimated $dN/dy$ entries~\cite{KAPUSTA_1992}. 
\begin{table}
\centering      
\begin{tabular}{|c|c|c|c|c|c|c|c|}
\hline
Accelerator / & Experiment & $A$   & $\sqrt{s}$    & $y_{nucl}$ & $dN/dy$ & $\tau_0$ & $T_0$ \\
Collider      &            &       & [A$\cdot$GeV] &            &         & [fm]     & [MeV] \\ 
\hline\hline
SPS           & WA98       & $208$ & 17            & 2.8        &  825    & 1.0      & 190   \\ 
\hline\hline
RHIC          & PHENIX     & $208$ & 200           & 5.3        & 1734    & 0.5      & 310   \\
LHC           & ALICE      & $208$ & 5500          & 8.6        & 5625    & 0.5      & 450   \\
\hline
\end{tabular}
\caption[Accelerator Specific Quantities Used by Srivastava for the Extraction of Thermal Photon Spectra]{Accelerator Specific Quantities Used by Srivastava for the Extraction of Thermal Photon Spectra~\cite{SRIVASTAVA_1999}. The $dN/dy$ entries were estimated as proposed in~\cite{KAPUSTA_1992} and the $\tau_0$ assumptions were made in light of parton cascade model results~\cite{GEIGER_1992_GEIGER_1995,SRIVASTAVA_PCM_1999}. Equation~(\ref{initial_conditions}) with $a_k = a_q = 47.5\,\pi^2/90$ was used to compute the initial temperatures from the corresponding values of $dN/dy$ and $\tau_0$.}
\label{Tab_initial_conditions}
\end{table}
To compute comparable thermal photon spectra in our approach, modifications in the QGP thermal photon production rates are necessary due to the assumed three-flavored QGP: the sum of the squared electrical charge numbers includes now also the strange quark and the difference of the $N_f$-dependent integrals $J_T$ and $J_L$ becomes~\cite{SRIVASTAVA_1999}
\be
        (J_T-J_L) = 9.32.
\ee
With these modifications, the accelerator specific values listed in Tab.~\ref{Tab_initial_conditions}, and the remaining model parameters set to
\begin{eqnarray}
           g_q & = & 47.5       \;\;\;\;\mbox{(three-flavored QGP)},
\nonumber\\
           g_h & = & 3          \;\;\;\;\mbox{(ideal massless pion gas)},
\nonumber\\
           T_c & = & 160\; \mbox{MeV},
\nonumber\\
           T_f & = & 100\; \mbox{MeV}.
\label{Srivastava_Comparison}
\end{eqnarray}
we receive the thermal photon spectra shown in the lower halfs of Figs.~\ref{Fig_SPS_Comparison}, \ref{Fig_RHIC_Comparison}, and~\ref{Fig_LHC_Comparison}. The upper halfs of these figures present the results of~\cite{SRIVASTAVA_1999} to allow a direct comparison. At SPS energies, the thermal spectra from the QGP state of matter are again basically identical for $p_{\perp} \ge 1\;\GeV$, while the thermal spectra from the HHG state of matter show some differences especially in the high $p_{\perp}$-range. This difference becomes even more pronounced at RHIC and LHC energies and is a consequence of the different transverse behavior of the fireball: the transverse expansion modeled in~\cite{SRIVASTAVA_1999} causes a Doppler effect that mimics a higher temperature in the HHG. Also for the QGP state of matter, a slight blueshift in the spectra can be seen at RHIC and LHC energies. 
\befig
        \centerline{\psfig{figure=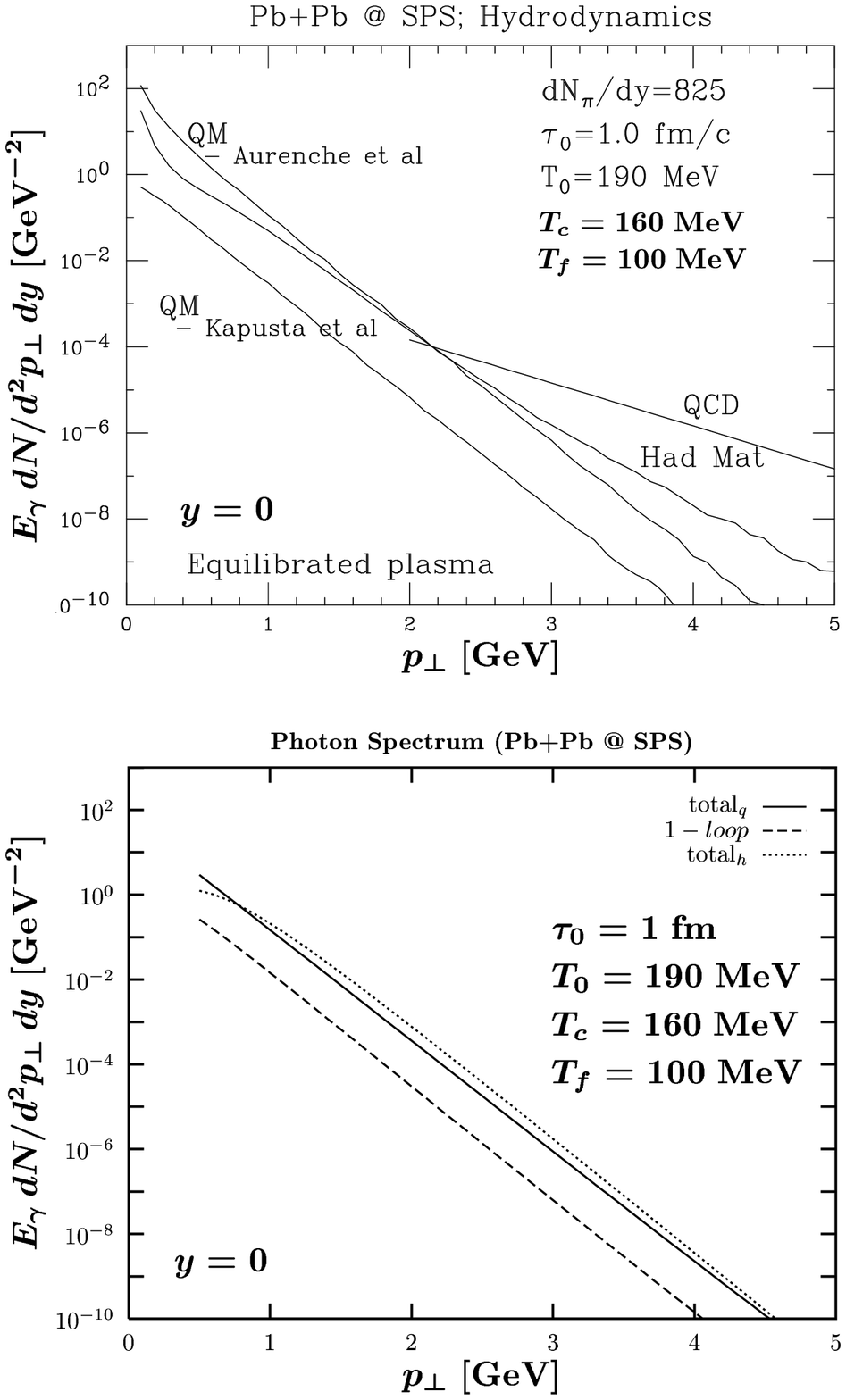,clip=,width=4.in}}
\caption[Thermal Photon Spectra from Central $Pb + Pb$ Collisions at SPS Energies]
{%\thispagestyle{empty}
Thermal Photon Spectra from Central $Pb + Pb$ Collisions at SPS Energies. The upper diagram is taken from~\cite{SRIVASTAVA_1999} (Upper Diagram), while the lower diagram has been calculated in our approach. In both diagrams, the contributions from the HHG (``Had Mat'' in the upper diagram and ``$total_h$'' (dotted line) in the lower diagram), the QGP with bremsstrahlung processes (``QM$_{Aurenche\;et\;al}$'' in the upper diagram and ``$total_q$'' (solid line) in the lower diagram), and the QGP without bremsstrahlung processes (``QM$_{Kapusta\;et\;al}$'' in the upper diagram and by ``$1-loop$'' (dashed line) in the lower diagram) are illustrated. In addition, the upper diagram shows also an expectation for the prompt photon yield which is labeled by ``QCD''.} 
%The thermal photon yield from HHG is labeled by ``$Had Mat$'' in the upper diagram and by ``$total_h$'' (dotted line) in the lower diagram. If QGP bremsstrahlung processes are included, the spectra labeled by ``$QM_{Aurenche et al}$'' in the upper diagram and by ``$total_q$'' (solid line) in the lower diagram indicate the thermal photon yield from QGP. If QGP bremsstrahlung processes are neglected, the spectra labeled by ``$QM_{Kapusta et al}$'' in the upper diagram and by ``$1-loop$'' (dashed line) in the lower diagram indicate the thermal photon yield from QGP.
\label{Fig_SPS_Comparison}
\efig
\befig
        \centerline{\psfig{figure=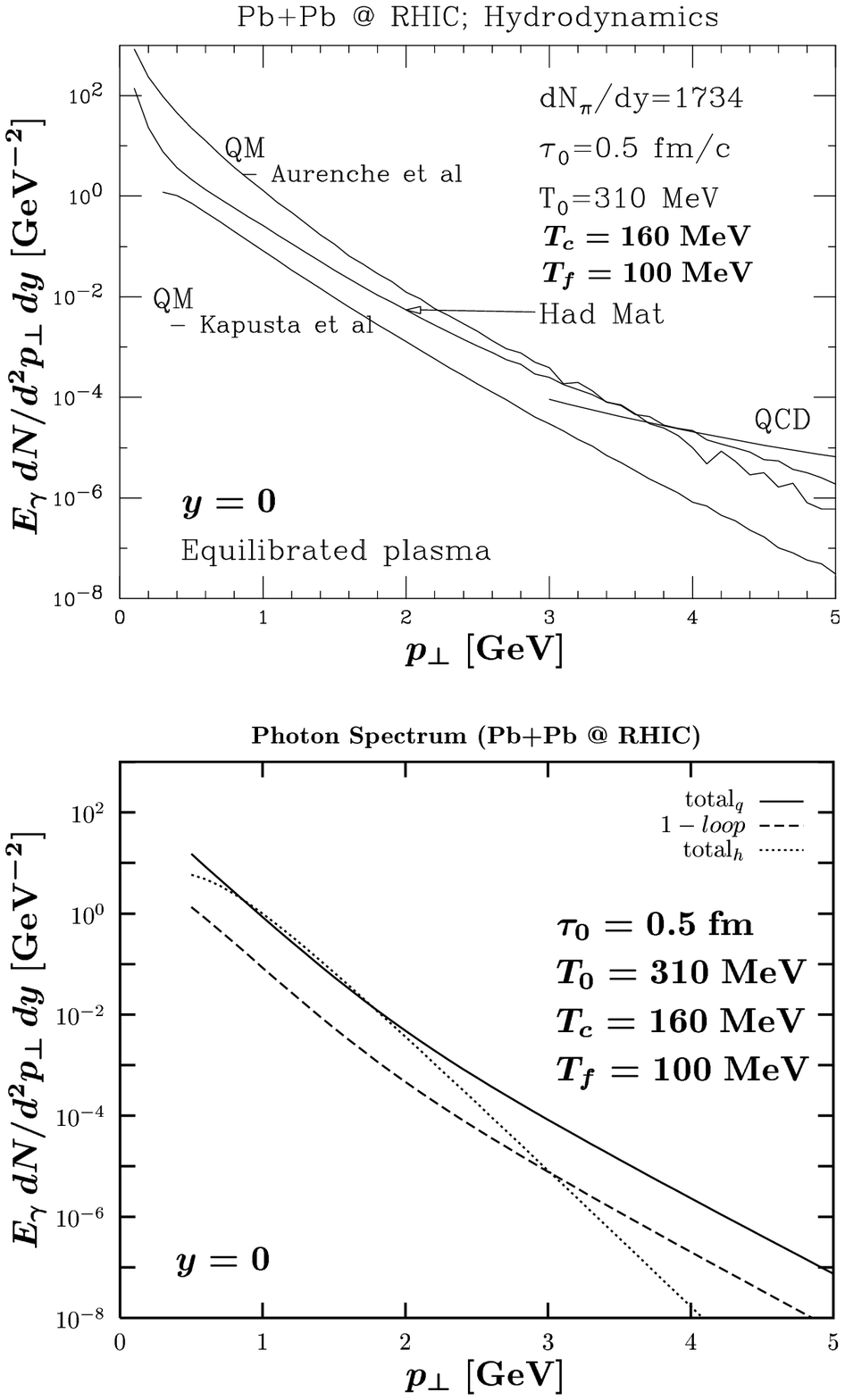,clip=,width=4.in}}
\caption[Thermal Photon Spectra from Central $Pb + Pb$ Collisions at RHIC Energies]{Thermal Photon Spectra from Central $Pb + Pb$ Collisions at RHIC Energies. Same as Fig.~\ref{Fig_SPS_Comparison} but for RHIC energies.}
\label{Fig_RHIC_Comparison}
\efig
\befig
        \centerline{\psfig{figure=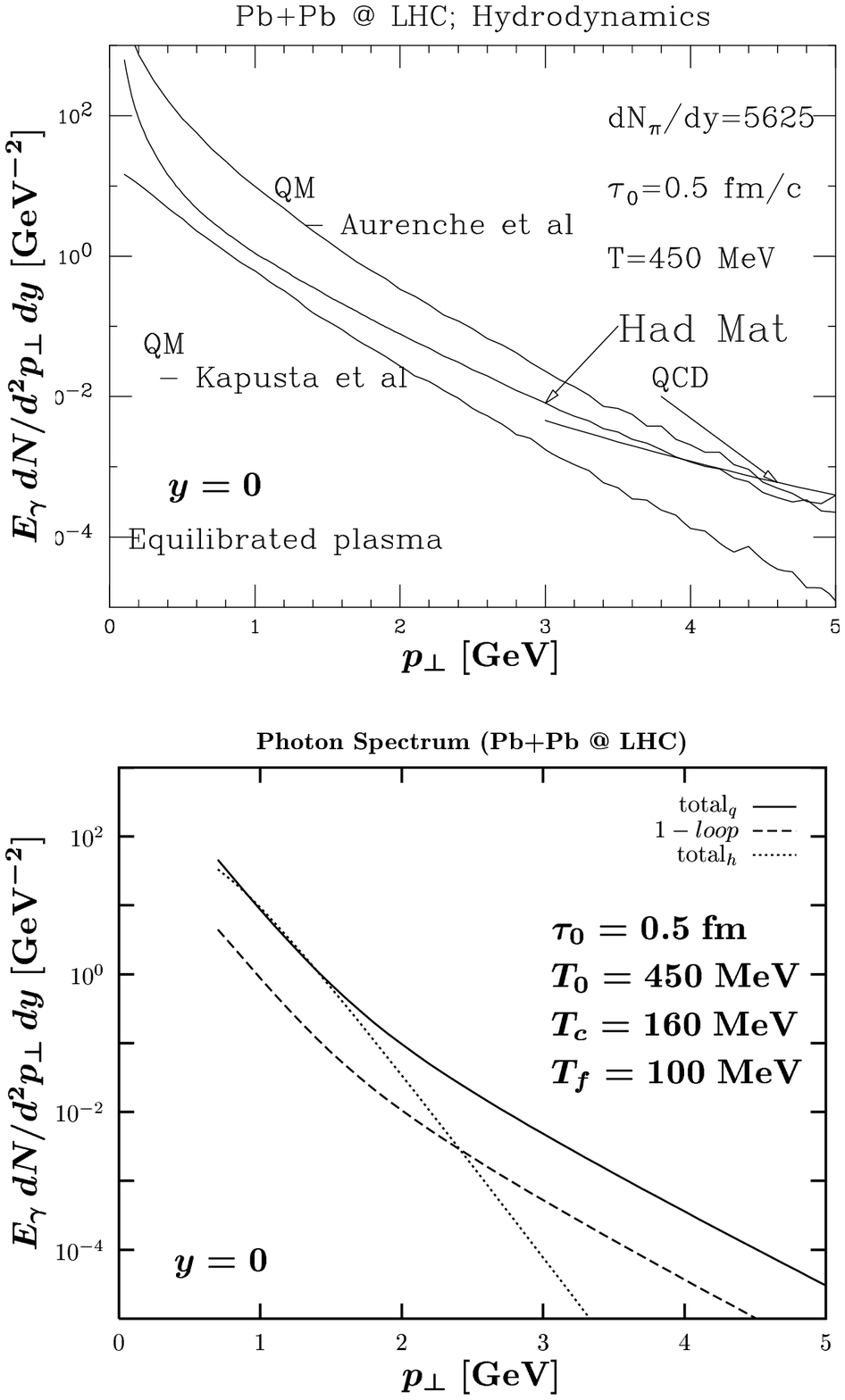,clip=,width=4.in}}
\caption[Thermal Photon Spectra from Central $Pb + Pb$ Collisions at LHC Energies]{Thermal Photon Spectra from Central $Pb + Pb$ Collisions at LHC Energies. Same as Fig.~\ref{Fig_SPS_Comparison} but for LHC energies.}
\label{Fig_LHC_Comparison}
\efig

There are many other studies of thermal photon production in ultra-relativistic heavy ion collisions besides the two cited above. However, they were performed without taking the QGP bremsstrahlung processes into account. In~\cite{ALAM_1996} and~\cite{CLEYMANS_1997}, the 2+1 hydrodynamical model based on the SHASTA algorithm was used to study the transverse expansion and the effects of the different HHG EOS's on the thermal photon spectra, where the initial conditions were obtained as in~\cite{SRIVASTAVA_1999,SRIVASTAVA_SINHA_1999}. A different approach to the initial conditions was followed in~\cite{SOLLFRANK_1997}: Sollfrank et al. extracted also hadron spectra and fitted them with the choice of the initial conditions to the experimentally measured spectra. Then, these initial conditions were used in the calculation of the thermal photon spectra. In addition, the influence of different HHG EOS's including finite baryon density was checked. The underlying model of this study was again the 2+1 hydrodynamical calculation performed with the SHASTA algorithm. An alternative algorithm which solves 3+1 hydrodynamical equations with the EOS is HYLANDER~\cite{HYLANDER_1990} that was applied in~\cite{ARBEX_1995}. Within this simulation, no initial conditions need to be specified because the initial compression (or collision) stage is also modeled. Also in three-fluid hydrodynamical models, the compression stage is simulated and no initial conditions must be specified. Three-fluid hydrodynamics describes both of the colliding nuclei and the central region of secondary particles with separate fluids. In this spirit, thermal photons have been used as a measure for the rapidity dependence of the temperature~\cite{DUMITRU_1995}.

Most of the above investigations have been performed for the experiments at the CERN SPS, WA80 and WA98, in which the assumption of a chemically equilibrated plasma should be a good approximation, while the assumption of a vanishing baryon density should be only a crude approximation. A complementary picture seems to arise for RHIC and LHC, where one should face a central region of zero baryon density but no chemically equilibrated plasma. Therefore, methods have been developed to describe a chemically non-equilibrated QGP~\cite{BIRO_1993}, which have already been applied in extractions of thermal photon spectra at RHIC and LHC~\cite{STRICKLAND_1994,TRAXLER_1996,SRIVASTAVA_1997}. An alternative equilibration scenario is the one of the increasingly strongly interacting parton plasma (ISIPP) in which photon production has also been studied~\cite{WONG_1998}.

%The above studies give hints on interesting physics that should be covered in a future extension of this systematic investigation. ... SPS data ... more realistic HHG EOS ... finite baryon density ... more realistic QGP EOS ... RHIC and LHC data ... chemically non-equilibrated QGP ... dilepton rates ... hadron rates ...

\cleardoublepage
\chapter{Conclusion}
\label{Conclusion}
A systematic investigation of hard thermal photon spectra from ultra-relativistic heavy ion reactions was presented with emphasis on the effects of bremsstrahlung processes in the quark-gluon plasma (QGP). 
The thermal photon spectra were computed from thermal photon production rates within a simple, well understood model for central ultra-relativistic heavy ion collisions in the transparent energy region. This model, which is based on the Bjorken model, describes the fireball evolution in one-fluid 1+1 hydrodynamics as an only longitudinally expanding tube in which entropy is conserved (adiabatic expansion). With the implemented equation of state (EOS) that is an ingredient in every hydrodynamic calculation, a phase transition and a no phase transition scenario were simulated for the central region with vanishing baryon density. In the phase transition scenario, a first-order phase transition from QGP to hot-hadronic gas (HHG) was implemented, where the QGP and the HHG were modeled respectively by an ideal massless parton gas and an ideal massless pion gas. The matching at the critical boundary was achieved with the Gibbs criteria and Maxwell construction. The parameters that determined the evolution in this scenario were the thermalization time~$\tau_0$, the initial temperature~$T_0$, the transition temperature~$T_c$, and the freeze-out temperature~$T_f$. In the no phase transition scenario, only the HHG state of matter was assumed and modeled using the ideal massless pion gas. Here, the parameters~$\tau_0$, $T_0$, and~$T_f$ described the fireball evolution completely. 
Two additional parameters had to be specified in the extraction of the thermal photon spectra, the mass number $A$ of the projectile determining the transverse size of the fireball and the projectile rapidity~$y_{nucl}$ determining the longitudinal limits of the fireball. Finally, the thermal photon production rates entered as the most critical input in the computation of the thermal photon spectra. For the QGP, the 1-loop production rates, Compton scattering and $q\bar{q}$-annihilation, and the recently derived 2-loop production rates describing ``ordinary'' bremsstrahlung ($bremss$) and $q\bar{q}$-annihilation with an additional scattering in the medium ($q\bar{q}-aws$) were considered, while for the HHG, the rate describing the processes $\pi\rho \rightarrow \pi\gamma$ and $\pi\rho \rightarrow a_1 \rightarrow \pi\gamma$ was implemented. Our prime interest here and throughout the complete thesis was on the 2-loop production rates in the QGP that contain the bremsstrahlung processes. In fact, the astonishing result that these rates are in the same order of the coupling constants as the 1-loop rates triggered the research for this thesis. The main issue was then on the impact of these bremsstrahlung processes on the thermal photon spectrum. It was attacked in a systematic investigation, in which the influence of every model parameter was checked. Already the photon spectra at fixed temperatures amplified the interest in the bremsstrahlung processes because the $q\bar{q}-aws$ process appeared as the dominant thermal photon source over the 1-loop processes.

A first look at the thermal photon spectra additionally affirmed this observation. In the phase transition sample scenario with the parameters set to $\tau_0 = 1\;\fm$, $T_0 = 250\;\MeV$, $T_c = 170\;\MeV$, $T_f = 150\;\MeV$, $y_{nucl} = 8.6$, and $A = 208$, a significant enhancement of the thermal photon yield was found in the high-$p_{\perp}$ region, $p_{\perp} > 3\;\GeV$, which could be traced back to an enhancement of about one order in magnitude over the complete considered $p_{\perp}$-range in the yield from the QGP phase due to the two-loop processes taken into account. This enhancement of the yields from QGP state of matter was found in the subsequent systematic investigation to be independent of specific parameter settings. Thus, the strength of this two-loop effect coincides with the strength of the QGP contribution, which goes up for high initial temperatures~$T_0$ and low transition temperatures~$T_c$. In summary, the following dependences were found by considering each model parameter, mass number $A$ of projectile, projectile rapidity~$y_{nucl}$, thermalization time~$\tau_0$, initial temperature~$T_0$, transition temperature~$T_c$, and freeze-out temperature~$T_f$, separately, with the above phase transition sample scenario serving as the basis for the investigation. The spectra turned out to be proportional to~$A^{2/3}$ and~$\tau_0^2$ and nearly independent of~$y_{nucl}$ and~$T_f$.
%, while only a negligible dependence was found for~$y_{nucl}$ and~$T_f$. 
For~$T_0$ and~$T_c$, the dependence could not be expressed in terms of a simple proportionality. Instead, the different collision phase lifetimes and the computed spectra were closely examined. The strongest dependence of the thermal photon yields with respect to their magnitude and their slope was found for the initial temperature~$T_0$. Because the collision phase lifetimes show basically a cubic dependence on~$T_0$ and the photon spectra show basically a quadratic dependence on~$T$, this is clearly understood. In fact, $T_0$ is the crucial parameter for the total thermal photon yields. The dependence on~$T_c$ is not so significant but also interesting: for low values of~$T_c$, the pure QGP phase is more dominant than for high values of~$T_c$. This was directly observed in the lifetimes of the collision phases, where a low value of $T_c$ supports a relatively long lived QGP phase and relatively short lived mixed and pure HHG phases. An influence on the total spectrum has also been observed: higher values of $T_c$ result in higher thermal photon yields. This behavior could be traced back to a mean temperature that is higher for high values of~$T_c$ than for low values of~$T_c$. 

The comparison of the thermal photon spectra from the phase transition scenario with those from the no phase transition scenario, which should exhibit the quality of thermal photons as a potential signature for the QGP formation, was only briefly considered since the ideal massless pion gas was found inappropriate to describe the purely hadronic scenario. Thus, this important comparison is recommended for a future extension of this work, in which a more realistic EOS should be used to describe the HHG state of matter.

The look at experimental data was unfortunately limited by the availability of direct photon data. So far, only upper limits for the direct photon production have been extracted from fixed target $200\;A\cdot\GeV$ $S + Au$ data at the CERN SPS. They were used to identify maximum values for the initial temperature, where the value found with bremsstrahlung processes was 15 MeV (for high $T_c$'s) to 20 MeV (for low $T_c$'s) below the value found without bremsstrahlung processes. These are no severe restrictions on the model parameters, but they illustrate the effect of the bremsstrahlung processes in terms of a difference in the initial temperature. For more severe restrictions on the parameters, one must wait for the WA98, PHENIX, and ALICE data which should be more explicit because of the advanced experimental instrumentation. 

While waiting for the WA98 results on direct photon production, other theoretical works were considered, where the comparison with the investigation of Srivastava and Sinha was particularly informative. In parallel to this thesis, Srivastava and Sinha inspected the 2-loop bremsstrahlung processes in a hydrodynamical model that describes also transverse expansion (2+1 hydrodynamics) and implements a more realistic HHG EOS. Interestingly, their results for the phase transition scenario corresponding to the $200\;A\cdot\GeV$ $S + Au$ collisions at the CERN SPS were almost identical with ours. This means that in the considered scenario, the simple 1+1 hydrodynamical model is as competent as the more sophisticated 2+1 hydrodynamical model. For different scenarios describing $Pb + Pb$ collisions at SPS, RHIC, and LHC energies, another recent study~\cite{SRIVASTAVA_1999} was found on the preprint server that used again the 2+1 hydrodynamical model with the more realistic HHG EOS to explore the effects of the bremsstrahlung processes. In comparison to this study, some significant differences were observed especially for the HHG photon contribution in the high-$p_{\perp}$ region, which were traced back to the different transverse behavior. These differences indicate that the domain of the late collision stage in a long lived fireball is sensible to transverse behavior. Other theoretical investigation, which did not include the QGP bremsstrahlung processes, were also briefly reviewed. They provide hints on further interesting physics aspects that should be examined systematically in a future extension of this investigation.
%At RHIC and LHC energies, only comparisons with other works have been performed. Many of these models go beyond the simple model employed in this thesis, as also transverse expansion and more realistic EOS for the HHG are considered. However, the thermal photon production rates for bremsstrahlung processes were not considered. However, the other investigations exhibit that, e.g., transverse expansion, is very important. Therefore, we propose a step-by-step extension of this systematic investigation with close comparisons to other investigations and close comparisons with the WA98 data. 

In prospect of the upcoming experimental data, we strongly recommend a step-by-step extension of this investigation, in which the very systematical approach should be kept as the first principle. First, more realistic EOS's for the QGP and, as already mentioned, for the HHG state of matter should be implemented. For the QGP, one could use parameterizations of lattice QCD data or take into account effective parton masses~\cite{PESHIER_1994,PESHIER_1996}. For the HHG, finite hadron masses and a higher number of hadrons should be included. This will allow the important comparison of the results from the phase transition scenario with those from the no phase transition scenario, which has been beyond the scope of this thesis due to the inappropriate HHG EOS of the ideal massless pion gas. Second, the EOS and also the photon production rates should be extended to the case of finite baryon density. In view of the expected WA98 data for $Pb + Pb$ collisions at the CERN SPS, it will be important to understand the influence of finite baryon density.  For RHIC and LHC, finite baryon density calculations will not be necessary since the degree of stopping is predicted much weaker than for SPS, but on the other hand, chemical equilibrium will not be fulfilled. Thus, the third step should be the extension of the hydrodynamical model and the photon production rates to the case of chemical non-equilibrium, which could be done by using fugacities. When a solid understanding of the above phenomena and their effects on the thermal photon spectra has been collected, we propose to include transverse expansion as the final step, where one could use the SHASTA algorithm (2+1), the HYLANDER algorithm (3+1) or develop a different new approach. Once the systematic investigation is driven up to this point for thermal photons, it should be easily transferable to dileptons and hadrons.

With the heavy ion physics program underway at RHIC and LHC, the above study would definitively be of great interest over the next decade and could participate in the potential discovery of the QGP formation at RHIC or LHC. The C++ program developed for this thesis could serve as a solid basis for the above studies since it was written in an object oriented style ready for extension.

%The first step should be the implementation of more realistic EOS's for the QGP and, more important, for the HHG state of matter. For the QGP, one should use parameterizations of lattice QCD data or take into account effective parton masses. For the HHG, many more hadrons with finite masses should be included in the derivation of the EOS. This will then allow the important comparison of the phase transition scenario with the no phase transition scenario, which was stopped in this thesis due to the inappropriate HHG EOS of the ideal massless pion gas. For SPS energies, one should then also consider finite baryon density, which would also postulate an extension of the rates to the case of finite baryon density. For RHIC and LHC, finite baryon density calculations will not be necessary since the degree of stopping is predicted much weaker than for SPS. Instead, chemical equilibrium will not be fulfilled. Thus, the model should be extended to chemical non-equilibrium which could be done by using fugacities. Not only the model, also the rates will need to be modified under the assumption of chemical equilibrium. When a solid understanding of the above phenomena and their influences have been collected, we would recommend to include also transverse expansion, where one could use the SHASTA algorithm (2+1) or the HYLANDER algorithm (3+1) or develop a different new approach. Once the systematic investigation is driven up to this point for thermal photons, it should be easily transferable to dileptons and hadrons.

\cleardoublepage
%
%
% ___ Bibliography __________________________________________________
%
%

\cleardoublepage
%
% ___ Acknowledgements ______________________________________________
%
\pagestyle{empty}
%
%
% acknowledgments.tex
%
\begin{Large}
\begin{center}
{\bf ACKNOWLEDGMENTS}
\end{center}
\end{Large}
Foremost, I thank Prof. U. Mosel, the head of the Institute f\"ur Theoretische Physik I, for the friendly admission to the institute and for giving me the opportunity to participate in the search for the quark-gluon plasma. I also appreciated very much his lecture ``Introduction to Path Integrals in Field Theory'' that turned out as an almost private course on fascinating physics.

Next, I want to thank Dr.~Markus Thoma for the instructive research topic and his friendly supervision. Whenever I had a question, he helped in a calm and patient manner. Further, he earns an extra merit because through his addiction to Thai food, I enjoyed some delicious meals and experienced challenging hot moments.

While writing this thesis, I enjoyed a very close and fruitful collaboration with Dr. Munshi Mustafa, who is not only an excellent teacher but now also a good friend. I thank him very much for the interesting discussions on heavy ion physics. The suggestions he made after reading the preliminary versions of this work were extremely valuable.

I am also grateful to Dr.~Stefan Leupold, Frank Hofmann, Gunnar Martens, and Dr.~Christoph Traxler for their interest in questions that came up during the research for this thesis. Their readiness to help in mathematical problems and in mastering the art of C++ programming was very helpful.

Dr. Klaus Schertler deserves a special thank since he showed me how to backup my files just two hours before I accidentally deleted the complete code for this thesis. Without his advice and help in recovering the files, I would have lost two months of work.

Further, I am thankful to Prof. H. Lenske since he helped me out with his copy card and his transparency markers in the most critical moments. He thus broke the laws of Murphy several times. In addition, it was always pleasant not to be the only person in the institute at two o'clock in the morning. 

Prof. W. Cassing should be acknowledged for his insightful questions also during my talk in the Friday afternoon entertainment session. 

Elke Jung earns gratitude for her administrative support. It was a pleasure to associate with her since she has always been very cheerful and helpful. The institute is a friendlier place because of her.

I also thank the other members of the institute and especially the crew of the fifth floor for providing an entertaining and pleasant atmosphere. 

Zhe Xu should be mentioned separately. I thank him for being a very special friend and a loyal companion during the ten semesters I spent in Giessen. His engagement in the final project for Numerical Mathematics II is just one example for the many things that I appreciated very much.

Now, I come to the world outside of the institute. The climate in the Unterhof WG was really fun and helped a lot in relaxing from physics. I am especially grateful to Anja Richmond for her proofreading of this thesis and to Anta\c{c} S\"uerkan for inviting me to TV total. Janine Fritz is thanked for the many beautiful letters that I just could not answer appropriately because of this work. My gratitude also to Prof.~Bill Reay for all the encouraging and motivating e-mails. Because of him, I tend to look at things in a different way.

My sincere thanks to the Friedrich-Naumann-Stiftung for supporting me since April 1996. Visiting their seminars and meeting other fellows was enriching and an ideal complement to the study of physics. Here, my regards to Prof. D. Manz for the many interesting and pleasant discussions.

In a very special way, I want to thank my parents since they always provided me with care, love, and advice. Marlis Kunert deserves also a special thank for her warm-hearted friendship and hospitality. 

Finally, I am extremely thankful to Natascha Kunert for being the sparkling star in my life. I thank her very, very much for all the wonderful moments, her advice, and especially her love. I dedicate this thesis to her.

\end{document}